\documentclass[a4paper,11pt]{article}
\pdfoutput=1
\usepackage{jheppub}
\usepackage{amssymb}
\usepackage{amsmath}
\usepackage{mathtools}
\usepackage{amsfonts}
\usepackage{dsfont}
\usepackage{young}
\usepackage[vcentermath]{youngtab}
\usepackage{bm}
\usepackage{braket}
\usepackage{simplewick}
\usepackage[offset=1.25em]{simpler-wick}
\usepackage{bm}
\usepackage[most]{tcolorbox}

\tcbset{colback=yellow!10!white, colframe=red!50!black,
	highlight math style= {enhanced, 
		colframe=red,colback=red!10!white,boxsep=0pt}
}

\usepackage{mathrsfs}
\usepackage{tikz}
\usetikzlibrary{shapes}
\usetikzlibrary{arrows.meta}
\usetikzlibrary{positioning}
\usetikzlibrary{positioning}
\usetikzlibrary{decorations.markings}
\usetikzlibrary{decorations.pathmorphing}

\usepackage{comment}

\numberwithin{equation}{section}

\DeclareMathAlphabet{\mathpzc}{OT1}{pzc}{m}{it}

\newcommand{\mrm}[1]{\mathrm{#1}}
\newcommand{\mcl}[1]{\mathcal{#1}}
\newcommand{\mbb}[1]{\mathbb{#1}}

\newcommand{\mfr}[1]{\mathfrak{#1}}
\newcommand{\mscr}[1]{\mathscr{#1}}
\newcommand{\msf}[1]{\mathsf{#1}}

\newcommand{\wt}[1]{\tilde{#1}}
\newcommand{\wh}[1]{\widehat{#1}}

\newcommand{\cint}[1]{\underset{#1}{\oint}}
\newcommand{\llangle}{\left\langle}
\newcommand{\rrangle}{\right\rangle}

\newcommand{\be}{\begin{equation}}
	\newcommand{\ee}{\end{equation}}

\title{Beyond the Tensionless Limit: \\ \hspace*{1.6cm} Integrability in the Symmetric Orbifold}
\author{Matthias R.\ Gaberdiel$^a$, Rajesh Gopakumar$^b$, and Beat Nairz$^a$}
\affiliation{$^a$ Institut f\"ur Theoretische Physik, ETH Zurich, \\
	\hspace*{0.3cm}Wolfgang Pauli Strasse 27, CH-8093 Z\"urich, Switzerland}
\affiliation{$^b$ International Centre for Theoretical Sciences-TIFR, \\
	\hspace*{0.3cm}Shivakote, Hesaraghatta Hobli, \\
	\hspace*{0.3cm}Bengaluru North, India 560 089}
\emailAdd{gaberdiel@itp.phys.ethz.ch, rajesh.gopakumar@icts.res.in, nairzb@student.ethz.ch}

\allowdisplaybreaks[2]

\author{}
\abstract{The symmetric orbifold of $\mathbb{T}^4$ is exactly dual to string theory on ${\rm AdS}_3\times {\rm S}^3 \times \mathbb{T}^4$ \newline with minimal ($k=1$) NS-NS flux. In this paper we study the perturbation of the symmetric orbifold that is dual to switching on R-R flux, and hence to deforming the theory away from the tensionless point. More specifically, we determine systematically the action of a centrally extended supersymmetry algebra on the CFT states, and deduce from it the anomalous conformal dimensions. In the $w$-twisted sector with large $w$ the structure is similar to what was found for ${\cal N}=4$ SYM: the basic excitations are multi-magnons whose individual dispersion relation is fixed by symmetry, and the comparison with the BMN answer suggests that the result is true to all orders in perturbation theory. Finally we show that the multi-magnon states interact via an integrable $S$-matrix and possess a natural family of bound states.}
\emailAdd{}

\begin{document}

\maketitle

\section{Introduction}\label{sec:introduction}

In the parameter space of gauge-string dualities \cite{Maldacena:1997re}, there are two natural points around which one can make a controlled expansion. These correspond to either the free ($\lambda=0$) gauge theory (the small radius or tensionless limit of the dual string theory) or its ultra-strong coupling ($\lambda \rightarrow \infty$) limit (weakly curved AdS spacetime). The latter gives rise to the conventional  $\alpha'$ expansion of the string theory which translates into a novel strong coupling expansion in the field theory. The former, on the other hand, is the starting point for the familiar perturbative QFT expansion. This is, however, relatively unexplored from the bulk point of view, when viewed as an expansion around $\alpha'=\infty$. It is likely that we will learn about the fundamental nature of string theory through such an expansion. For instance, the tensionless point has a large unbroken higher spin symmetry and is special in many other ways \cite{Vasiliev:1989re,Sundborg:2000wp,Witten,Klebanov:2002ja}.

A laboratory where we might address this question systematically is in the case of ${\rm AdS}_3/{\rm CFT}_2$. In recent work, the dual to the free symmetric orbifold CFT $\text{Sym}^N\bigl(\mbb{T}^4\bigr)$ was identified as string theory on $\text{AdS}_3\times {\rm S}^3\times \mbb{T}^4$ with one unit of NS-NS flux \cite{Gaberdiel:2018rqv,Eberhardt:2018ouy}. Furthermore, it was possible to make progress in deriving the duality by making manifest that the correlators on both sides agree \cite{Eberhardt:2019ywk,Dei:2020zui,Gaberdiel:2020ycd}.\footnote{This was recently further elaborated on in \cite{McStay:2023thk,Dei:2023ivl,Knighton:2023mhq}.} This agreement relied upon the Lunin-Mathur idea \cite{Lunin:2000yv,Lunin:2001pw} of computing symmetric orbifold correlators by lifting them to a covering space which was then identified with the worldsheet. It turns out that worldsheet correlators at $k=1$ localise precisely on those Riemann surfaces which admit covering maps to the boundary ${\rm S}^2$ with specified branching data \cite{Eberhardt:2019ywk,Dei:2020zui,Gaberdiel:2020ycd, Eberhardt:2020akk, Knighton:2020kuh}.

With this understanding of the free symmetric orbifold CFT and its dual, we can perturb the theory by exactly marginal operators which preserve the ${\cal N}=(4,4)$ supersymmetry. It is known that the space of these operators is twenty dimensional (see e.g.\ \cite{David:2002wn} for a review). Of these, sixteen moduli originate from the untwisted sector of the orbifold theory and are relatively trivial, corresponding to deforming the moduli of the $\mbb{T}^4$.  There are four additional ones coming from the 2-cycle twisted sector which are more interesting. These organise as a triplet and singlet under an internal $\mathfrak{su}(2)$. Here we will focus on the singlet or `blow-up' mode of the orbifold $\Phi,$ given in eq.~(\ref{pertop}) which, as we will see, is physically the most interesting from the bulk point of view as well. The strength of this perturbation denoted by $g$, will play the same role that the 't~Hooft coupling $\lambda$ does in ${ \cal N}=4$ super Yang-Mills.\footnote{This perturbation has also been studied in the context of a topological twisting of the  orbifold CFT \cite{Lerche:2023wkj}. A similar perturbation also plays a role for string theories with more general NS-NS flux \cite{Eberhardt:2019qcl,Eberhardt:2021vsx}, see also the recent discussion in \cite{Hikida:2023jyc}.}

A striking feature of perturbative ${\cal N}=4$ super Yang-Mills is the hidden integrability \cite{Minahan:2002ve} that governs the spectrum (and interactions) of the theory in the planar limit, see e.g.\ \cite{Beisert:2010jr} for a review. In the case of large operators, the spectrum is particularly simple to characterise. They are given by (1+1) dimensional multi magnon excitations around a BPS vacuum state. The energy, i.e.\ conformal dimension, of a single magnon is given by an all order in $\lambda$ dispersion relation
\be\label{4dmag}
E_1(p) = \sqrt{1+\frac{\lambda}{\pi^2}\sin^2{\pi p}} \ ,
\ee
where the states are characterised by their 1d `momentum' $p$.

The individual magnons interact with each other via a 2-body $S$-matrix. Integrability then implies that multi magnons have energies given by the sum of the individual magnon energies of eq.~(\ref{4dmag}), and their interactions are characterised by $S$-matrices which factorise into 2-body scatterings. Furthermore, the $S$-matrix reveals that there are additional bound states of $Q$ multi-magnons which describe asymptotic states with energies \cite{Dorey:2006dq}
\be\label{4dmag-bd}
E_Q(p) = \sqrt{Q^2+\frac{\lambda}{\pi^2}\sin^2{\pi p}} \ .
\ee
There can be further bound states of these excitations, and in turn they generate the full tower of states.
The momentum dependence in eq.~(\ref{4dmag}) is fixed by a centrally extended superalgebra symmetry which acts on the magnons \cite{Beisert:2005tm}. It turns out that it is possible for individual magnons to be in representations of this extended algebra with non-zero central charge, while allowing for multi-magnon states with net zero momentum to have vanishing central charge, as must be the case for any physical state. The $\lambda$ dependence is then rendered plausible by matching with both weak and strong coupling expansions, in particular the supergravity expansion in the BMN limit \cite{Berenstein:2002jq}.
\smallskip

In this paper, we will establish the corresponding results for the symmetric product orbifold CFT when perturbed by the operator $g\, \Phi$. Broadly, we will find a similar picture to the one sketched above, as we will outline shortly. However, the  $\text{AdS}_3\times {\rm S}^3\times \mbb{T}^4$ background exhibits a number of novel features relative to $\text{AdS}_5\times {\rm S}^5$, see e.g.\ \cite{Berenstein:2002jq,Babichenko:2009dk,Hoare:2013lja}:
\begin{itemize}
	\item[(i)] the dispersion relation of the magnons is modified since the perturbed background has both R-R and NS-NS three form flux., c.f.\ eq.~(\ref{2dmag}) below to (\ref{4dmag});
	\item[(ii)] the $\mbb{T}^4$ factor gives rise to gapless excitations;
	\item[(iii)] the unperturbed CFT exhibits separate left- and right-moving magnons, which however start to interact with one another once the perturbation is switched on --- in particular they exhibit a non-trivial relative $S$-matrix.
\end{itemize}

Much of our results rely on a first principles conformal perturbation theory calculation of the anomalous conformal dimension of a single magnon excitation. This is carried out in the $w$'th twisted cycle sector at finite $w$. The calculation is simplified considerably by first looking at how the action of a suitable supercharge on the magnon excitations is affected by the perturbation. The advantage is that this is already non-zero at first order in $g$, and thus one needs to only consider a correlator with three twisted sector operators inserted. This is then sufficient to obtain the dispersion relation to the first non-vanishing (quadratic) order in $g$. The twisted sector correlators are themselves computed using the lift to the covering space where they reduce to free field computations given the covering map, which is easy enough to explicitly write down for a 3-point function. (We will be working in the large $N$ limit, and thus will only consider sphere coverings.)

The finite $w$ results already exhibit quite an intricate structure. In particular, one finds, as expected on general grounds, a non-zero central term in an extended supersymmetry algebra like in the 4d case, though the specific form is, at first sight, quite different. It nevertheless vanishes, as it must, on physical states, i.e.\ those which obey the orbifold invariance condition.

Taking the large $w$ limit of these results is quite subtle. This, in effect, opens out a circle --- representing the $w$-cycle for finite $w$ --- to a straight line for $w=\infty$. In the process one introduces an ordering into the excitations which is crucial like in higher dimensions. Doing this carefully, one finds that the central charges now take on the form found in \cite{Beisert:2005tm} for ${\cal N}=4$ super Yang-Mills. This verifies their momentum dependence by an explicit calculation to leading order in $g^2$. Combined with the general argument for the central charge based on the closure of the supersymmetry algebra allows one to write down the analogue of eq.~(\ref{4dmag})
\be\label{2dmag}
\epsilon_1(p) = \sqrt{(1-p)^2+4g^2\sin^2{\pi p}} \ .
\ee
As mentioned, the difference in the first term in the square root arises from the presence of NS-NS three-form flux in the unperturbed theory, see (i) above. In fact, a comparison with the dispersion relation in the BMN limit tells us that eq.~(\ref{2dmag}) corresponds to a background with $k= q_{NS}=1$ unit of NS-NS flux, and $q_{R}$ units of R-R-flux,  where $2\pi g=g_sq_{R}$, see \cite{Berenstein:2002jq,Hoare:2013lja}. This also makes precise the folklore that the $\Phi$ perturbation corresponds to the radius or RR-flux deformation in the bulk. Quite surprisingly, though, we find that the fractional torus modes exhibit the dispersion relation associated to the  \emph{massive} excitations associated to ${\rm AdS}_3 \times {\rm S}^3$.\footnote{We have also performed a similar analysis for the fractional ${\cal N}=4$ generators that were originally proposed to be dual to the massive ${\rm AdS}_3 \times {\rm S}^3$ excitations, see e.g.\ \cite{Lunin:2002fw,Gomis:2002qi,Gava:2002xb}, and they \emph{also} seem to have the same dispersion relation.\label{foot1}}

We also see, at infinite $w$, that the action of the supercharges on multi-magnon states only permutes their momenta. Thus the anomalous dimensions of multi-magnon excitations is simply the sum of the individual terms given in eq.~(\ref{2dmag}). These facts suggest an underlying integrability. Postulating, as in \cite{Beisert:2005tm}, that there exists a 2-body $S$-matrix which commutes with the supersymmetry generators leads to an overdetermined problem, which however has a solution that is unique up to an overall phase factor. The resulting $S$-matrix satisfies the necessary conditions to be the $S$-matrix of an integrable system: in particular, it obeys the Yang-Baxter equation and is unitary (for a suitable choice of phase factors).  The integrability we find meshes nicely with the fact that the bulk theory is believed to be integrable \cite{Babichenko:2009dk}, see also \cite{Borsato:2013qpa,Lloyd:2014bsa} for subsequent work.\footnote{After this paper was posted on the {\tt arXiv}, \cite{Frolov:2023pjw} appeared in which it is argued that our $S$-matrix agrees in fact with that given in \cite{Lloyd:2014bsa}.}

Given the $S$-matrix, one can examine its poles for additional asymptotic (bound) states. One finds a set of $Q$-magnon bound states with the simple dispersion relation
\be\label{2dmag-bd}
\epsilon_Q(p) = \sqrt{(Q-p)^2+4g^2\sin^2{\pi p}} \ ,
\ee
which, in some sense, fill out the different Brillouin zones in momentum space. We also note that the excitations for $p\in \mathbb{Z}$ do not acquire an anomalous conformal dimension, and are therefore potentially dual to the `massless' $\mathbb{T}^4$ excitations --- in fact, they are in natural one-to-one correspondence with the massless BMN modes associated to the $\mathbb{T}^4$, see (ii) above, and satisfy exactly the same commutation relations.

Our results therefore constitute good evidence that the perturbation of the symmetric orbifold is indeed integrable, and that it fits nicely into what was expected based on general grounds. Previous attempts at uncovering the integrable structure of the symmetric orbifold include \cite{David:2008yk,Pakman:2009zz,Pakman:2009mi}, and our approach follows in spirit what they had attempted. However, as far as we can tell, the details are somewhat different.

One of the very suggestive aspects of our explicit perturbative analysis is that it is done by lifting the spacetime CFT correlator to the covering space which is to be identified with the worldsheet of the dual string theory \cite{Eberhardt:2019ywk}. Thus what we are actually evaluating is, in a concrete sense, a worldsheet correlator. Moreover, as we saw, this corresponds to turning on R-R flux.\footnote{It may be worth pointing out that for $k=1$ the perturbing field arises on the worldsheet from the $w=2$ spectrally flowed sector, and hence does not agree with the usual R-R perturbation for generic NS-NS level $k$ \cite{Berkovits:1999im}.} We expect that a close analysis of these putative worldsheet correlators will teach us about perturbing away from the tensionless limit in the dual string theory --- lessons which we expect to prove useful in general \cite{Cho:2018nfn}, and also more specifically for the $\text{AdS}_5/\text{CFT}_4$ duality \cite{Gaberdiel:2021qbb,Gaberdiel:2021jrv}.
\smallskip

The paper is organised as follows. The basic structure of the symmetric orbifold theory and the relevant perturbation is reviewed in Section~\ref{sec:orbifoldCFT}. The explicit first order perturbation calculation is explained in detail in Section~\ref{strategy}, and the exact finite $w$ results are derived in eqs.~(\ref{eq:ferm-bos_contraction}), (\ref{eq:bos-bos_contraction}), and (\ref{eq:ferm-ferm_contraction}). In Section~\ref{sec:finitew} we confirm that these explicit results indeed lead to a vanishing central charge on physical states, and we determine the anomalous conformal dimensions (to second order in preturbation theory) for some simple two-magnon states. Section~\ref{sec:effective_description} explains how these results simplify in the large $w$ limit, and in particular, derive the large $w$ limits of the above equations, see eq.~(\ref{5.11}) and (\ref{boscon}), respectively. We also generalise the analysis to the full supersymmetry algebra (and the full set of magnon operators) there --- in Section~\ref{strategy} we had only studied a suitable subsector. This enables us to obtain the exact form of the dispersion relation in the large $w$ limit. In Section~\ref{sec:BMN} we relate our results to the BMN limit, and use this to argue that the dispersion relation of eq.~(\ref{epsp}) is actually correct to all orders in the perturbative expansion. Section~\ref{sec:s-matrix} determines the $S$-matrix of our system, shows that it satisfies the Yang-Baxter equation, and determines the bound states. Our conclusions are given in Section~\ref{sec:conclusion}, and there are six appendices where some technical results are described.

\section{The symmetric product orbifold}\label{sec:orbifoldCFT}

In this section we summarise the relevant aspects of the symmetric product orbifold CFT of $\mbb{T}^4$, assembling all the ingredients needed for the perturbative calculation in the next section. Thus, we discuss the magnon like excitations that arise from the fractional modes of the $\mbb{T}^4$ theory, the covering map approach to calculating correlators \cite{Lunin:2000yv}, the BPS vacuum around which we expand \cite{Lunin:2001pw}, and the perturbation dual to turning on the R-R flux through $\text{AdS}_3\times {\rm S}^3$ \cite{Gaberdiel:2015uca,Fiset:2022erp,Apolo:2022fya,Benjamin:2021zkn}.

\subsection{General orbifolds and fractional modes}\label{sec:orbifold_introduction}

Suppose $M$ describes a `seed' CFT --- in our case $M$ will be associated to $\mbb{T}^4$, i.e.\ it will be a theory of four free bosons and fermions. The symmetric product orbifold CFT of the seed theory is then defined to be
\begin{equation}
\text{Sym}^N\bigl(M\bigr) = M^N/S_N\ ,
\end{equation}
where $S_N$ acts by permuting the copies or `colours' of $M$.

As is familiar with orbifold theories, the theory consists of an untwisted sector that comprises the orbifold invariant states of the product theory. In addition, we have twisted sectors associated to the conjugacy classes of the orbifold group $S_N$. The different conjugacy classes are labelled by cycle shapes, and the sectors that will be important for us are the single-cycle sectors that consists of those permutations that are conjugate to a cyclic permutation of $w<N$ colours; a representative element of this conjugacy class is thus the permutation $\sigma_w = (12\cdots w)$. In the sector twisted by $\sigma_w$, the fields  $S^{(i)}$, associated to the $i$'th copy of $M$, are mapped into one another as one goes around the corresponding twist field $\phi_{\sigma_w}(x)$, i.e.\ the fields satisfy the boundary conditions
\begin{equation}\label{twist}
S^{(i)}\bigl(e^{2\pi i }x+x_0\bigr)\,\phi_{\sigma_w}(x_0) = S^{(i+1)}\bigl(x+x_0\bigr)\,\phi_{\sigma_w}(x_0)\ , \qquad i\in\{1,2,\ldots, w\} \ ,
\end{equation}
where $e^{2\pi i}x+x_0$ is short-hand for a path encircling $x_0$ once, and we identify $S^{(w+1)}\equiv S^{(1)}$. (The fields associated to the other copies $S^{(j)}$ with $j>w$ are unaffected by the twist field, i.e.\ are periodic when taken around $\phi_{\sigma_w}(x_0)$.)

In order to find good mode expansions for these fields, it is therefore convenient to consider the linear combinations
\begin{equation}\label{Sk}
\mcl{S}^{[k]}(x) = \frac{1}{\sqrt{w}}\, \sum_{j=1}^{w} e^{-2\pi i\frac{k}{w}j}\,S^{(j)}(x)\ ; \qquad (k=0,1\ldots , (w-1)) ,
\end{equation}
which satisfy
\begin{equation}
\mcl{S}^{[k]}\bigl(e^{2\pi i}x+x_0\bigr)\,\phi_{\sigma_w}(x_0) = e^{2\pi i \frac{k}{w}}\,\mcl{S}^{[k]}\bigl(x+x_0\bigr)\,\phi_{\sigma_w}(x_0)\ .
\end{equation}
This means that $(x-x_0)^{-\frac{k}{w}+n}\,\mcl{S}^{[k]}(x)$, where $n\in\mathbb{Z}$, is single-valued in the presence of the twist, and we can define the fractional modes
\begin{equation}\label{eq:general_fractional_mode}
\mcl{S}_{-\frac{k}{w} - h +n}\, \phi_{\sigma_w}(x_0) = \cint{C(x_0)}\!\!dx\,(x-x_0)^{-\frac{k}{w}+n-1}\,\mcl{S}^{[k]}(x)\,\phi_{\sigma_w}(x_0)\ .
\end{equation}
Here, $h$ is the conformal dimension of $S^{(i)}$, and $C(x_0)$ denotes a small circle around $x_0$. Note that we have dropped the label $k$ on the left-hand-side, and it is indeed convenient to combine all of these modes (for all the different values of $k$) into one set of modes with mode numbers that differ from $-h$ by a $w$-fractional number. These modes are usually referred to as the `fractional' modes of the symmetric orbifold theory. In our case, we will primarily concentrate on the fractional modes associated with the torus bosons and fermions.

\subsection{Correlators and covering maps}

A convenient method for the calculation of correlators of twist fields is by  means of covering maps \cite{Lunin:2000yv}. The basic idea is to lift the correlator to the covering surface in order to undo the twists described by (\ref{twist}). We shall always take the symmetric orbifold to be defined on the Riemann sphere $\mbb{CP}^1$, and we shall denote the coordinates there by $x_i$. On the other hand, the coordinates on the covering surface $\Sigma$ (which can have higher genus, although we shall concentrate on the case when it is also a sphere --- this describes the leading contributions at large $N$) will be denoted by $z_i$. For a correlator of $n$ single-cycle twist fields with twist $w_i$ at positions $x_i$, $(i=1,\dots,n)$, a (branched) covering map is then a meromorphic map\footnote{We use a bold $\bm{\Gamma}$ to distinguish the covering map from the Gamma function which will appear in many expressions.}
\begin{equation}
\bm{\Gamma}\,:\,\,\Sigma\longrightarrow \mbb{CP}^1\ ;\qquad  z\longmapsto\bm{\Gamma}(z)\ ,
\end{equation}
such that $\bm{\Gamma}(z_i)=x_i$ and
\begin{equation}\label{covering}
\bm{\Gamma}(z)=x_i+a_i\,\bigl(z-z_i\bigr)^{w_i}+\mcl{O}\bigl((z-z_i)^{w_i+1}\bigr)\ ,\qquad z\to z_i\ .
\end{equation}
Furthermore, $\partial\bm{\Gamma}$ is non-vanishing except at the punctures $z_i$. Intuitively, this means that locally around a twist field insertion, $\Sigma$ looks like a $w_i$-fold cover that undoes the twist in the sense that as $z$ goes around $z_i$ once, $\bm{\Gamma}(z)$  goes around $x_i$ $w_i$-many times, and hence returns the fields to their original configuration. The twisted sector ground state $\phi_{\sigma_w}$ is lifted to the NS vacuum if $w$ is odd (and the R vacuum, i.e.\ a spin field, if $w$ is even \cite{Lunin:2001pw}), and we shall denote the corresponding state by $\hat{\phi}_{\sigma_w}$. One also needs to account for the  conformal anomaly that leads to some $\bm{\Gamma}$ dependent factors multiplying the correlator on the covering surface, see \cite{Lunin:2000yv,Dei:2019iym}. In order to understand the lift of a fractional mode we consider the lift of the field $S^{(i)}$ (for some choice of $i$),
\be
\hat{{S}}(z) = S^{(i)} \bigl( \bm{\Gamma}(z)\bigr) \ .
\ee
By construction, the analytic continuation of the lift of the field $S^{(i)}$ gives rise to all the other copies and combines to give a single-valued field $\hat{S}(z)$ on the covering space. The lift of the fractional descendants becomes
\begin{align}
\mcl{S}_{-\frac{k}{w} - h +n}\phi_{\sigma_w}(x_0) & = \cint{C(x_0)}\!\!dx\,(x-x_0)^{-\frac{k}{w}+n-1}\,\mcl{S}^{[k]}(x) \,\phi_{\sigma_w}(x_0) \\
& = \frac{1}{\sqrt{w}}\, \cint{C(z_0)} \!\! dz\, \bigl( \partial\bm{\Gamma}(z)\bigr)^{1-h} (\bm{\Gamma}(z)-x_0)^{-\frac{k}{w}+n-1}\,\hat{S}(z) \,\hat{\phi}_{\sigma_w}(z_0)
\end{align}
on the covering surface. Since $\bm{\Gamma}(z_0)=x_0$ and $\bm{\Gamma}$ is ramified of degree $w$ at $z_0$, see eq.~(\ref{covering}), the contour integral makes sense. By expanding $\bm{\Gamma}$ in a power series, one sees that the fractional mode will give rise to a sum of ordinary descendants on the covering surface. For some more details on how this lifting works in practice, see for example \cite{Gaberdiel:2022oeu}.

The field $\phi_{\sigma_w}$, associated to the specific permutation $\sigma_w=(1\cdots w)$, is not orbifold invariant, since under conjugation by a group element in $S_N$, the colours that appear inside the cycle typically change. In order to consider orbifold invariant correlators, we need to consider the normalised sum over the whole conjugacy class, see e.g.~\cite{Dei:2019iym}. We denote the orbifold invariant field associated to the $w$-cycle by $\phi_w$, and we will be working with these fields from now on. Then, to calculate a correlator of $\phi_{w_i}$ fields, one has to sum over all covering maps which satisfy the defining conditions.

\subsection{The \texorpdfstring{$\mbb{T}^4$}{T4} orbifold}

We shall be interested in the symmetric orbifold of $\mbb{T}^4$, i.e.\ the theory of four left- and right-moving bosons and fermions. The theory of a single $\mbb{T}^4$ has $\mcl{N}=(4,4)$ superconformal symmetry with central charge $c=6$. We shall denote the left-moving fields by
\begin{equation}
\alpha^i\ ,\qquad \bar{\alpha}^i\ ,\qquad \psi^\pm\ ,\qquad \bar{\psi}^\pm\ ,
\end{equation}
where $\alpha^i$ are bosons with $i\in\{1,2\}$ and $\psi^\pm$ are fermions. For the fermions, the plus-minus index labels their R-symmetry charge $\pm \frac{1}{2}$. These fields are free and in our conventions satisfy the OPEs
\begin{equation}
\bar{\alpha}^i(x)\,\alpha^j(y)\sim \frac{\epsilon^{ij}}{(x-y)^2}\ ,\qquad \bar{\psi}^\pm(x)\,\psi^\mp(y) \sim \frac{\pm 1}{x-y}\ ,
\end{equation}
while all other OPEs are trivial. The right-moving fields will be denoted by a tilde,
\begin{equation}
\wt{\alpha}^i\ ,\qquad \wt{\bar{\alpha}}{}^i\ ,\qquad \wt{\psi}^\pm\ ,\qquad \wt{\bar{\psi}}{}^\pm\ ,
\end{equation}
and behave analogously. The generators of the $\mcl{N}=(4,4)$ superconformal algebra can be expressed in terms of bilinears of these fields, and we have summarised our conventions in Appendix~\ref{app:algebra}. We will, in particular, consider the four supercharges
\begin{equation}
G^\pm\ ,\qquad G'^\pm\ ,
\end{equation}
as well as the corresponding right-movers, see eq.~(\ref{N4fields}). These fields have R-charge $\pm \frac{1}{2}$ as well.

In the planar limit, $N\to\infty$, the symmetric orbifold of $\mbb{T}^4$,
\begin{equation}
\text{Sym}^N\bigl(\mbb{T}^4\bigr) = \bigl(\mbb{T}^4\bigr)^N/S_N
\end{equation}
is dual to string theory on $\text{AdS}_3\times S^3\times \mbb{T}^4$ with $k=1$ units of NS-NS flux and no R-R flux \cite{Eberhardt:2018ouy,Eberhardt:2019ywk,Dei:2020zui}.
Indeed, as was shown in \cite{Eberhardt:2019ywk,Dei:2020zui,Eberhardt:2020akk,Knighton:2020kuh}, the worldsheet plays the role of the covering surface, and thus performing symmetric orbifold calculations on the covering space is de facto the same as working on the worldsheet. In particular, the  orbifold correlators admit a genus expansion in the sense that the contribution of a covering surface of genus $\msf{g}$ scales (to leading order in $N$) as \cite{Lunin:2000yv}
\begin{equation}
N^{1-\msf{g}}\ .
\end{equation}
This then fits with the string worldsheet expansion since the string coupling constant is related to $N$ via \cite{Pakman:2009zz}
\begin{equation}
g_s^2\sim \frac{1}{N}\ .
\end{equation}
In the large $N$ limit\,\footnote{Note that the $N$ here behaves like the {\it square} of the rank of the gauge group in the Yang-Mills theory.}, we will therefore only take covering surfaces with $\msf{g}=0$, into account, i.e.~we consider covering spaces which are themselves a sphere; in this case $\bm{\Gamma}$ is a rational map.

\subsection{The perturbing field}

We are interested in the perturbation of the symmetric orbifold theory by an exactly marginal ${\cal N}=(4,4)$ preserving perturbation. Apart from the obvious exactly marginal deformations that deform the shape of the torus, the interesting perturbations come from the $2$-cycle twisted sector. They will therefore break the symmetric orbifold structure, and thus deform the theory away from the symmetric orbifold point. We shall consider a specific perturbation from the $2$-cycle twisted sector which is characterised by the condition that it is a singlet with respect to the diagonal action of the residual $\mfr{su}(2)$, see e.g.\ \cite{Fiset:2022erp},
\begin{equation}\label{pertop}
\Phi = \frac{i}{\sqrt{2}}\, \bigl(G^-_{-\frac{1}{2}}\wt{G}'^-_{-\frac{1}{2}}-G'^-_{-\frac{1}{2}}\wt{G}^-_{-\frac{1}{2}}\bigr)\ket{\mrm{BPS}_-}_2\ .
\end{equation}
Here, $\ket{\mrm{BPS}_-}_2$ is the BPS state in the 2-twisted sector with dimension and charge $h=\wt{h}=j=\wt{j}=\frac{1}{2}$, see appendix \ref{app:states} for more details. Hence, $\Phi$ has left and right conformal dimension $1$ and R-charge $0$.

The theory is then deformed by adding the term
\begin{equation}
g\,\int d^2y\,\,\Phi(y,\bar{y})
\end{equation}
to the action. Practically, this means that a correlator $\langle O_1\cdots O_n\rangle_g$ in the deformed theory has an expression in terms of a power series in $g$ as
\begin{align}
\left\langle O_1\cdots O_n\right\rangle_g =& \left\langle O_1\cdots O_n\right\rangle_0 +g \, \int d^2y\,\left\langle O_1\cdots O_n \Phi(y,\bar{y})\right\rangle_0 \nonumber\\
&+ \frac{g^2}{2}\,\int d^2y_1 d^2y_2\,\left\langle O_1\cdots O_n\Phi(y_1,\bar{y}_1)\Phi(y_2,\bar{y}_2)\right\rangle_{0,\text{connected}}+\dots\ .
\end{align}
As it turns out, it will be sufficient to study only the first order perturbation of a two-point function, and for this the renormalisation will be essentially trivial. Furthermore, since to first order in perturbation theory we are considering a $3$-point function, the relevant covering map will have a simple form.

\section{The perturbative calculation at finite \texorpdfstring{$w$}{w}}\label{strategy}

In order to understand the perturbation of the symmetric orbifold by $\Phi$ we shall follow a similar strategy as was first used for ${\cal N}=4$ SYM in \cite{Beisert:2005tm}, see also \cite{Gava:2002xb,David:2008yk}. Starting with a suitable reference BPS state in each twisted sector, we shall generate all states by the action of the fractional torus modes, which we interpret as being `magnons'. On these magnon excitations we shall then study the action of those supercharge generators that annihilate the BPS state. This action can be calculated to first order in perturbation theory, and we shall explicitly work this out for finite $w$.

As we shall see in the following sections, the answer simplifies nicely in the large $w$ limit, and this will determine the leading order correction to the anomalous dimension.  Furthermore, the form of the action of the supercharges on arbitrary states is consistent with simple expressions one can write down on general grounds, reflecting an extended supersymmetry algebra. The vanishing of the additional central terms on physical states then fixes the momentum dependence which also matches that from our explicit perturbative calculation. Finally, the self-consistency of the action of the supercharges allows us to deduce from this the anomalous conformal dimensions to {\it all orders} in the perturbation.

\subsection{The \texorpdfstring{$w$}{w}-cycle twisted sector}

Let us start by fixing our conventions in the $w$-cycle twisted sector. It will be convenient to take as our reference BPS state $|{\rm BPS}\rangle_w$ the `top' BPS state with $h=\tilde{h}=j=\tilde{\jmath} = \frac{w+1}{2}$, see Appendix~\ref{app:states} for our conventions. The other states in the $w$-cycle twisted sector are then obtained from this state by the action of the magnon operators, which we define via
\be\label{modes}
\begin{aligned}
& \psi^-( \tfrac{n}{w}) \equiv  \psi^-_{-\frac{1}{2} + \frac{n}{w}} \ , \ \
\psi^+( \tfrac{n}{w}) \equiv \psi^+_{-\frac{3}{2} + \frac{n}{w}} \ ,\qquad
\alpha^i( \tfrac{n}{w}) \equiv \tfrac{1}{\sqrt{ 1-\frac{n}{w}}} \, \alpha^i_{-1 + \frac{n}{w}}  \ \ (i=1,2)  \ ,
\\
& \bar{\psi}^-( \tfrac{n}{w}) \equiv \bar{\psi}^-_{-\frac{1}{2} + \frac{n}{w}} \ , \ \
\bar{\psi}^+( \tfrac{n}{w}) \equiv \bar{\psi}^+_{-\frac{3}{2} + \frac{n}{w}} \ ,\qquad
\bar{\alpha}^i( \tfrac{n}{w}) \equiv \tfrac{1}{\sqrt{1-\frac{n}{w}}} \, \bar{\alpha}^i_{-1 + \frac{n}{w}}  \ \ (i=1,2) \ ,
\end{aligned}
\ee
and similarly for the corresponding right-movers. The normalisations are motivated by the fact that these magnons then become `unit-normalised' since
\be
\|\alpha^i_{-1+\frac{n}{w}}\sigma_w\|^2 =
\llangle(\alpha^i_{-1+\frac{n}{w}}\sigma_w)^\dagger(\infty)\,\alpha^i_{-1+\frac{n}{w}}\sigma_w(0)\rrangle =
1- \tfrac{n}{w} \ ,
\ee
as can be calculated using the covering map $z\mapsto z^w$.

The modes in eq.~(\ref{modes}) have the property that they all increase the eigenvalue of ${\cal C} = L_0 - K^3_0$ by $(1 - \tfrac{n}{w})$, i.e.\
\begin{align}
& [{\cal C},\psi^-(\tfrac{n}{w})] = (1 - \tfrac{n}{w})\, \psi^-(\tfrac{n}{w}) \ , \qquad
[{\cal C},\psi^+(\tfrac{n}{w})] = (1 - \tfrac{n}{w})\, \psi^+(\tfrac{n}{w}) \ , \\
& \qquad \qquad \qquad  [{\cal C},\alpha^i( \tfrac{n}{w})] =(1 - \tfrac{n}{w})\,  \alpha^i( \tfrac{n}{w}) \ ,
\end{align}
and similarly for the barred modes. A similar statement also holds for the right-movers, except that the relevant charge for them is $\tilde{\cal C} = \tilde{L}_0 - \tilde{K}^3_0$. Here the ${\cal C} $ and $\tilde{{\cal C}}$ eigenvalues of the BPS state vanish
\be
{\cal C} \, |{\rm BPS}\rangle_w = \tilde{{\cal C}} \, |{\rm BPS}\rangle_w = 0 \ .
\ee
The supercharge generators of the ${\cal N}=(4,4)$ algebra that annihilate the BPS states consist of
\be\label{N4super}
G^\mp_{\pm \frac{1}{2} } \ , \quad G^{'\mp}_{\pm \frac{1}{2} } \ , \qquad
\tilde{G}^\mp_{\pm \frac{1}{2} } \ , \quad \tilde{G}^{'\mp}_{\pm \frac{1}{2} } \ .
\ee
It will be convenient to study at first\footnote{We shall later generalise the discussion to the full set of supercharge generators in (\ref{N4super}), see Section \ref{sec:fullsup} below.} only the subalgebra that is generated by the modes\footnote{A mnemonic is that the $S$'s are the $-\frac{1}{2}$ raising modes while the $Q$'s are the  $+\frac{1}{2}$ lowering modes, the subscripts $(1,2)$ refer to the unprimed and primed generators respectively.}  \cite{David:2008yk,Hoare:2013lja,Borsato:2013qpa}
\be\label{N4subset}
Q_1 \equiv G^-_{+\frac{1}{2}} \ , \qquad S_1 \equiv G^+_{-\frac{1}{2}} \ , \qquad
\tilde{Q}_2 \equiv \tilde{G}^{'-}_{+\frac{1}{2}} \ , \qquad \tilde{S}_2 \equiv \tilde{G}^{'+}_{-\frac{1}{2}} \ ,
\ee
and which satisfy the algebra relations
\be
\begin{array}{rclrcl}
\{ Q_1, S_1 \} & \equiv & {\cal C} = (L_0 - K^3_0) \ , \quad  &
\{ \tilde{Q}_2, \tilde{S}_2 \} & \equiv & \tilde{\cal C} = (\tilde{L}_0 - \tilde{K}^3_0)\  , \\
\{ Q_1, Q_1 \} & = & \{ S_1, S_1 \} = 0 \ , \quad  & \{ \tilde{Q}_2, \tilde{Q}_2 \} & = & \{ \tilde{S}_2, \tilde{S}_2 \} = 0 \ , \\
\{Q_1,\tilde{Q}_2\} & = & \{Q_1,\tilde{S}_2\}= 0 \ , \quad & \{S_1,\tilde{Q}_2\} & = & \{S_1,\tilde{S}_2\}= 0 \ .
\end{array}
\ee
Furthermore, we can then restrict ourselves to the magnon generators
\be\label{magnonsub}
\psi^-( \tfrac{n}{w}) \ , \ \ \alpha^2( \tfrac{n}{w}) \ , \ \ \tilde{\psi}^-( \tfrac{n}{w}) \ , \ \ \tilde{\alpha}^1( \tfrac{n}{w}) \ .
\ee

\subsection{The physical state condition}

For the following it will be important that not all magnon descendants of the BPS states are indeed present in the symmetric orbifold theory. The reason is that a state needs to be orbifold invariant; in the $w$-cycle twisted sector this is the condition that the state that is obtained by the action of the fractional modes is invariant under the centraliser of $\sigma_w=(1\cdots w)$, which is the cyclic group $\mbb{Z}_w$. In terms of the magnon operators, this condition can be easily formulated: a state $|\Psi \rangle_w$ of the form
\be
|\Psi \rangle_w = \prod_{i=1}^{r} A(\tfrac{n_i}{w}) \prod_{j=1}^{s} \tilde{A}(\tfrac{m_j}{w}) \, |{\rm BPS}\rangle_w \ ,
\ee
where each $A(\frac{n_i}{w}) $ is a mode of the form (\ref{modes}), and similarly for the right-movers $\tilde{A}(\frac{m_j}{w})$, is orbifold invariant provided that
\be\label{momentumcons}
\sum_{i=1}^{r} \frac{n_i}{w} - \sum_{j=1}^{s} \frac{m_j}{w} \in  \mathbb{Z} \ .
\ee
Given the form of (\ref{Sk}), we can think of $\frac{n}{w}$ as the momentum of the corresponding mode, and thus (\ref{momentumcons}) describes `momentum conservation' of the associated magnon states, see also \cite{Gava:2002xb}.

\subsection{The structure of the perturbation}

Let us first calculate how the above supercharge generators act on the single magnon states before perturbation. This just follows from the commutation relations of the free fields, and we find
\be
\begin{array}{llll}
Q_1\, \alpha^2( \tfrac{n}{w}) \, |{\rm BPS}\rangle_w & = \sqrt{1 - \tfrac{n}{w}}\, \psi^-( \tfrac{n}{w}) \, |{\rm BPS}\rangle_w  \qquad \qquad
& \tilde{Q}_2\,  \alpha^2( \tfrac{n}{w}) \, |{\rm BPS}\rangle_w & =  0  \\[2pt]
Q_1\, \psi^-( \tfrac{n}{w})  \, |{\rm BPS}\rangle_w & = 0 \qquad \qquad
& \tilde{Q}_2\,  \psi^-( \tfrac{n}{w}) \, |{\rm BPS}\rangle_w & =  0 \\[2pt]
S_1\, \alpha^2( \tfrac{n}{w}) \, |{\rm BPS}\rangle_w & = 0  \qquad \qquad
& \tilde{S}_2\,  \alpha^2( \tfrac{n}{w}) \, |{\rm BPS}\rangle_w & =  0  \\[2pt]
S_1\, \psi^-( \tfrac{n}{w}) \, |{\rm BPS}\rangle_w & =  \sqrt{1 - \tfrac{n}{w}}\, \alpha^2( \tfrac{n}{w}) \, |{\rm BPS}\rangle_w  \qquad \qquad
& \tilde{S}_2\,  \psi^-( \tfrac{n}{w}) \, |{\rm BPS}\rangle_w & =  0 \ ,
\end{array}
\ee
and similarly for the right-moving magnon operators. We now want to understand how these equations are modified under the perturbation. Before doing any explicit computations, we can already guess what the answer will be like: the non-trivial coefficients above may get modified --- we shall denote these by $a^n_n$ and $d^n_n$ below --- and there are additional terms that may appear, namely\footnote{There can be other transitions, which involve states with additional excitations on the right-hand-side. These transitions scale by a relative factor of $1/w$ in the large $w$ limit, and we therefore disregard them for our analysis.\label{fn:other_transitions}}
\be\label{deformedaction}
\begin{array}{llll}
Q_1\, \alpha^2( \tfrac{n}{w}) \, |{\rm BPS}\rangle_w & = a^n_n\, \psi^-( \tfrac{n}{w}) \, |{\rm BPS}\rangle_w   \quad
& \tilde{Q}_2\,  \alpha^2( \tfrac{n}{w}) \, |{\rm BPS}\rangle_w & =  0  \\[2pt]
Q_1\, \psi^-( \tfrac{n}{w})  \, |{\rm BPS}\rangle_w & = 0  \quad
& \tilde{Q}_2\,  \psi^-( \tfrac{n}{w})  \, |{\rm BPS}\rangle_w & =  b^m_n \,
\alpha^2( \tfrac{m}{w-1}) \, {\cal Z}_- \, |{\rm BPS}\rangle_w  \\[2pt]
S_1\, \alpha^2( \tfrac{n}{w}) \, |{\rm BPS}\rangle_w & = 0   \quad
& \tilde{S}_2\,  \alpha^2( \tfrac{n}{w}) \, |{\rm BPS}\rangle_w & =  c^m_n \,
\psi^-( \tfrac{m}{w+1}) \, {\cal Z}_+ \, |{\rm BPS}\rangle_w \\[2pt]
S_1\, \psi^-( \tfrac{n}{w}) \, |{\rm BPS}\rangle_w & = d^n_n \, \alpha^2( \tfrac{n}{w}) \, |{\rm BPS}\rangle_w  \quad
& \tilde{S}_2\,  \psi^-( \tfrac{n}{w})  \, |{\rm BPS}\rangle_w & =  0 \ ,
\end{array}
\ee
where ${\cal Z}_\pm$ maps the reference BPS states of different twisted sectors into one another,
\be\label{Zdef}
{\cal Z}_\pm \, |{\rm BPS}\rangle_w = |{\rm BPS}\rangle_{w\pm 1} \ ,
\ee
and we expect $\frac{n}{w} \approx \frac{m}{w+1}$ in the two terms of the second column. Here we have used that to first order in perturbation theory the introduction of the $2$-cycle twisted sector state will map the $w$-cycle state to a state in the $(w\pm 1)$-cycle twisted sector.\footnote{Actually, to first order in perturbation theory, $a^n_n$ and $d^n_n$ will not get modified, but will remain equal to $a^n_n = d^n_n=\sqrt{1 - \tfrac{n}{w}}$.
We also note that there can be intermediate states in a multi-cycle twisted sector, which, however, are suppressed in $\frac{1}{\sqrt{N}}$ by the genus counting argument of \cite{Pakman:2009zz}. They therefore do not contribute in the planar limit that we are considering here.\label{foot5}} Furthermore, we have imposed that for large $w$ and $m\approx n$, the conformal dimension and $K^3_0$ charges are not modified --- for the case of the conformal dimension we will see this explicitly in Section~\ref{genpert}. For example, the state $\tilde{S}_2\,  \alpha^2( \tfrac{n}{w}) \, |{\rm BPS}\rangle_w$ has the charges
\be
\tilde{S}_2\,  \alpha^2( \tfrac{n}{w}) \, |{\rm BPS}\rangle_w : \quad
h = \tfrac{w+1}{2} + (1 - \tfrac{n}{w}) \ , \quad \tilde{h} = \tfrac{w+1}{2} +\tfrac{1}{2} \ , \quad
j = \tfrac{w+1}{2} \ , \quad \tilde{j} = \tfrac{w+1}{2} + \tfrac{1}{2} \ ,
\ee
where $j$ and $\tilde{j}$ are the eigenvalues of $K^3_0$ and $\tilde{K}^3_0$, respectively. This then agrees with the charges of the state on the right-hand-side,
\be
\psi^-( \tfrac{m}{w+1})  \, |{\rm BPS}\rangle_{w+1} : \quad
h = \tfrac{w+2}{2} + (\tfrac{1}{2} - \tfrac{m}{w+1}) \ , \quad \tilde{h} = \tfrac{w+2}{2}  \ , \quad
j = \tfrac{w+2}{2} - \tfrac{1}{2} \ , \quad \tilde{j} = \tfrac{w+2}{2} \  ,
\ee
provided that $\frac{n}{w} \approx \frac{m}{w+1}$. The analysis for the term proportional to $b_n^m$ is similar, and these two are the only terms that can appear at this order. This general structure was already anticipated in \cite{Gava:2002xb,David:2008yk}, where, however, the magnon excitations considered were the fractional modes of the ${\cal N}=4$ superconformal algebra, and the relevant coefficients were not completely determined.

In the following we shall find explicit expressions for these coefficients at finite $w$ (and $m$ and $n$); we shall then explain that these coefficients have a natural and simple large $w$ limit that we shall identify, see Section~\ref{sec:effective_description}.

\subsection{The general form of the perturbation calculation}\label{genpert}

Before we go to the specifics of the individual coefficients, it is instructive to explain the general structure of the perturbation calculation. To be concrete, let us consider the correlator involving $\tilde{S}_2 =\tilde{G}'^+_{-1/2}$  (the other cases work similarly)
\begin{equation}
\underset{C(0)}{\oint}d\bar{x}\int d^2y\, {}_{w+1}\big\langle \Psi_2 | \tilde{G}'^+(\bar{x})\,\Phi(y,\bar{y})\, | \Psi_1 \bigr\rangle_w \ ,
\end{equation}
where the states $\ket{\Psi_{1,2}}$ have right-moving conformal dimensions $\tilde{h}_1=\frac{w}{2}$, $\tilde{h}_2=\frac{w+1}{2}$ as indicated by their subscripts, and the left-moving conformal dimensions $h_j$ are arbitrary. (Thus $\ket{\Psi_{1,2}}$ are left-moving magnon descendants of the BPS states, which we may assume to be quasiprimary.) Since both $| \Psi_{1,2}  \rangle$'s are BPS with respect to the right-movers, we can only get a non-trivial correlator provided that $\tilde{G}'^+$ has a non-trivial OPE with $\Phi$, and thus only one term in $\Phi$ contributes, namely
\begin{equation}
G^-_{-\frac{1}{2}}\widetilde{G}'^-_{-\frac{1}{2}}\ket{\mrm{BPS}_-}_2\ ,
\end{equation}
for which the relevant OPE is
\begin{align}
\tilde{G}'^+(\bar{x})\,\Phi(y,\bar{y}) &=\! -\frac{1}{(\bar{x}-\bar{y})^2}\,V\big( G^-_{-\frac{1}{2}}\ket{\mrm{BPS}_-}_2,y,\bar{y}\big) -\frac{1}{\bar{x}-\bar{y}}\,V\big( G^-_{-\frac{1}{2}}\tilde{L}_{-1}\ket{\mrm{BPS}_-}_2,y,\bar{y}\big) \!+ \!\text{reg}\nonumber\\
&= \!-\Big(\frac{1}{(\bar{x}-\bar{y})^2} + \frac{1}{\bar{x}-\bar{y}}\frac{\partial}{\partial \bar{y}}\Big)\,V\big( G^-_{-\frac{1}{2}}\ket{\mrm{BPS}_-}_2,y,\bar{y}\big) +\text{reg}\ ,
\end{align}
where `reg' stands for terms that are regular in $(\bar{x}-\bar{y})$, and $V(\ket{\phi},y,\bar{y})$ denotes the vertex operator associated to the state $\ket{\phi}$. The correlator with the insertion of
\be
V\big( G^-_{-1/2}\ket{\mrm{BPS}_-}_2,y,\bar{y}\big)
\ee
is a three-point function with $y$ dependence
\begin{equation}
y^{h_2-h_1-1}\bar{y}^0 \ ,
\end{equation}
and thus the derivative with respect to $\bar{y}$ vanishes. As a consequence we are left with an integral of the form
\begin{equation}
\langle\,\cdots\rangle\times \underset{C(0)}{\oint}d\bar{x}\int d^2y\,\,y^{h_2-h_1-1}(\bar{x}-\bar{y})^{-2}\ ,
\end{equation}
where $ \langle\,\cdots\rangle$ is the three-point function $\langle \Psi_2 |\, V\bigl(G^-_{-1/2}\ket{\mrm{BPS}_-}_2,y,\bar{y}\bigr)\, | \Psi_1\rangle$  for $y=\bar{y}=1$.
To calculate the integral, we now rescale $y = ux$ so that
\begin{equation}
\underset{C(0)}{\oint}d\bar{x} \, \frac{x^{h_2-h_1}}{\bar{x}}\int d^2u\,\,u^{h_2-h_1-1}(1-\bar{u})^{-2}\ .
\end{equation}
Next, following \cite{Pakman:2009mi}, we make use of the analytic continuation
\begin{equation}\label{integral}
\int d^2u \,\,u^a\,\bar{u}^{\bar{a}}\,(1-u)^b\,(1-\bar{u})^{\bar{b}} = \pi\,  \frac{\Gamma(1+a)\Gamma(1+b)\Gamma(-\bar{a}-\bar{b}-1)}{\Gamma(-\bar{a})\Gamma(-\bar{b})\Gamma(a+b+2)}\ .
\end{equation}
Since in our case $\bar{a} =0$, the factor $\frac{1}{\Gamma(-\bar{a})}=0$, and the integral can only be non-zero provided that $1+a = h_2-h_1$ is a non-positive integer, in which case $\Gamma(1+a)=\infty$. Then, the contour integral selects further
\begin{equation}\label{conservation}
\underset{C(0)}{\oint}d\bar{x}\,\, \frac{x^{h_2-h_1}}{\bar{x}} = \delta_{h_2,h_1}\ ,
\end{equation}
i.e.\ $1+a=0$. Since furthermore $b=0$ and $\bar{b} = -2$, the integral in (\ref{integral}) thus gives $\pi$. The main point to take away from this is that the $y$-integration ensures that only states for which $\ket{\Psi_{1,2}}$ have the same left-moving conformal dimension contribute.

We should mention that for a single magnon excitation the condition $h_1=h_2$ implies that $\frac{n}{w} = \frac{m}{w+1}$, which is only possible if both are integers. However, since we only have a single $y$-integral (at first order), for a multi-magnon state, the condition $h_1=h_2$ will also allow for solutions where the individual magnons do not satisfy individually $\frac{n}{w} \in \mathbb{Z}$. (The case of a single magnon excitation is therefore a bit degenerate, but it is nevertheless instructive to study it first.)

\subsection{Determining the individual coefficients at finite \texorpdfstring{$w$}{w}}\label{sec:finite_twist}

With these preparations we can now evaluate the different coefficients in eq.~(\ref{deformedaction}). Let us start with the coefficient corresponding to
\begin{align}
c^{m}_n & = \underset{C(0)}{\oint}d\bar{x}\int d^2y\,  {}_{w+1}\!\langle {\rm BPS} |  \, \bigl(\psi^-(\tfrac{m}{w+1})\bigr)^\dagger\, \, \tilde{G}'^+(\bar{x})\,\Phi(y,\bar{y})\, \, {\alpha}^2(\tfrac{n}{w}) | {\rm BPS}\rangle_w \nonumber \\
& = N_w\cdot \pi \cdot \delta_{\frac{m}{w+1},\frac{n}{w}}  \times c^{'m}_n \ , \label{3.27}
\end{align}
where $N_w$ is an overall normalisation factor that is determined in Appendix~\ref{app:orbifold_averaging}, and the next two factors come from the $y$-integral, see eqs.~(\ref{integral}) and (\ref{conservation}), while the remaining overlap $c'_{m,n}$ is
\be\label{c'}
c^{'m}_n =  {}_{w+1}\!\langle {\rm BPS} |  \, \bigl(\psi^-(\tfrac{m}{w+1})\bigr)^\dagger\, \,
V\bigl(G^-_{-1/2}\ket{\mrm{BPS}_-}_2,1,\bar{1}\bigr)\, \, {\alpha}^2(\tfrac{n}{w}) | {\rm BPS}\rangle_w  \ .
\ee
Here the bra-state is the conjugate of the state we want to extract. The amplitude in (\ref{c'}) is effectively  left-moving --- the right-moving part is just the overlap of the three BPS states --- and we can therefore ignore the right-movers. To evaluate it, we now go to the covering surface, using the covering map
\be\label{coveringww2}
\bm{\Gamma}(z) = (w+1)z^w-wz^{w+1}\ ,
\ee
that maps $ \bm{\Gamma}(0)=0$, $ \bm{\Gamma}(1)=1$, and $ \bm{\Gamma}(\infty)=\infty$, and has ramification indices $w$, $2$, and $w+1$ at $0$, $1$ and $\infty$, respectively. The lift of the state at zero, $ {\alpha}^2(\frac{n}{w}) | {\rm BPS}\rangle_w$ becomes
\begin{equation}
\frac{1}{\sqrt{w-n}} \,
\cint{C(0)}\!\!dz\,\bm{\Gamma}(z)^{-1+\frac{n}{w}}\,\hat{\alpha}{}^2(z)\,\ket{\wh{\mrm{BPS}}}_w
\end{equation}
on the covering surface, where, as before, the fields (states) on the covering surface are distinguished by a hat. Similarly, the out-state at infinity can be written as
\be
\frac{1}{\sqrt{w+1}} \,  \cint{C(\infty)}\!\!d\zeta\,\bigl(\partial\bm{\Gamma}(\zeta)\bigr)^{\frac{1}{2}}\bm{\Gamma}(\zeta)^{-\frac{m}{w+1}}\, \,  {}_{w+1}\bigl\langle \wh{\rm BPS}| \, \hat{\bar{\psi}}^+(\zeta)  \ .
\ee
Let us denote by $\hat{\Phi}_L(1)$ the lift of the (left-moving part of the) perturbing field, i.e.\ of $V\bigl(G^-_{-1/2}\ket{\mrm{BPS}_-}_2,1,\bar{1}\bigr)$ to the covering surface. Then
\begin{align}
c^{'m}_n = & \, \tfrac{1}{\sqrt{(w+1)(w-n)}}\, \,  \cint{C(\infty)}\!\!d\zeta\,\bigl(\partial\bm{\Gamma}(\zeta)\bigr)^{\frac{1}{2}}\, \bm{\Gamma}(\zeta)^{-\frac{m}{w+1}}\, \,  \cint{C(0)}\!\!dz\,\bm{\Gamma}(z)^{-1+\frac{n}{w}}\, \times  \nonumber\\
& \qquad \qquad \qquad \times  {}_{w+1}\bigl\langle \wh{\rm BPS}| \, \hat{\bar{\psi}}^+(\zeta) \, \hat{\Phi}_{L}(1) \, \hat{\alpha}{}^2(z)\,\ket{\wh{\mrm{BPS}}}_w \ .\label{cintegral} 
\end{align}
The amplitude in the second line is now a correlator for a single $\mbb{T}^4$ CFT on the covering space. It can thus be calculated using standard CFT techniques, see Appendix~\ref{app:amp1}, and one finds
\be\label{amp1}
{}_{w+1}\bigl\langle \wh{\rm BPS}| \, \hat{\bar{\psi}}^+(\zeta) \, \hat{\Phi}_{\rm L}(1) \, \hat{\alpha}{}^2(z)\,\ket{\wh{\mrm{BPS}}}_w =
\bigl(\zeta-1\bigr)^{-\frac{1}{2}}\, \zeta^{\frac{w+1}{2}}\, (1-z)^{-2} \ .
\ee
Finally, the double integral in (\ref{cintegral}) can be performed, see Appendix~\ref{cmnint} for more details, and one finds
\begin{tcolorbox}[ams align]\label{eq:ferm-bos_contraction}
c^{'m}_n \equiv
\begin{tikzpicture}[baseline=-.5ex]
\node[](n1)at(0,0){$\bar{\psi}^+_{m}$};
\node[](n2)[right=-1ex of n1]{$\alpha^2_{n}$};
\draw[](n1.south)--([yshift=-1ex]n1.south)--([yshift=-1ex]n2.south)node[midway]{$\times$}--(n2.south);
\end{tikzpicture}
& = \frac{(w+1)^{\frac{n}{w}}}{w^{\frac{m}{w+1}}}\big(\tfrac{w}{w+1}\big)^{m-n} \, \times\frac{\Gamma\bigl((w+1)(1 - \frac{n}{w})\bigr)\, \Gamma\bigl(w(1 - \tfrac{m}{w+1})\bigr)}{ \Gamma \bigl((w+1)(1 - \tfrac{m}{w+1})\bigr)\, \Gamma \bigl(w(1 - \tfrac{n}{w})\bigr)}\nonumber\\
&\quad \,    \times \frac{e^{-i\pi \frac{m}{w+1}}}{\pi^2 \sqrt{1-\frac{n}{w}}} \,\Gamma(1-\tfrac{m}{w+1})\,\Gamma(\tfrac{n}{w})\,\sin\big(\pi\tfrac{n}{w}\big)\, \sin\big(\pi\tfrac{m}{w+1}\big)\ .
\end{tcolorbox}
\noindent Here we have introduced the crossed contraction to denote this correlator, which will be useful below. The cross indicates that a supercharge acts on the boson to allow the transition. We also remind the reader that $\Gamma$ here refers to the Gamma function, not to be confused with the covering map $\bm{\Gamma}$ of eq.~(\ref{coveringww2}).

The coefficient $b^m_n$ is simply the complex conjugate of $c^m_n$, but if we want to consider also multi-magnon states we will need additional contractions. For example, for the action of $\tilde{S}_2$ on the $2$-magnon state
\begin{align}
\tilde{S}_2\,\alpha^2(\tfrac{n_1}{w})\,\alpha^2(\tfrac{n_2}{w})\,\ket{\mrm{BPS}}_w &=\nonumber\\
&\hspace{-2.8cm}=\sum_{m_1,m_2} c^{m_1,m_2}_{n_1,n_2} \psi^-(\tfrac{m_1}{w+1})\, \alpha^2(\tfrac{m_2}{w+1})\,\ket{\mrm{BPS}}_{w+1}\label{eq:2-magnon_charge_action0}\ ,
\end{align}
the relevant covering space integral turns out to be
\begin{align}
c^{m_1,m_2}_{n_1,n_2}   & = N_w\cdot \pi \cdot \delta_{\frac{m_1+m_2}{w+1},\frac{n_1+n_2}{w}}\times \tfrac{1}{\sqrt{(w+1) (w+1-m_2) (w-n_1)(w-n_2)}} \nonumber \\
&  \hspace{-0.5cm} \times \cint{C(\infty)}\!\!d\zeta_1\,\bigl(\partial\bm{\Gamma}(\zeta_1)\bigr)^{\frac{1}{2}}\bm{\Gamma}^{-\frac{m_1}{w+1}}(\zeta_1)\cint{C(\infty)}\!\!d\zeta_2\,\bm{\Gamma}^{1-\frac{m_2}{w+1}}(\zeta_2)\,\cint{C(0)}\!\!dz_1\,\bm{\Gamma}^{-1+\frac{n_1}{w}}(z_1)\,\cint{C(0)}\!\!dz_2\,\bm{\Gamma}^{-1+\frac{n_2}{w}}(z_2) \nonumber \\
& \hspace{-0.5cm} \times  {}_{w+1}\bigl\langle \widehat{\mrm{BPS}}| \,\,\hat{\bar{\psi}}{}^+(\zeta_1)\,\hat{\bar{\alpha}}{}^1(\zeta_2)\,\,\hat{\Phi}^\mrm{L}(1)\,\,\hat{\alpha}^2(z_1)\,\hat{\alpha}^2(z_2)\,\,|\wh{\mrm{BPS}}\big\rangle_w\ ,
\end{align}
where $N_w$ is the overall normalisation factor of Appendix~\ref{app:orbifold_averaging}, and $\pi\cdot \delta_{*,*}$ comes again from the $y$-integral, see eqs.~(\ref{integral}) and (\ref{conservation}). The correlator in the last line can now be calculated relatively easily since the fields are free, and we find
\begin{align}
& {}_{w+1}\bigl\langle \widehat{\mrm{BPS}}| \,\,\hat{\bar{\psi}}{}^+(\zeta_1)\,\hat{\bar{\alpha}}{}^1(\zeta_2)\,\,\hat{\Phi}^\mrm{L}(1)\,\,\hat{\alpha}^2(z_1)\,\hat{\alpha}^2(z_2)\,\,|\wh{\mrm{BPS}}\big\rangle_w \nonumber \\
& \qquad  = \bigl(\zeta_1-1\bigr)^{-\frac{1}{2}}\, \zeta_1^{\frac{w+1}{2}}\, \Bigl[   \frac{1}{(\zeta_2-z_1)^2}\frac{1}{(1-z_2)^2}+\frac{1}{(\zeta_2-z_2)^2}\frac{1}{(1-z_1)^2}  \Bigr]\ .
\end{align}
Thus there are two terms, and they can be identified with the different `Wick contractions'
\begin{equation}\label{eq:2-magnon_c_contr}
c^{m_1,m_2}_{n_1,n_2} = g\Big(
\begin{tikzpicture}[baseline=-.5ex]
\node[](n1)at(0,0){$\bar{\psi}^+_{m_1}$};
\node[](n2)[right=-1ex of n1]{$\bar{\alpha}^1_{m_2}$};
\node[](n3)[right=-1ex of n2]{$\alpha^2_{n_1}$};
\node[](n4)[right=-1ex of n3]{$\alpha^2_{n_2}$};
\draw[](n1.south)-- ([yshift=-1ex]n1.south)--([yshift=-1ex]n3.south)node[midway]{$\times$}--(n3.south);
\draw[](n2.north)-- ([yshift=1ex]n2.north)--([yshift=1ex]n4.north)--(n4.north);
\end{tikzpicture}
+
\begin{tikzpicture}[baseline=-.5ex]
\node[](n1)at(0,0){$\bar{\psi}^+_{m_1}$};
\node[](n2)[right=-1ex of n1]{$\bar{\alpha}^1_{m_2}$};
\node[](n3)[right=-1ex of n2]{$\alpha^2_{n_1}$};
\node[](n4)[right=-1ex of n3]{$\alpha^2_{n_2}$};
\draw[](n1.south)-- ([yshift=-1ex]n1.south)--([yshift=-1ex]n4.south)node[midway]{$\times$}--(n4.south);
\draw[](n2.north)-- ([yshift=1ex]n2.north)--([yshift=1ex]n3.north)--(n3.north);
\end{tikzpicture}
\Big)\times N_w \cdot \pi\cdot \delta_{\frac{n_1+n_2}{w},\frac{m_1+m_2}{w+1}}+\mcl{O}\bigl(g^2\bigr)\ ,
\end{equation}
where the `boson-boson' contraction equals
\begin{align}
\wick{\c {\bar{\alpha}^1_m}\,\,\c{\alpha^2_n}} &\equiv \tfrac{1}{\sqrt{(w+1-m)(w-n)}}\cint{C(\infty)}\!\!d\zeta\,\bm{\Gamma}^{1-\frac{m}{w+1}}(\zeta)\,\cint{C(0)}\!\!dz\,\bm{\Gamma}^{-1+\frac{n}{w}}(z)\, \times \nonumber\\
& \qquad \times
{}_{w+1}\bigl\langle \wh{\rm BPS}| \, \hat{\bar\alpha}{}^1(\zeta)\,  \hat{\phi}_2(1) \, \hat{\alpha}{}^2(z)\,\ket{\wh{\mrm{BPS}}}_w \\
& = \tfrac{1}{\sqrt{(w+1-m)(w-n)}}\cint{C(\infty)}\!\!d\zeta\,\bm{\Gamma}^{1-\frac{m}{w+1}}(\zeta)\,\cint{C(0)}\!\!dz\,\bm{\Gamma}^{-1+\frac{n}{w}}(z)\, \times  \frac{1}{(\zeta-z)^2} \ .
\end{align}
Here $\hat{\phi}_2$ is the lift of the ground state in the $2$-cycle twisted sector (that remains after the Wick contractions involving the fermions and one boson from the state at zero have been performed, see the analysis for $c_{m,n}$), and we have used that $\hat{\phi}_2$ behaves like the vacuum with respect to the bosons to go to the last line. We have introduced the usual notation for a Wick contraction for this correlator, to distinguish it from the fermion-boson contraction denoted by a crossed contraction. Evaluating the integral we then find
\begin{tcolorbox}[ams align]\label{eq:bos-bos_contraction}
\wick{\c {\bar{\alpha}^1_m}\,\,\c{\alpha^2_n}}
&=   \frac{(w+1)^{\frac{n}{w}}}{w^{\frac{m}{w+1}}}\big(\tfrac{w}{w+1}\big)^{m-n} \,  \,
\times\sqrt{\frac{{1-\frac{m}{w+1}}}{{1-\frac{n}{w}}}}\, \frac{\Gamma\left((w+1)(1 - \frac{n}{w})\right)\, \Gamma\bigl(w(1 - \tfrac{m}{w+1})\bigr)}{ \Gamma \bigl((w+1)(1 - \tfrac{m}{w+1})\bigr)\, \Gamma \bigl(w(1 - \tfrac{n}{w})\bigr)} \nonumber \\
& \quad \times \frac{e^{-i\pi \frac{m}{w+1}}}{\pi^2 \sqrt{w (w+1)}} \,  \frac{(-1)^{m-n}}{\frac{m}{w+1}- \frac{n}{w}} \,
\sin\bigl( \pi \tfrac{n(w+1)}{w} \bigr) \,\sin\bigl( \pi \tfrac{mw}{w+1} \bigr) \, \Gamma \bigl(\tfrac{n}{w}\bigr)\, \Gamma \bigl(1-\tfrac{m}{w+1}\bigr)\ .
\end{tcolorbox}
\noindent Finally, there is also a similar `fermion-fermion' contraction defined by
\begin{align}
\wick{\c {\bar{\psi}^+_m}\,\,\c{\psi^-_n}} &:= \tfrac{1}{\sqrt{w(w+1)}}\cint{C(\infty)}\!\!d\zeta\,
\bigl(\partial\bm{\Gamma}(\zeta)\bigr)^{\frac{1}{2}}\bm{\Gamma}^{-\frac{m}{w+1}}
\cint{C(0)}\!\!dz\, \bigl(\partial\bm{\Gamma}(z)\bigr)^{\frac{1}{2}}\bm{\Gamma}^{-1+\frac{n}{w}}
\, \times \nonumber\\
& \qquad \times
{}_{w+1}\bigl\langle \wh{\rm BPS}| \, \hat{\bar\psi}{}^+(\zeta)\,  \hat{\phi}_2(1) \, \hat{\psi}{}^-(z)\,\ket{\wh{\mrm{BPS}}}_w \ . \label{amp2}
\end{align}
The amplitude of the second line requires some care and is evaluated in Appendix~\ref{app:amp2}. The integral can then be performed, and it leads to a very similar expression as for the boson-boson contraction in eq.~(\ref{eq:bos-bos_contraction})
\begin{tcolorbox}[ams align]\label{eq:ferm-ferm_contraction}
\wick{\c {\bar{\psi}^+_m}\,\,\c{\psi^-_n}}
&=   \frac{(w+1)^{\frac{n}{w}}}{w^{\frac{m}{w+1}}}\big(\tfrac{w}{w+1}\big)^{m-n} \,  \,
\times\frac{\Gamma\bigl((w+1)(1 - \frac{n}{w})\bigr)\, \Gamma\bigl(w(1 - \tfrac{m}{w+1})\bigr)}{ \Gamma \bigl((w+1)(1 - \tfrac{m}{w+1})\bigr)\, \Gamma \bigl(w(1 - \tfrac{n}{w})\bigr)}\\
& \quad \times \frac{e^{-i\pi \frac{m}{w+1}}}{\pi^2 \sqrt{w(w+1)}} \, \frac{(-1)^{m-n}}{\frac{m}{w+1}- \frac{n}{w}} \,
\sin\bigl( \pi \tfrac{n(w+1)}{w} \bigr) \,\sin\bigl( \pi \tfrac{mw}{w+1} \bigr) \, \Gamma \bigl(\tfrac{n}{w}\bigr)\, \Gamma \bigl(1-\tfrac{m}{w+1}\bigr)\ . \nonumber
\end{tcolorbox}
\noindent Thus these same-species contractions only differ by the normalisation factor
\be
\wick{\c {\bar{\alpha}^1_m}\,\,\c{\alpha^2_n}} = \sqrt{\frac{{1-\frac{m}{w+1}}}{{1-\frac{n}{w}}}} \,\wick{\c {\bar{\psi}^+_m}\,\,\c{\psi^-_n}} \ .
\ee
The main difference between the same-species contractions eqs.~(\ref{eq:bos-bos_contraction}) and (\ref{eq:ferm-ferm_contraction}) on the one hand, and the fermion-boson contraction eq.~(\ref{eq:ferm-bos_contraction}) on the other, is the additional term
\begin{equation}
\frac{1}{\frac{m}{w+1}-\frac{n}{w}}
\end{equation}
in the second line of eq.~(\ref{eq:bos-bos_contraction}) or (\ref{eq:ferm-ferm_contraction}). This factor implies that transitions where the mode numbers are approximately conserved are dominant. In the large twist $w$ limit, this will essentially lead to pairwise momentum conserving scattering, see section \ref{ssec:anomalous_dim} and section \ref{sec:effective_description}.

The Wick contraction like structure, present at finite $w$, which ultimately follows from the corresponding contractions on the covering space, is critical in ensuring that the multi-magnon state is in a tensor product representation of the individual magnon states, as far as the action of the supercharges go.

\section{Consistency conditions at finite \texorpdfstring{$w$}{w}}\label{sec:finitew}

The previous considerations now allow us to define the action of the four supercharges of eq.~(\ref{N4subset}) on all states that can be created by the left-moving magnon operators of eq.~(\ref{magnonsub}) from the BPS states to first order in perturbation theory. We can similarly work out the action of the supercharges on states that involve also right-moving magnon excitations -- and we shall do so below in more detail --- but actually, there are already interesting consequences within this restricted set-up.
More specifically, we shall show that the $\mcl{N}=(4,4)$ superalgebra is indeed preserved, and no central extension occurs. Secondly, we will write down an expression for the anomalous dimension of a two-magnon state which has a simple and suggestive form for large $w$.

\subsection{Absence of central terms}\label{sec:central}

As we have seen, the right-moving supercharges act non-trivially on these purely left-moving excitations, and so do the left-moving supercharges --- their action is unmodified to first order in perturbation theory. In particular, we can therefore ask how the anticommutator of left- and right-moving supercharges acts on these states, e.g.\ the anti-commutator
\be
\{G^+_{-\frac{1}{2}}, \tilde{G}^{'+}_{-\frac{1}{2}}\} = \{S_1 , \tilde{S}_2 \}  \stackrel{!}{=} 0\ .
\ee
Since the perturbation preserves ${\cal N}=(4,4)$ superconformal symmetry, this anti-commu\-ta\-tor must vanish. However, on the face of it, it does not, and reflects  a centrally extended algebra. We shall show in the following that these additional terms do indeed vanish on orbifold-invariant (physical) states. This is obviously as anticipated, see e.g.\ \cite{Beisert:2005tm,Gava:2002xb}, but it is reassuring (and instructive!) to see that this actually works out explicitly.

To be concrete let us concentrate in the following on the states of the form\footnote{As mentioned before, we could consider more excitation modes, a mixture of left- and right-movers and of bosons and fermions. We will remark briefly at the end of the section on how the results extend to these general states, but defer the full discussion to the large $w$ analysis of Section~\ref{sec:effective_description}.}
\begin{equation}\label{eq:2-magnon_state}
\alpha^2(\tfrac{n_1}{w})\,\alpha^2(\tfrac{n_2}{w})\,\ket{\mrm{BPS}}_w\ ,
\end{equation}
for which we have already derived the action of $\tilde{S}_2 = \tilde{G}^{'+}_{-1/2}$, see eq.~(\ref{eq:2-magnon_charge_action0}). These states are orbifold invariant (or `physical') if
\begin{equation}\label{orbinv2}
\frac{n_1}{w}+\frac{n_2}{w} \in \mbb{Z}\ .
\end{equation}
Using eq.~(\ref{eq:2-magnon_charge_action0}) as well as (\ref{deformedaction}), the action of the anti-commutator $\{G^+_{-\frac{1}{2}} ,\tilde{G}^{'+}_{-\frac{1}{2}}\}$ on the state (\ref{eq:2-magnon_state}) gives
\begin{equation}
\sum_{m_1,m_2} f^{m_1,m_2}_{n_1,n_2}\, \alpha^2(\tfrac{m_1}{w+1})\,\alpha^2(\tfrac{m_2}{w+1})\,\ket{\mrm{BPS}}_{w+1}\ ,
\end{equation}
where
\begin{equation}
f^{m_1,m_2}_{n_1,n_2} = c^{m_1,m_2}_{n_1,n_2}\,\sqrt{1-\tfrac{m_1}{w+1}}+c^{m_2,m_1}_{n_1,n_2}\,\sqrt{1-\tfrac{m_2}{w+1}}\ ,
\end{equation}
and $c^{m_1,m_2}_{n_1,n_2}$ was determined in eq.~(\ref{eq:2-magnon_c_contr}), and the square root factors come from the $d_m^m$ coefficient, see footnote~\ref{foot5}.

We expect that this coefficient must vanish on physical states, and we show in Appendix~\ref{app:vanishing_central_extension} that this is indeed the case. In fact, we will see that this is a consequence of the conformal dimension matching condition
\begin{equation}\label{confmatch}
\frac{n_1+n_2}{w}=\frac{m_1+m_2}{w+1}\ ,
\end{equation}
see the Kronecker delta in the definition of $c^{m_1,m_2}_{n_1,n_2}$. For finite $w$, eq.~(\ref{confmatch}) implies that both sides are separately integer, and hence, in particular, that the state is physical, i.e.\ eq.~(\ref{orbinv2}).

We have checked that the same conclusion also holds for arbitrary states, and the mechanism is always the same: the relevant coefficient is zero provided that the conformal dimension matching condition is satisfied, and in turn, this implies, at finite $w$, that the state in question is physical. While this confirms that our calculation is consistent --- the perturbation, if correctly implemented, must preserve ${\cal N}=(4,4)$ superconformal symmetry on physical states --- the mechanism by which this is achieved is a bit different than what had been expected in the literature before, see e.g.\ \cite{Gava:2002xb,Lloyd:2014bsa}.

\subsection{Anomalous conformal dimension}\label{ssec:anomalous_dim}

For states that only have left-moving excitations, the calculation of the anomalous dimension simplifies drastically, see also \cite{Gava:2002xb}.\footnote{Usually, the calculation of the anomalous conformal dimension requires considering a four-point function, see for example \cite{Gaberdiel:2015uca,Benjamin:2021zkn,Apolo:2022fya}. However, for the case at hand, this is equivalent to the calculation we are about to explain \cite{Keller:2019suk}.} The basic reason is that it is enough to determine their right-moving conformal dimension, which can be directly obtained from the anti-commutator
\be\label{anomalouspureleft}
\Delta \tilde{h}\, \Psi = \bigl(\tilde{L}_0 - \tilde{K}^3_0 \bigr) \, \Psi = \{\tilde{G}^{'+}_{-\frac{1}{2}}, \tilde{G}'^-_{+\frac{1}{2}} \} \, \Psi \equiv \{\tilde{Q}_{2}, \tilde{S}_{2} \} \, \Psi\ ,
\ee
where we have used that before the perturbation $\bigl(\tilde{L}_0 - \tilde{K}^3_0 \bigr) \, \Psi=0$, as well as the fact that the eigenvalues of $\tilde{K}^3_0$ are quantised (and hence cannot become anomalous). Because of locality, the left-moving anomalous conformal dimension must be equal, i.e.
\be
\Delta h = \Delta \tilde{h} \ ,
\ee
and thus we can directly deduce the total anomalous conformal dimension,
\be
\Delta h + \Delta \tilde{h}
\ee
from the anti-commutator in eq.~(\ref{anomalouspureleft}). As we will see, the anomalous conformal dimensions will be of order $g^2$, since the action of both right-moving supercharges is proportional to $g$. As a consequence the argument therefore does not directly generalise to arbitrary states since the coefficients $a_n^n$ and $d_n^n$, see eq.~(\ref{deformedaction}), will have ${\cal O}(g^2)$ corrections, that will also have to be taken into account in order to determine the anomalous conformal dimension. However, as will be explained below, see Section~\ref{sec:effective_description},  there is actually a way to find the general formula for the anomalous conformal dimensions, at least for large $w$.

Given the structure of the action of the right-moving supercharges on the left-moving magnon states\footnote{As discussed in Footnote \ref{fn:other_transitions}, we disregard certain transitions which are subleading in the large $w$ limit. It would be good to confirm, by a direct calculation of the anomalous dimension (involving a four-point function), that this omission does not change the anomalous dimensions at large $w$.},
\begin{subequations}\label{eq:2-magnon_charge_action}
\begin{align}
\wt{Q}_{2}\cdot\,\alpha^2(\tfrac{n_1}{w})\,\alpha^2(\tfrac{n_2}{w})\,\ket{\mrm{BPS}}_w &= 0\ ,\\[4pt]
\wt{S}_{2}\cdot\,\alpha^2(\tfrac{n_1}{w})\,\alpha^2(\tfrac{n_2}{w})\,\ket{\mrm{BPS}}_w &=\sum_{m_1,m_2}c^{m_1, m_2}_{n_1, n_2}\,\psi^-(\tfrac{m_1}{w+1})\,\alpha^2(\tfrac{m_2}{w+1})\,\ket{\mrm{BPS}}_{w+1}\ , \\
\wt{Q}_{2}\cdot\,\psi^-(\tfrac{m_1}{w+1})\,\alpha^2(\tfrac{m_2}{w+1})\,\ket{\mrm{BPS}}_{w+1} &=\sum_{n_3,n_4}\,\, b^{n_3, n_4}_{m_1, m_2}\,\alpha^2(\tfrac{n_3}{w})\,\alpha^2(\tfrac{n_4}{w})\,\ket{\mrm{BPS}}_{w}\ ,
\end{align}
\end{subequations}
it is clear that the anti-commutator in (\ref{anomalouspureleft}) will be of the form
\begin{align}\label{eq:2-magnon_anti-comm}
\{\wt{Q}_{2},\wt{S}_{2}\}\,\,\alpha^2(\tfrac{n_1}{w})\,\alpha^2(\tfrac{n_2}{w})\,\ket{\mrm{BPS}}_w = \sum_{n_3,n_4} \wt{\gamma}^{n_3,n_4}_{n_1,n_2}\,\,\alpha^2(\tfrac{n_3}{w})\,\alpha^2(\tfrac{n_4}{w})\,\ket{\mrm{BPS}}_w\ ,
\end{align}
where
\begin{equation}
\wt{\gamma}^{n_3,n_4}_{n_1,n_2} = \sum_{m_1,m_2} \bigl(c^{m_1,m_2}_{n_3,n_4}\bigr)^*\, c^{m_1,m_2}_{n_1,n_2}\ ,
\end{equation}
and we have used that the $b$ coefficients are simply the complex conjugates of the $c$ coefficients. In order to determine the anomalous conformal dimensions we therefore need to diagonalise the matrix
$\wt{\gamma}$.

The sum over $m_1$ and $m_2$ is restricted by the conformal dimension constraint, see the Kronecker delta in eq.~(\ref{eq:2-magnon_c_contr}), and thus only the terms with
\begin{equation}\label{constraint}
\frac{n_1}{w}+\frac{n_2}{w} = \frac{m_1}{w+1}+\frac{m_2}{w+1}
\end{equation}
contribute. At finite $w$, in particular, this means that both sides have to be integers (since $\mrm{gcd}(w,w+1)=1$), i.e.\ that the states will be physical (i.e.\ orbifold invariant).

\begin{figure}[ht]
\centering

\definecolor{graphorange}{HTML}{f7a919}
\definecolor{graphblue}{HTML}{1d79db}
\definecolor{graphgreen}{HTML}{31a30f}
\begin{tikzpicture}[scale=0.9]
\draw[black,-{Latex[length=2mm, width=1mm]}](0,0)--(10,0)node[anchor=north]{$\frac{n_1}{w}$};
\draw[black,-{Latex[length=2mm, width=1mm]}](0,0)--(0,6.5)node[anchor=east]{$\frac{\widetilde{\gamma}}{g^2}$};
\foreach \y in {0.296624,0.593247,0.889871,1.18649,1.77974,2.07636,2.37299,2.66961,3.26286,3.55948,3.85611,4.15273,4.74598,5.0426,5.33922,5.63585,6.22909}
{
	\draw[](0,\y)--(0.075,\y);
}
\foreach \a/\y in {1/1.48312, 2/2.96624, 3/4.44935, 4/5.93247}
{
	\draw[](0,\y)--(0.15,\y);
	\node[]at(-0.25,\y){{\small\a}};
}
\foreach \x in {0.444444,0.888889,1.33333,1.77778,2.66667,3.11111,3.55556,4.,4.88889,5.33333,5.77778,6.22222,7.11111,7.55556,8.,8.44444,9.33333}
{
	\draw[](\x,0)--(\x,0.075);
}
\foreach \a/\x in {$\frac{10}{w}$/2.22222, $\frac{20}{w}$/4.44444, $\frac{30}{w}$/6.66667, $\frac{40}{w}$/8.88889}
{
	\draw[](\x,0)--(\x,0.15);
	\node[]at(\x,-0.35){{\small\a}};
}
\draw[line width = 1.5,graphgreen](0.0,0.0)--(0.222222,0.178797)--(0.444444,0.361576)--(0.666667,0.547941)--(0.888889,0.737479)--(1.11111,0.929765)--(1.33333,1.12436)--(1.55556,1.3208)--(1.77778,1.51863)--(2.,1.71738)--(2.22222,1.91656)--(2.44444,2.11568)--(2.66667,2.31426)--(2.88889,2.51179)--(3.11111,2.70777)--(3.33333,2.9017)--(3.55556,3.09309)--(3.77778,3.28144)--(4.,3.46625)--(4.22222,3.64705)--(4.44444,3.82335)--(4.66667,3.99469)--(4.88889,4.1606)--(5.11111,4.32065)--(5.33333,4.47439)--(5.55556,4.62142)--(5.77778,4.76134)--(6.,4.89375)--(6.22222,5.0183)--(6.44444,5.13465)--(6.66667,5.24248)--(6.88889,5.34148)--(7.11111,5.43138)--(7.33333,5.51194)--(7.55556,5.58293)--(7.77778,5.64414)--(8.,5.69542)--(8.22222,5.73662)--(8.44444,5.76762)--(8.66667,5.78833)--(8.88889,5.7987);
\foreach \x/\y in {0.222222/0.177497,0.444444/0.358856,0.666667/0.543714,0.888889/0.731675,1.11111/0.922322,1.33333/1.11523,1.55556/1.30994,1.77778/1.50601,2./1.70296,2.22222/1.90033,2.44444/2.09762,2.66667/2.29436,2.88889/2.49005,3.11111/2.68419,3.33333/2.8763,3.55556/3.06588,3.77778/3.25244,4./3.43549,4.22222/3.61455,4.44444/3.78916,4.66667/3.95885,4.88889/4.12316,5.11111/4.28166,5.33333/4.43391,5.55556/4.57951,5.77778/4.71806,6./4.84918,6.22222/4.97252,6.44444/5.08773,6.66667/5.1945,6.88889/5.29253,7.11111/5.38155,7.33333/5.46131,7.55556/5.5316,7.77778/5.59222,8./5.64299,8.22222/5.68378,8.44444/5.71447,8.66667/5.73498,8.88889/5.74525}
{
	\filldraw[graphblue](\x+0.075,\y)--(\x,\y+0.105)--(\x-0.075,\y)--(\x,\y-0.105)--(\x+0.075,\y);
}
\foreach \x/\y in {0.222222/0.176477,0.444444/0.356681,0.666667/0.540274,0.888889/0.726886,1.11111/0.91613,1.33333/1.1076,1.55556/1.30089,1.77778/1.49557,2./1.69119,2.22222/1.88732,2.44444/2.08349,2.66667/2.27925,2.88889/2.47412,3.11111/2.66764,3.33333/2.85932,3.55556/3.04869,3.77778/3.23527,4./3.41858,4.22222/3.59816,4.44444/3.77354,4.66667/3.94426,4.88889/4.10987,5.11111/4.26992,5.33333/4.424,5.55556/4.57168,5.77778/4.71257,6./4.84628,6.22222/4.97245,6.44444/5.09074,6.66667/5.20081,6.88889/5.30236,7.11111/5.39512,7.33333/5.47882,7.55556/5.55323,7.77778/5.61815,8./5.67339,8.22222/5.7188,8.44444/5.75425,8.66667/5.77964,8.88889/5.79491}
{
	\filldraw[graphorange](\x,\y)circle(0.45mm);
}
\end{tikzpicture}
\caption{A plot of the anomalous dimension for two bosonic fractional modes as a function of $n_1$. Plotted are the eigenvalues of $\wt{\gamma}$ (orange dots), its diagonal elements (blue diamonds), and the function of eq.~(\ref{anofunc})  (green line) for $w=81$. }
\label{fig:anom_dim}
\end{figure}
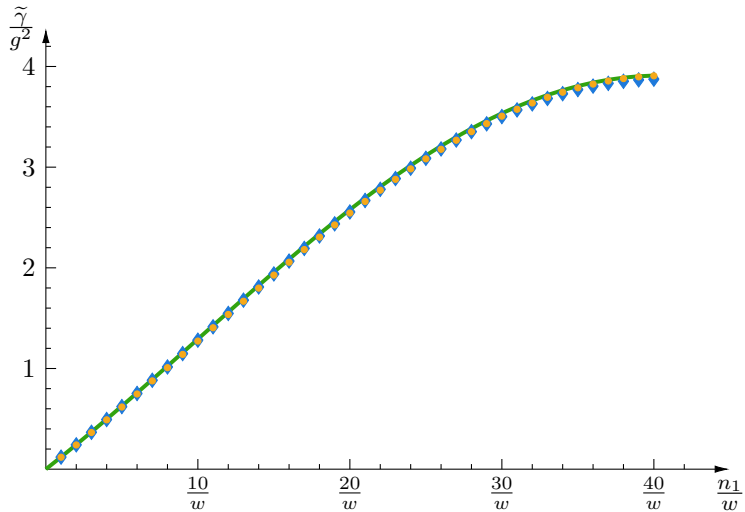

However, even taking this constraint into account, the matrix $\wt{\gamma}$ is not diagonal at finite $w$. On the other hand, because of the factor in the denominator of eq.~(\ref{eq:bos-bos_contraction})
\be
\frac{m}{w+1}-\frac{n}{w} \ ,
\ee
the dominant transitions are those for which $\frac{m_i}{w+1}$ is close to $\frac{n_j}{w}$ --- unless $n_j$ is a multiple of $w$ and $m_i$ a multiple of $w+1$, the denominator can never be zero since $w$ and $w+1$ are coprime. One should therefore expect that as we take the large $w$ limit, $\wt{\gamma}$ will become effectively diagonal. Furthermore, by taking a careful large $w$ limit of the individual coefficients, see Section~\ref{sec:effective_description} for more details, one can actually determine these diagonal values to be
\begin{equation}\label{eq:anom_dim_limit}
\wt{\gamma}^{n_1,n_2}_{n_1,n_2} \to g^2\,\left(\frac{\sin^2\big(\pi\frac{n_1}{w}\big)}{1-\frac{n_1}{w}}+\frac{\sin^2\big(\pi\frac{n_2}{w}\big)}{1-\frac{n_2}{w}}\right)\ ,\quad w\to\infty \ .
\end{equation}
We have checked that this is indeed a very good large $w$ approximation; for example, we have plotted in Figure~\ref{fig:anom_dim} the eigenvalues and the diagonal entries of $\wt{\gamma}$ for $w=81$ against the large $w$ prediction of eq.~(\ref{eq:anom_dim_limit}), and the agreement is already excellent. Note that because of eq.~(\ref{constraint}) we have only considered states with $n_2 = w - n_1$, and for them the function in eq.~(\ref{eq:anom_dim_limit}) becomes
\be\label{anofunc}
\frac{\sin^2\big(\pi\frac{n_1}{w}\big)}{1-\frac{n_1}{w}}+\frac{\sin^2\big(\pi\frac{w-n_1}{w}\big)}{1-\frac{w-n_1}{w}} = \sin^2\big(\pi\tfrac{n_1}{w}\big)  \Bigl( \frac{1}{1-\frac{n_1}{w}} + \frac{1}{\frac{n_1}{w}}  \Bigr) =
\frac{ \sin^2\big(\pi\frac{n_1}{w}\big)}{(1-\frac{n_1}{w}) \frac{n_1}{w}} \ .
\ee

As will be explained below in Section~\ref{sec:effective_description}, our large $w$ prediction actually fits perfectly with the expectations based on the BMN analysis, see in particular \cite{Hoare:2013lja}. One can also perform a similar analysis for states that are made from more than two magnons and that involve  fermionic magnons, etc., and the structure is always the same. We will not elaborate more on these finite $w$ results, but instead now study how the large $w$ limit can be taken systematically.

\section{The large \texorpdfstring{$w$}{w} limit}\label{sec:effective_description}

We are interested in the spectrum of operators with large $w$.
As we have seen in the previous section, the action of the various supercharges simplifies in the large $w$ limit. In this section we want to explain how this large $w$ limit can be taken systematically. More specifically, we shall consider the limit where we take $w$, as well as the magnon quantum numbers $n_i$ simultaneously to infinity, while keeping the `momentum'
\be
p_i = \frac{n_i}{w}  \qquad \hbox{fixed} \ ,
\ee
and we shall therefore write the modes as $\alpha^i(\frac{n}{w}) = \alpha^i(p)$, etc. To simplify notation we shall also denote the BPS reference state simply by $ |w\rangle \equiv |{\rm BPS}\rangle_w $.

Suppose now we want to calculate the action of $\tilde{S}_2$ on a multi-magnon state to first order in the coupling constant. As should be clear from the discussion in the previous section, this will amount to a sum over products of contractions, where each contraction is either given by (\ref{eq:ferm-bos_contraction}), or (\ref{eq:bos-bos_contraction}) or (\ref{eq:ferm-ferm_contraction}). In order to understand the large $w$ limit of this action, we therefore need to understand the large $w$ behaviour of each of these contraction terms individually.

\subsection{The same species contraction (\ref{eq:bos-bos_contraction}) or (\ref{eq:ferm-ferm_contraction})}

It is instructive to first start with either (\ref{eq:bos-bos_contraction}) or (\ref{eq:ferm-ferm_contraction}) -- their functional form is the same. As we explained above, these contractions are dominated by the configurations for which the denominator $\frac{m}{w+1} - \frac{n}{w}$ becomes small. We now want to show that, in the large $w$ limit, we have
\begin{tcolorbox}[ams equation]\label{boscon}
\wick{\c {\bar{\alpha}^1(q)}\,\,\c{\alpha^2(p)}} =  \wick{\c {\bar{\psi}^+(q)}\,\,\c{\psi^-(p)}} =
\delta(p-q) \times \left\{ \begin{array}{cc} 1 \\ e^{-2\pi i p} \end{array} \right. \  ,
\end{tcolorbox}
\noindent where $q=\frac{m}{w+1}$, $p=\frac{n}{w}$ are continuous parameters for infinite $w$. To see this, let us write $p=\frac{n}{w}$, where $n$ and $w$ are integers, and analytically continue eq.~(\ref{eq:bos-bos_contraction}) in $q=\frac{m}{w+1}$,
\be\label{5.3}
(\ref{eq:bos-bos_contraction}) =    \frac{e^{-i \pi \frac{m}{w+1}} \, (-1)^{(w+1)q}}{\sqrt{w(w+1)}}\,\frac{\sin(\pi q\,w)}{\pi\,\bigl(q-\frac{n}{w}\bigr)}\times \text{rest$(q)$}\ .
\ee
where the $\lim_{p\rightarrow q}\text{rest$(q)$} = 1$ contains in particular the first line of (\ref{eq:bos-bos_contraction}), as well as the factor
\be
(-1)^{-n}\, \frac{\sin\bigl( \pi \tfrac{n(w+1)}{w} \bigr)}{\pi} \,\, \Gamma \bigl(\tfrac{n}{w}\bigr)\, \Gamma \bigl(1-\tfrac{m}{w+1}\bigr) \rightarrow 1 \qquad \hbox{for $q=\tfrac{m}{w+1}\rightarrow p=\frac{n}{w}$} \ .
\ee
Eq.~(\ref{5.3}) does not have a pole on the real axis since the apparent pole in eq.~(\ref{5.3}) is cancelled by the sine factor in the numerator. In the large $w$ limit the sine factor is highly oscillating, and we can therefore set $q=\frac{n}{w}$ everywhere except in the sine and the denominator. Written as a distribution, we therefore find
\begin{equation}
\wick{\c {\bar{\alpha}^1(q)}\,\,\c{\alpha^2(p)}}\bigl[\phi\bigr] = \sqrt{\frac{w+1}{w}}\,e^{-i\pi \frac{n}{w}}(-1)^{\frac{n\,(w+1)}{w}}\,\int dq\,f_w(q)\,\phi(q) \ ,
\end{equation}
where
\begin{equation}
f_w(q) = \frac{\sin(\pi q\,w)}{\pi\,\bigl(q-\frac{n}{w}\bigr)}\ ,
\end{equation}
and the additional factor of $(w+1)$ comes from the change of measure $dm\mapsto dq$. On the other hand, if we take the test functions $\phi$ to be in the Schwartz space, see e.g.\ \cite{BB}, it follows that
\be
\lim_{w\rightarrow \infty} \frac{\sin(\pi q\,w)}{\pi\,\bigl(q-\frac{n}{w}\bigr)} = (-1)^n\,\delta\bigl(q-\tfrac{n}{w}\bigr) \ ,
\ee
where the $(-1)^n$ appears because of
\begin{equation}
f_w\bigl(\tfrac{n}{w}\bigr) = (-1)^n\,w\ .
\end{equation}
Finally, we note that the phase factor multiplying the delta function depends on whether we write $(-1)$ as $e^{i\pi}$ or $e^{-i\pi}$, in which case we either get
\begin{equation}
e^{-i\pi \frac{n}{w}}(-1)^{\frac{n}{w}} = 1\qquad \text{or}\qquad e^{-i\pi \frac{n}{w}}(-1)^{\frac{n}{w}}=e^{-2\pi i \frac{n}{w}}\ ,
\end{equation}
thus establishing (\ref{boscon}). Obviously, it will be important to understand how we should choose the phase, and we shall come back to this issue once we have analysed the large $w$ limit of the contraction (\ref{eq:ferm-bos_contraction}), see Section~\ref{sec:phasechoice} below.

\subsection{The boson-fermion contraction (\ref{eq:ferm-bos_contraction})}

The fermion-boson contraction (\ref{eq:ferm-bos_contraction}) is on the face of it not localised near $q=p$, because it does not contain the above denominator term. However, to first order in $g$, for each term in the expansion, see e.g.\ eqs.~(\ref{eq:2-magnon_charge_action0}) and (\ref{eq:2-magnon_c_contr}), there will always only be one fermion-boson contraction term, while the remaining terms are same-species contractions, i.e.\ either (\ref{eq:bos-bos_contraction}) or (\ref{eq:ferm-ferm_contraction}) --- this simply follows from the fact that the fermion-boson contraction (\ref{eq:ferm-bos_contraction}) requires the specific descendants of the perturbing field, see eq.~(\ref{amp1}) where both the in-boson and the out-fermion are contracted with the descendants of the perturbing field. (This is to be contrasted with the same-species contractions (\ref{eq:bos-bos_contraction}) and (\ref{eq:ferm-ferm_contraction}) for which only $\sigma_2$ plays a role.)

Now, as we explained in Section~\ref{genpert}, the $y$-integral will always lead to overall momentum conservation, i.e.
\be
\sum_i p_i = \sum_i q_i \ .
\ee
Together with the fact that all the other contractions individually perserve momentum, see eq.~(\ref{boscon}), it follows that also for the fermion-boson contraction momentum is conserved! This is the first signature of a hidden quantum integrability in the system whereby the interactions of the magnons is `diffractionless' i.e.\ only permutes the different momenta. As we shall see later in Section~\ref{sec:s-matrix} an ansatz for the factorised scattering of $n$-magnons in terms of a $2$-$2$  $S$-matrix obeys the nontrivial Yang-Baxter consistency condition.

With this simplification in the large $w$ limit, we can simply evaluate (\ref{eq:ferm-bos_contraction}) for $q=p$, and we find\footnote{Here we have also used that in the large $w$ limit, the factor $N_w$ of Appendix~\ref{app:orbifold_averaging} becomes one, see eq.~(\ref{B.35}).}
\begin{tcolorbox}[ams equation]\label{5.11}
\begin{tikzpicture}[baseline=-.5ex]
\node[](n1)at(0,0){$\bar{\psi}^+(p)$};
\node[](n2)[right=-1ex of n1]{$\alpha^2(p)$};
\draw[](n1.south)-- ([yshift=-1ex]n1.south)--([yshift=-1ex]n2.south)node[midway]{$\times$}--(n2.south);
\end{tikzpicture}= e^{-i\pi p}\frac{\sin(\pi p)}{\pi \sqrt{1-p}}=\frac{1}{2\pi i}\frac{1-e^{-2\pi i p}}{\sqrt{1-p}}\ .
\end{tcolorbox}

As a simple example, we can now use these results to reproduce (\ref{eq:anom_dim_limit}). Writing $p_i = \frac{n_i}{w}$ and $q_j=\frac{m_j}{w+1}$, the $c_{q_1,q_2}^{p_1,p_2}$ coefficient becomes in the large $w$ limit
\begin{align}
c_{q_1,q_2}^{p_1,p_2} &= g\Big(
\begin{tikzpicture}[baseline=-.5ex]
\node[](n1)at(0,0){$\bar{\psi}^+(q_1)$};
\node[](n2)[right=-1ex of n1]{$\bar{\alpha}^1(q_2)$};
\node[](n3)[right=-1ex of n2]{$\alpha^2(p_1)$};
\node[](n4)[right=-1ex of n3]{$\alpha^2(p_2)$};
\draw[](n1.south)-- ([yshift=-1ex]n1.south)--([yshift=-1ex]n3.south)node[midway]{$\times$}--(n3.south);
\draw[](n2.north)-- ([yshift=1ex]n2.north)--([yshift=1ex]n4.north)--(n4.north);
\end{tikzpicture}
+
\begin{tikzpicture}[baseline=-.5ex]
\node[](n1)at(0,0){$\bar{\psi}^+(q_1)$};
\node[](n2)[right=-1ex of n1]{$\bar{\alpha}^1(q_2)$};
\node[](n3)[right=-1ex of n2]{$\alpha^2(p_1)$};
\node[](n4)[right=-1ex of n3]{$\alpha^2(p_2)$};
\draw[](n1.south)-- ([yshift=-1ex]n1.south)--([yshift=-1ex]n4.south)node[midway]{$\times$}--(n4.south);
\draw[](n2.north)-- ([yshift=1ex]n2.north)--([yshift=1ex]n3.north)--(n3.north);
\end{tikzpicture}
\Big) \nonumber\\
& \qquad \qquad \times\pi \, \delta(p_1+p_2-q_1-q_2)\nonumber\\
&= \frac{g}{2i}\,\frac{1-e^{-2\pi i p_1}}{\sqrt{1-p_1}}e^{\nu_1 i\pi p_2}\,\delta(p_1-q_1)\delta(p_2-q_2)\nonumber\\
&\quad+ \frac{g}{2i}\,\frac{1-e^{-2\pi i p_2}}{\sqrt{1-p_2}}e^{\nu_2 i\pi p_1}\,\delta(p_1-q_2)\delta(p_2-q_1)\ ,
\end{align}
where $\nu_i\in\{0,-2\}$. (Note that the `momentum' $p_i$ is indeed individually conserved, as explained above.) This then directly gives the anomalous dimension of eq.~(\ref{eq:anom_dim_limit}),
\begin{equation}
\wt{\gamma} = g^2\,\frac{\sin^2(\pi p_1)}{1-p_1}+g^2\,\frac{\sin^2(\pi p_2)}{1-p_2} \ .
\end{equation}

\subsection{Choosing the phase}\label{sec:phasechoice}

It remains to understand how the phase in (\ref{boscon}) needs to be chosen. For this we recall from Section~\ref{sec:central}, that at finite $w$, the anti-commutator $\{S_1,\wt{S}_2\}$ vanished automatically on physical states. However, with the above large $w$ expressions, this will only continue to be the case provided we choose the phases in (\ref{boscon})  appropriately.

In order to understand how this comes about, let us consider the large $w$ result for the action on the two boson excitations,
\begin{align}\label{SStilde}
\{S_1,\wt{S}_2\}\,  \alpha^2(p_1)\alpha^2(p_2)\ket{w} = & \frac{g}{2i}\,\Bigl(\bigl(1-e^{-2\pi i p_1}\bigr)e^{\nu_2 i \pi p_2}+\bigl(1-e^{-2\pi i p_2}\bigr)e^{\nu_1 i \pi p_1}\Bigr) \times \nonumber \\
& \qquad  \times \alpha^2(p_1)\alpha^2(p_2)\ket{w+1}\ .
\end{align}
(The factors of $\tfrac{1}{\sqrt{1-p}}$ in $c$ have cancelled against $d=\sqrt{1-p}$ to this order.)
This is clearly not always zero, and thus the $\mcl{N}=(4,4)$ algebra is generically broken. However, in the actual orbifold theory only the orbifold invariant states with $p_1+p_2\in \mbb{Z}$ survive. We should only require the anti-commutator to vanish on these states, and this is the case provided we take $\nu_1=0$, $\nu_2=-2$, since then
\begin{equation}\label{ans1}
\bigl(1-e^{-2\pi i p_1}\bigr)e^{-2 i \pi p_2}+\bigl(1-e^{-2\pi i p_2}\bigr) = 1-e^{-2\pi i(p_1+p_2)}\ ,
\end{equation}
which vanishes precisely on physical states.\smallskip

To see the systematic procedure of how to choose the phases consistently, we look at a three-magnon state. One can check that the central extension vanishes on physical states if we take the action of $\wt{S}_2$ to be
\begin{align}
\wt{S}_2\,\alpha^2(p_1)\alpha^2(p_2)\alpha^2(p_3)\ket{w} =& \frac{g}{2i}\,\frac{1-e^{-2\pi i p_1}}{\sqrt{1-p_1}}e^{-2i\pi (p_2+p_3)}\,\psi^-(p_1)\alpha^2(p_2)\alpha^2(p_3)\ket{w+1}\nonumber\\
&+ \frac{g}{2i}\,\frac{1-e^{-2\pi i p_2}}{\sqrt{1-p_2}}e^{-2i\pi p_3}\,\alpha^2(p_1)\psi^-(p_2)\alpha^2(p_3)\ket{w+1}\nonumber\\
&+ \frac{g}{2i}\,\frac{1-e^{-2\pi i p_3}}{\sqrt{1-p_3}}\,\alpha^2(p_1)\alpha^2(p_2)\psi^-(p_3)\ket{w+1}\ .
\end{align}
This should be interpreted as follows. The supercharge $\wt{S}_2$ acts on only one of the magnons at a time, corresponding to a fermion-boson contraction for this magnon, while the other two magnons have a boson-boson contraction, leading to a phase. We choose the phase such that it is trivial if the supercharge acts to the \textit{right} of the magnon, and $e^{-2\pi i p}$ if it acts to the \textit{left}. This necessitates an ordering of the magnons, which is not present for finite $w$, where all the different magnon operators commute with one another. With this convention, the anti-commutator $\{S_1,\wt{S}_2\}$ on the three-magnon state is then proportional to
\begin{equation}
\bigl(1-e^{-2\pi i p_1}\bigr)e^{-2 i \pi (p_2+p_3)}+ \bigl(1-e^{-2\pi i p_2}\bigr)e^{-2i\pi p_3}+ \bigl(1-e^{-2\pi i p_3}\bigr) = 1-e^{-2\pi i (p_1+p_2+p_3)}\ .
\end{equation}
It is not difficult to see that this argument generalises to $n$ magnons --- in fact the systematics is very much the same as that in \cite{Beisert:2005tm}, and it relies on the telescoping identity
\be\label{telescope}
\sum_{k=1}^n(1-e^{-2\pi i p_k}) \prod_ {j=k+1}^n e^{-2\pi i p_j} = (1-e^{-2\pi i \sum_k p_k})\ .
\ee

The dependence of the states on the ordering of the magnons also arose in the 4D case \cite{Beisert:2005tm}, and it is a consequence of the strict infinite $w$ limit in which the cyclic symmetry that was present for finite $w$ is lost and we need to keep the linear ordering in mind. In order to reflect this also in our notation, we denote in the following the multi-magnon states in the infinite $w$ limit by
\be
| A_1(p_1), A_2(p_2), \cdots, A_l(p_l)\rangle \equiv \lim_{w\to \infty} A_1(p_1)\, A_2(p_2)\ \cdots A_l(p_l) \ket{w} \ ;
\ee
this is inspired by the notation that was, for example, used in \cite{Ahn:2010ka}.

\subsection{Algebraic description}

Given that each supercharge basically acts on one magnon at a time, we can now write the action compactly in terms of commutators (and anti-commutators), except that we need to implement the above phase choice consistently. It is not difficult to see that the above results amount to the prescription that
\begin{equation}\label{tSa2}
[\wt{S}_2,\alpha^2(p)]=\frac{g}{2 i}\frac{1-e^{-2\pi i p}}{\sqrt{1-p}} \,\psi^-(p)\,\mcl{Z}_+\ ,
\end{equation}
where $\mcl{Z}_+$ is the operator of eq.~(\ref{Zdef}). Furthermore, in order to implement the correct choice of phases we require that
\begin{equation}\label{Zal}
\mcl{Z}_\pm\,\alpha^2(p) = e^{\mp 2\pi i p}\,\alpha^2(p)\,\mcl{Z}_\pm\ ,
\end{equation}
which is the natural analogue of \cite{Beisert:2005tm}, where the $\mcl{Z}_\pm$ operators insert or remove a site from the spin-chain.

As a very simple example, let us check that this reproduces (\ref{ans1}) on our archetypal two-magnon state. Setting
\begin{equation}\label{cp}
c_1(p)= \frac{g}{2 i}\, \frac{1-e^{-2\pi i p}}{\sqrt{1-p}} \ ,
\end{equation}
where the index `$1$' refers to the fact that this is the result to first order in perturbation theory,
\begin{align}
\wt{S}_2\, |\alpha^2(p_1),\alpha^2(p_2)\rangle &= \bigl| [\wt{S}_2,\alpha^2(p_1)],\alpha^2(p_2)\bigr\rangle +\bigl|\alpha^2(p_1),[\wt{S}_2,\alpha^2(p_2)] \bigr\rangle\nonumber\\
& \hspace*{-1cm} = c_1(p_1)\, \bigl| \psi^-(p_1)\mcl{Z}_+, \alpha^2(p_2)\bigr\rangle
+ c_1(p_2)\,\bigl|\alpha^2(p_1),\psi^-(p_2)\mcl{Z}_+\bigr\rangle \\
& \hspace*{-1cm}  = c_1(p_1)e^{-2 \pi i p_2}\, \bigl| \psi^-(p_1),\alpha^2(p_2)\,\mcl{Z}_+\bigr\rangle
+ c_1(p_2)\, \bigl| \alpha^2(p_1),\psi^-(p_2)\,\mcl{Z}_+\bigr\rangle \ , \nonumber
\end{align}
thereby implementing precisely the phases from (\ref{ans1}). The other cases work similarly.

\subsection{The exact dispersion relation}\label{subsec:dispersion}

Abstracting from our leading order (in perturbation theory) answer, we can now consider the general form of the action of the supercharges on the magnons in the infinite $w$ limit. Following the logic of  \cite{Beisert:2005tm} we want to show that the consistency of the extended superalgebra determines the momentum dependence of the dispersion relation for each magnon state exactly.\footnote{We shall argue later by comparison with the BMN limit, see Section~\ref{sec:BMN}, that the coupling constant dependence is also exact.} Let us begin by rewriting the relations of eq.~(\ref{deformedaction}) in (anti)-commutator form, \begin{subequations}
\begin{align}
[Q_1,\alpha^2(p)] &= a(p)\,\psi^-(p)\ , & \{Q_1,\psi^-(p)\} &= 0\ ,\\
[S_1,\alpha^2(p)] &= 0\ , & \{S_1,\psi^-(p)\} &= d(p)\,\alpha^2(p)\ ,\\
[\wt{Q}_2,\alpha^2(p)] &= 0\ ,& \{\wt{Q}_2,\psi^-(p)\} &= b(p)\,\alpha^2(p)\,\mcl{Z}_-\ ,\\
[\wt{S}_2,\alpha^2(p)] &= c(p)\,\psi^-(p)\,\mcl{Z}_+\ , & \{\wt{S}_2,\psi^-(p)\} &= 0 \ .
\end{align}
\end{subequations}
To first order in perturbation theory, we found that the coefficients satisfy --- see Section~\ref{sec:phasechoice}, e.g.\ eq.~(\ref{SStilde})
\begin{equation}\label{eq:coeff_sine_cond}
\bigl( a(p)b(p) \bigr)^* = c(p)d(p) = \frac{g}{2i}\,\bigl(1-e^{-2\pi i p}\bigr)\ .
\end{equation}
However, we also saw that this momentum dependence (and choice of phase) was essentially fixed by requiring that $\{S_1,\wt{S}_2\}$ vanishes on physical states, thanks to the identity eq.~(\ref{telescope}).
Thus as in \cite{Beisert:2005tm}, we expect on general grounds
\begin{equation}\label{eq:coeff_sine_cond2}
\bigl( a(p)b(p) \bigr)^* = c(p)d(p) = \frac{f(g)}{2i}\, \bigl(1-e^{-2\pi i p}\bigr)\ .
\end{equation}
Here $f(g)$ is not fixed  {\it a priori}, except that it is equal to $g$ in a perturbative expansion, as in eq.~(\ref{eq:coeff_sine_cond}). We will however, see that there is strong reason to believe that $f(g)=g$ to all orders in perturbation theory.

By a similar analysis we also know the corresponding action of these supercharges on the right-moving modes in (\ref{magnonsub}), which take the form
\begin{subequations}
\begin{align}
[Q_1,\tilde{\alpha}^1(p)] &= 0\ , & \{Q_1,\tilde{\psi}^-(p)\} &= \bigl(b(p)\bigr)^*\,\tilde{\alpha}^1(p)\,\mcl{Z}_- \ ,\\
[S_1,\tilde{\alpha}^1(p)] &= \bigl(c(p)\bigr)^*\,\tilde{\psi}^-(p)\,\mcl{Z}_+ \ , & \{S_1,\tilde{\psi}^-(p)\} &= 0\ ,\\
[\wt{Q}_2,\tilde{\alpha}^1(p)] &= -\bigl(a(p)\bigr)^*\,\tilde{\psi}^-(p) \ ,& \{\wt{Q}_2,\tilde{\psi}^-(p)\} &= 0 \ ,\\
[\wt{S}_2,\tilde{\alpha}^1(p)] &= 0 \ , & \{\wt{S}_2,\tilde{\psi}^-(p)\} &= -\bigl(d(p)\bigr)^*\,\tilde{\alpha}^1(p) \ .
\end{align}
\end{subequations}
This follows from left-right symmetry, leading to complex conjugates, as well as the OPE conventions for the free bosons,
\begin{equation}
\bar{\alpha}^1(x)\alpha^2(y) \sim \frac{1}{(x-y)^2}\ ,\qquad \bar{\alpha}^2(x)\alpha^1(y) \sim -\frac{1}{(x-y)^2}\ .
\end{equation}
Furthermore, as we have mentioned before, we can calculate the left and right conformal dimensions by evaluating the anti-commutators
\begin{equation}
\mcl{C} = \{Q_1,S_1\} = L_0 - K^3_0 \ ,\qquad \wt{\mcl{C}} = \{\wt{Q}_2,\wt{S}_2\} = \tilde{L}_0 - \tilde{K}^3_0\ .
\end{equation}
Let us now consider an arbitrary state
\be
\Psi = \bigl| A_1(p_1),\ldots, A_r(p_r), \tilde{A}_1(\tilde{p}_1),\ldots, \tilde{A}_s(\tilde{p}_s)\bigr\rangle \ ,
\ee
where each $A_i$ and $\tilde{A}_j$ is a left- and right-moving magnon mode from (\ref{magnonsub}), respectively. Then, the above relations imply that
\begin{align}
\mcl{C}\,\Psi &= \Bigl( \sum_{i=1}^{r} a(p_i) d(p_i)  + \sum_{j=1}^{s} b(\tilde{p}_j) c(\tilde{p}_j) \Bigr) \, \Psi \ , \nonumber \\
\wt{\mcl{C}}\,\Psi &= \Bigl( \sum_{i=1}^{r} b(p_i) c(p_i)  + \sum_{j=1}^{s} a(\tilde{p}_j) d(\tilde{p}_j) \Bigr) \, \Psi \ .
\end{align}
Because of locality, the left and right anomalous dimensions must be equal, and it follows that
\begin{equation}
\mcl{C}-\wt{\mcl{C}} = \sum_{i=1}^{r} (1-p_i) - \sum_{j=1}^{s} (1-\tilde{p}_j) \ ,
\end{equation}
from which we deduce (barring some non-local effects) that
\begin{equation}\label{eq:coeff_ccharge_cond}
a(p)d(p)-b(p)c(p) = 1-p\ .
\end{equation}
Combining eqs.~(\ref{eq:coeff_sine_cond2}) and (\ref{eq:coeff_ccharge_cond}), we therefore have for each $p$,
\begin{align}\label{calc1}
\bigl(a(p)d(p)+b(p)c(p)\bigr)^2 &= \bigl(a(p)d(p)-b(p)c(p)\bigr)^2 + 4\,a(p)b(p)c(p)d(p)\nonumber\\
&= (1-p)^2 + 4\, g^2\,\sin^2(\pi p) \equiv \epsilon(p)^2\ ,
\end{align}
where we have anticipated that $f(g)=g$ to all orders.
This gives us the magnon dispersion relation $\epsilon(p)$
\begin{tcolorbox}[ams equation] \label{epsp}
\epsilon(p) = \sqrt{ (1-p)^2 + 4 \, g^2 \, \sin^2(\pi p)} \ .
\end{tcolorbox}
\noindent Hence, we find for the total Hamiltonian the eigenvalue
\begin{align}\label{scaling}
L_0+\wt{L}_0-K^3_0-\wt{K}{}^3_0 & = \sum_{i=1}^{r} \sqrt{(1-p_i)^2+ 4\,g^2\,\sin^2(\pi p_i)}
+ \sum_{j=1}^{s} \sqrt{(1-\tilde{p}_j)^2+ 4\,g^2\,\sin^2(\pi \tilde{p}_j)} \nonumber \\
& = \sum_{i=1}^{r} \epsilon(p_i) + \sum_{j=1}^{s}  \epsilon(\tilde{p}_j)     \ .
\end{align}
This describes a spectrum of multi-magnon states reflecting the underlying integrability.

For the following it will be convenient to choose suitable normalisation conventions so that
\begin{align}\label{coeff-mom-dep}
a(p)=d(p) &=  \sqrt{\tfrac{1}{2}\bigl((1-p)+\epsilon(p)\bigr)}\ , \qquad
b(p)=\bigl(c(p)\bigr)^* = \frac{g}{2i}\,\frac{e^{2\pi i p}-1}{\sqrt{\tfrac{1}{2}\bigl((1-p) + \epsilon(p) \bigr)}}\ .
\end{align}
Obviously, this reproduces our previous results, see in particular eq.~(\ref{cp}), to first order in $g$. We also note that, to quadratic order in $g$, it is of the form
\begin{align}
\sqrt{(1-p)^2+ 4\,g^2\,\sin^2(\pi p)} & = (1-p) +  2\,g^2\,\frac{\sin^2(\pi p)}{1-p} + \mcl{O}\bigl(g^4\bigr) \\
& = (1-p) +2\,\wt{\gamma}  \ ,
\end{align}
thus agreeing with (\ref{eq:anom_dim_limit}). Actually, for the analysis of the $S$-matrix, see Section~\ref{sec:s-matrix} below, it is useful to rewrite these coefficients in terms of Zhukovski variables,
\begin{equation}
a(p) = d(p) = \eta_p\ ,\qquad b(p)=\bigl(c(p)\bigr)^*=\frac{\eta_p}{x_p}e^{i\pi p}\ ,
\end{equation}
where the Zhukovski variables satisfy the conditions
\begin{equation}
\eta_p^2 = g\,x_p\,\sin(\pi p)\ ,\qquad x_p-\frac{1}{x_p} = \frac{1-p}{g\,\sin(\pi p)}\ .
\end{equation}

\subsection{Full symmetry algebra}\label{sec:fullsup}

Before we come to the discussion of the $S$-matrix, we need to extend our analysis to the full set of modes. This is to say, we now consider the larger algebra annihilating the state $\ket{w}$, given by (see (\ref{N4super}) and (\ref{N4subset}))
\begin{equation}\label{eq:symm_alg_modes}
Q_i\ ,\quad S_i\ ,\quad \wt{Q}_i\ ,\quad \wt{S}_i\ ,\quad i=1,2\ .
\end{equation}
Furthermore, we consider both left- and right-moving magnons. Then, the algebra can mix all the states
\begin{equation}
\alpha^i(p)\ ,\quad \psi^\pm(p)\ ,\quad \wt{\alpha}{}^i(p)\ ,\quad \wt{\psi}{}^\pm(p)\ .
\end{equation}
In the same way as above, one can then determine that the general algebra relations are
\begin{subequations}\label{eq:magnon_commutators}
\begin{align}
[Q_1,\alpha^2(p)]&= \eta_p\,\psi^-(p)\ , & \{Q_1,\psi^+(p)\} &= -\eta_p\,\alpha^1(p)\ ,\\
[Q_1,\wt{\alpha}^2(p)] &= e^{-i\pi p}\frac{\eta_p}{x_p}\,\wt{\psi}^+(p)\,\mcl{Z}_-\ , & \{Q_1,\wt{\psi}^-(p)\} &=  e^{-i\pi p}\frac{\eta_p}{x_p}\,\wt{\alpha}^1(p)\,\mcl{Z}_-\ ,\\
&&&\nonumber\\
[Q_2,\alpha^1(p)]&= -\eta_p\,\psi^-(p)\ , & \{Q_2,\psi^+(p)\} &= -\eta_p\,\alpha^2(p)\ ,\\
[Q_2,\wt{\alpha}^1(p)] &= - e^{-i\pi p}\frac{\eta_p}{x_p}\,\wt{\psi}^+(p)\,\mcl{Z}_-\ , & \{Q_2,\wt{\psi}^-(p)\} &=  e^{-i\pi p}\frac{\eta_p}{x_p}\,\wt{\alpha}^2(p)\,\mcl{Z}_-\ ,\\
&&&\nonumber\\
[S_1,\alpha^1(p)]&= -\eta_p\,\psi^+(p)\ , & \{S_1,\psi^-(p)\} &= \eta_p\,\alpha^2(p)\ ,\\
[S_1,\wt{\alpha}^1(p)] &= e^{i\pi p}\frac{\eta_p}{x_p}\,\wt{\psi}^-(p)\,\mcl{Z}_+\ , & \{S_1,\wt{\psi}^+(p)\} &= e^{i\pi p}\frac{\eta_p}{x_p}\,\wt{\alpha}^2(p)\,\mcl{Z}_+\ ,\\
&&&\nonumber\\
[S_2,\alpha^2(p)]&= -\eta_p\,\psi^+(p)\ , & \{S_2,\psi^-(p)\} &= -\eta_p\,\alpha^1(p)\ ,\\
[S_2,\wt{\alpha}^2(p)] &= e^{i\pi p}\frac{\eta_p}{x_p}\,\wt{\psi}^-(p)\,\mcl{Z}_+\ , & \{S_2,\wt{\psi}^+(p)\} &= -e^{i\pi p}\frac{\eta_p}{x_p}\,\wt{\alpha}^1(p)\,\mcl{Z}_+\ .
\end{align}
\end{subequations}
All other commutators are zero. The action of the right-moving charges is analogous, but with the complex conjugate phases in the coefficients. These relations are already written in terms of the Zhukovsky variables
\begin{equation}\label{eq:zhukovski_rel}
\eta_p^2 = g\,x_p\,\sin(\pi p)\ ,\qquad x_p-\frac{1}{x_p} = \frac{1-p}{g\,\sin(\pi p)}
\end{equation}
for convenience; in this language it is also easy to see that the coefficients are essentially all equal (up to some signs). One can check that with these coefficients, the central extension vanishes on all physical two magnon states provided that we have, c.f.\  eq.~(\ref{Zal})
\be\label{Zal1}
\mcl{Z}_\pm\, A(p) = e^{\mp 2\pi i p}\, A(p)\,\mcl{Z}_\pm\ , \qquad
\mcl{Z}_\pm\,\tilde{A}(\tilde{p}) = e^{\pm 2\pi i \tilde{p}}\,\tilde{A}(\tilde{p})\,\mcl{Z}_\pm\ ,
\ee
where $A(p)$ and $\tilde{A}(\tilde{p})$ are arbitrary left- and right-moving magnons, respectively.

\subsection{The barred magnons}\label{sec:barred}

The analysis for the barred generators is essentially the same, and the corresponding contractions just differ by some signs. More specifically, we have
\be\label{barcont}
\begin{tikzpicture}[baseline=-.5ex]
\node[](n1)at(0,0){${\psi}^+_{m}$};
\node[](n2)[right=-1ex of n1]{$\bar\alpha^2_{n}$};
\draw[](n1.south)-- ([yshift=-1ex]n1.south)--([yshift=-1ex]n2.south)node[midway]{$\times$}--(n2.south);
\end{tikzpicture}   = -
\begin{tikzpicture}[baseline=-.5ex]
\node[](n1)at(0,0){$\bar{\psi}^+_{m}$};
\node[](n2)[right=-1ex of n1]{$\alpha^2_{n}$};
\draw[](n1.south)-- ([yshift=-1ex]n1.south)--([yshift=-1ex]n2.south)node[midway]{$\times$}--(n2.south);
\end{tikzpicture} \ , \qquad
\wick{\c {\bar{\alpha}^1_m}\,\,\c{\alpha^2_n}}  = \wick{\c {{\alpha}^1_m}\,\,\c{\bar\alpha^2_n}} \ , \qquad
\wick{\c {{\psi}^+_m}\,\,\c{\bar\psi^-_n}}  = \wick{\c {\bar{\psi}^+_m}\,\,\c{\psi^-_n}} \ ,
\ee
and similarly for the other terms. As a consequence also the large $w$ limit behaves the same as for the unbarred modes, i.e.\ we have as in (\ref{Zal1})
\begin{equation}\label{Zalb}
\mcl{Z}_\pm\,\bar\alpha^2(p) = e^{\mp 2\pi i p}\,\bar{\alpha}^2(p)\,\mcl{Z}_\pm\ .
\end{equation}
In particular, these magnons therefore satisfy the same dispersion relation as the unbarred modes, see eq.~(\ref{epsp}).

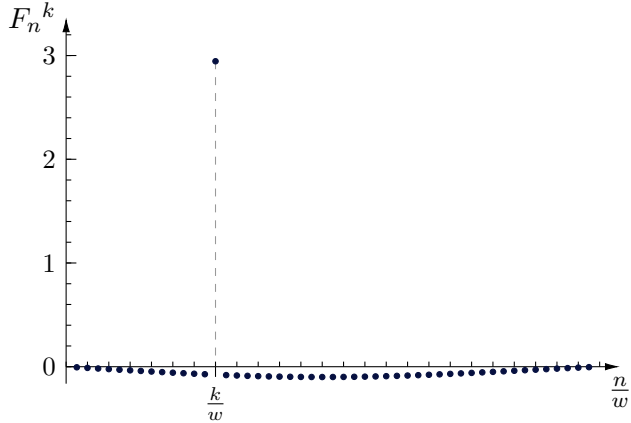
\begin{figure}[ht]
\centering

\definecolor{graphdarkblue}{HTML}{061040}
\begin{tikzpicture}[scale=0.9]
\draw[black,-{Latex[length=2mm, width=1mm]}](0,0.25)--(8.1,0.25)node[anchor=north]{$\frac{n}{w}$};
\draw[black,-{Latex[length=2mm, width=1mm]}](0,0)--(0,5.4)node[anchor=east]{$F_n{}^k$};
\foreach \y in {0.25,0.555614,0.861227,1.16684,1.47245,1.77807,2.08368,2.3893,2.69491,3.00052,3.30614,3.61175,3.91736,4.22298,4.52859,4.83421,5.13982}
{
	\draw[](0,\y)--(0.075,\y);
}
\foreach \a/\y in {0/0.25,1/1.77807,2/3.30614,3/4.83421}
{
	\draw[](0,\y)--(0.15,\y);
	\node[]at(-0.25,\y){{\small\a}};
}
\draw[color=black!40,dashed](2.184,4.75)--(2.184,0.25);
\draw[](2.184,0.25)--(2.184,0.1)node[anchor=north]{\small$\frac{k}{w}$};
\foreach \x in {0.,0.312,0.624,0.936,1.248,1.56,1.872,2.184,2.496,2.808,3.12,3.432,3.744,4.056,4.368,4.68,4.992,5.304,5.616,5.928,6.24,6.552,6.864,7.176,7.488,7.8}
{
	\draw[](\x,0.25)--(\x,0.325);
}
\foreach \x/\y in {0.156/0.24131,0.312/0.232487,0.468/0.223556,0.624/0.214558,0.78/0.205543,0.936/0.196562,1.092/0.187669,1.248/0.17892,1.404/0.17037,1.56/0.162073,1.716/0.154083,1.872/0.146454,2.028/0.139236,2.184/4.75,2.34/0.126219,2.496/0.120507,2.652/0.115379,2.808/0.110866,2.964/0.107,3.12/0.103805,3.276/0.101299,3.432/0.0994994,3.588/0.0984142,3.744/0.0980479,3.9/0.0983997,4.056/0.099463,4.212/0.101226,4.368/0.103673,4.524/0.106782,4.68/0.110526,4.836/0.114874,4.992/0.119792,5.148/0.12524,5.304/0.131176,5.46/0.137553,5.616/0.144323,5.772/0.151435,5.928/0.158836,6.084/0.166472,6.24/0.174287,6.396/0.182224,6.552/0.190228,6.708/0.198242,6.864/0.206211,7.02/0.214079,7.176/0.221792,7.332/0.229298,7.488/0.236542,7.644/0.243468}{
	\filldraw[color=graphdarkblue](\x,\y)circle(0.04);
}
\end{tikzpicture}

\caption{A plot of the coefficients $F_n{}^k$ from eq.~(\ref{Fdef}) for $w=50$ and $k=10$. They are strongly peaked at $n=k$ and small of order $1/w$ and negative otherwise. The sum over all $n$ gives zero.}
\label{fig:2}
\end{figure}

On the face of it this seems to lead to a contradiction. For example, we know that $K^-_0 = - (\bar\psi^- \psi^-)_0$ is part of the ${\cal N}=4$ superconformal algebra, and hence
\be\label{conserved}
\bigl( \tilde{L}_0 - \tilde{K}^3_0 \bigr)\, K^-_0 \,  |{\rm BPS}\rangle_w  = 0 \ ,
\ee
in the perturbed theory, i.e.\ for all values of $g$. On the other hand, in terms of the individual magnon modes we have, see eq.~(\ref{N4fields})
\be\label{sumex}
K^-_0 \,  |{\rm BPS}\rangle_w = - \sum_{n=1}^{w-1} \, :\bar{\psi}^-_{-\frac{1}{2} + \frac{n}{w}} \psi^-_{-\frac{1}{2} +(1 -\frac{n}{w})} : \, |{\rm BPS}\rangle_w \ ,
\ee
and in the large $w$ limit the eigenvalue of $\tilde{L}_0 - \tilde{K}^3_0$ equals
\begin{align}
&  \{\tilde{Q}_2 , \tilde{S}_2\}\,  \bar{\psi}^-_{-\frac{1}{2} + \frac{n}{w}} \psi^-_{-\frac{1}{2} +(1 -\frac{n}{w})}  \, |{\rm BPS}\rangle_w  \\
& \quad \cong \!\Bigl(\sqrt{(1-\tfrac{n}{w})^2 + 4 g^2 \sin^2(\pi \tfrac{n}{w})} +
\sqrt{(\tfrac{n}{w})^2 + 4 g^2 \sin^2(\pi \tfrac{n}{w})} - 1 \Bigr) \bar{\psi}^-_{-\frac{1}{2} + \frac{n}{w}} \psi^-_{-\frac{1}{2} +(1 -\frac{n}{w})}  \, |{\rm BPS}\rangle_w \,,  \nonumber
\end{align}
which does not seem to reproduce (\ref{conserved}). The resolution of this puzzle is that the sum in (\ref{sumex}) involves $w-1$ terms, and we cannot take the large $w$ limit of each term individually (since the number of terms also grows with $w$). In fact, evaluating the left-hand-side for finite $w$ leads to an expression of the form
\begin{align}\label{Fdef}
& \{\tilde{Q}_2 , \tilde{S}_2\}\, \bar{\psi}^-_{-\frac{1}{2} + \frac{n}{w}} \psi^-_{-\frac{1}{2} +(1 -\frac{n}{w})}  \, |{\rm BPS}\rangle_w =
\sum_{k} \, F_{n}{}^{k} \, \bar{\psi}^-_{-\frac{1}{2} + \frac{k}{w}} \psi^-_{-\frac{1}{2} +(1 -\frac{k}{w})}  \, |{\rm BPS}\rangle_w \ .
\end{align}
In the large $w$ limit, the coefficients $F_{n}{}^{k}$ are peaked at $k=n$, but the other terms are non-zero, see Figure~\ref{fig:2} for a specific example, and the sum over $n$ (for fixed $k$) actually turns out to vanish.

\section{Comparison to the BMN limit}\label{sec:BMN}

In this section we shall study how our symmetric orbifold results can be related to the BMN limit. This will allow us to argue that our dispersion relation, see eq.~(\ref{epsp}), is actually exact to all orders in $g$.

\subsection{The BMN spectrum}

In the BMN limit, the worldsheet theory has $4+4$ massive modes associated to the $\text{AdS}_3\times {\rm S}^3$ directions, and $4+4$ massless modes associated to the $\mathbb{T}^4$ \cite{Berenstein:2002jq}. (The latter behave just like in ordinary flat space string theory.)

Let us denote the massive boson modes\footnote{In the following, we shall be concentrating on the bosons associated to ${\rm AdS}_3 \times {\rm S}^3$; the analysis for the fermions is similar.} by $a^{i \, \dagger}_n$ and $b^{i \, \dagger}_n$, where $i\in\{1,2\}$ and $n$ runs over all the integers, $n\in\mathbb{Z}$, see Appendix~\ref{BMNdis} for our conventions. These modes satisfy the commutation relations with their corresponding annihilation operators
\be\label{acomm}
\bigl[\, a^{i}_n, a^{{j}\dagger}_m\, \bigr] = \delta^{ij}\, \delta_{nm} \ , \qquad \bigl[\, b^{{i}}_n, b^{{j}\dagger}_m\, \bigr] = \delta^{ij}\, \delta_{nm} \ .
\ee
Their dispersion relation was derived in \cite{Berenstein:2002jq,Hoare:2013lja}\footnote{In \cite{Berenstein:2002jq} only the first solution was given; it was subsequently realised in \cite{Hoare:2013pma} that this is only true for half the modes. We mention in passing that the dictionary of \cite{Gomis:2002qi,Gava:2002xb} was based on the old relations of \cite{Berenstein:2002jq}, but it is actually not quite consistent and one needs the modification of \cite{Hoare:2013pma} to make things work. For completeness we derive these modified dispersion relations in Appendix~\ref{BMNdis}.}
\be\label{eq:BMN_dispersion}
\omega_n = \sqrt{\sin^2\alpha + \Bigl( \cos\alpha+ \frac{n}{\alpha' p^+}  \Bigr)^2}  \ , \qquad
\bar{\omega}_n = \sqrt{\sin^2\alpha + \Bigl( \cos\alpha- \frac{n}{\alpha' p^+}  \Bigr)^2} \ ,
\ee
where $\alpha$ parametrises the amount of NS-NS flux ($q_{\rm NS}$) and R-R flux ($q_{\rm R}$) via,
\be
\cos\alpha = \frac{q_{\rm NS}}{\sqrt{q_{\rm NS}^2 + g_s^2 q_{\rm R}^2}} \ .
\ee
This is to say, the light-cone Hamiltonian (restricted to the massive modes of
the $\mrm{AdS}_3\times\mrm{S}^3$ directions) is given by
\be\label{lcH}
H_{\rm lc} = \sum_{i=1,2} \sum_{n\in \mathbb{Z} }  \bigl(  \omega_n\, a^{{i}\, \dagger}_n\, a^i_n  + \bar{\omega}_n {b}^{i\, \dagger}_n\, {b}^{{i}}_n \bigr)  \ .
\ee
The magnon excitations on the BMN worldsheet are constrained by the requirement that they need to satisfy a zero-momentum condition, and in terms of the above modes, it takes the form, see \cite[eq.~(3.3)]{Berenstein:2002jq}
\be\label{peri1}
\sum_{i=1,2} \sum_{n\neq 0 } n \bigl( a^{{i}\, \dagger}_n\, a^i_n  + {b}^{i\, \dagger}_n\, {b}^{{i}}_n \bigr) = 0 \ .
\ee
The flat $\mbb{T}^4$ directions are described by the Hamiltonian
\begin{equation}
H_{\mbb{T}^4} = \sum_{j=1}^4\sum_{n>0}\frac{n}{\alpha'
p^+}\,c^{j\,\dagger}_n\,c^j_n +
\text{right movers}\ .
\end{equation}

\subsection{Spectrum equivalence}

We now want to show that the spectrum generated by these BMN modes is reproduced by our deformed symmetric orbifold theory. In particular, this will allow us to identify the coupling constant $g$ with the R-R flux of the ${\rm AdS}_3 \times {\rm S}^3$ background. It will also suggest that our dispersion relation is actually exact to all orders in $g$.

For the purpose of showing that the spectra of the two descriptions coincide in the appropriate limit, we identify the massive ${\rm AdS}_3 \times {\rm S}^3$ BMN modes with the fractional ${\cal N}=4$ modes, following
\cite{Lunin:2002fw,Gomis:2002qi,Gava:2002xb}.\footnote{After this paper was posted on the {\tt arXiv}, \cite{Frolov:2023pjw} appeared in which the authors proposed that the fractional torus modes of the symmetric orbifold should be identified with the massless torus modes of the BMN description. They also argued that the massive ${\rm AdS}_3 \times {\rm S}^3$ BMN modes are invisible from the symmetric orbifold perspective. We agree with the first point, but not with the second. The following analysis was modified after their paper appeared.} More specifically, we identify
\be\label{a1mapfin}
\begin{array}{ll}
a^{1\, \dagger}_n \,  \longleftrightarrow \,  L_{-1-p(n)} \ , \qquad
& b^{1\, \dagger}_{n} \,  \longleftrightarrow \,  \tilde{L}_{-1+p(n)} \ ,  \\
a^{2\, \dagger}_n \,  \longleftrightarrow \,  K^-_{-p(n)} \ , \qquad
& b^{2\, \dagger}_{n} \,  \longleftrightarrow \,  \tilde{K}^-_{p(n)} \ ,
\end{array}
\ee
where the relation between $p(n) \sim n$ will be derived momentarily.  As was mentioned in Footnote~\ref{foot1}, we have also repeated the analysis of Section~\ref{strategy} for the fractional ${\cal N}=4$ modes (instead of the fractional torus modes), and the structure seems to be essentially the same, i.e.\ we find\footnote{Note that there is no contradiction in this since these ${\cal N}=4$ modes are probably not independent excitations, see also the discussion in Section~\ref{sec:barred}.}
\be
L_{-1+p}\ , \  \ \tilde{L}_{-1+p}\ , \ \ K^-_{p}\ , \ \ \tilde{K}^-_{p}: \qquad
\epsilon(p) = \sqrt{ (1-p)^2 + 4 \, g^2 \, \sin^2(\pi p)} \ .
\ee
In order to have the light-cone Hamiltonian match up with eq.~(\ref{scaling}), we therefore want to identify, see eq.~(\ref{epsp})
\be
\omega_n = \epsilon\bigl(-p(n)\bigr)\ , \qquad \bar{\omega}_n = \epsilon\bigl(p(n)\bigr) \ .
\ee
Since in the BMN regime
$n \ll \alpha' p_+ \mu$, we need to expand both sides in a power series of the respective parameters, and this then leads to the identifications
\be
2 \cos\alpha\, \frac{n}{\alpha' p_+ \mu} =   2 p(n) \ , \qquad
\Bigl( \frac{n}{\alpha' p_+ \mu} \Bigr)^2 = p(n)^2 (1 +  4\pi^2 g^2) \ ,
\ee
from which we deduce that we have the identification
\be
p(n) = \frac{1}{ \sqrt{1 +  4\pi^2 g^2}}\,  \frac{n}{\alpha' p_+ \mu}  \ , \qquad \hbox{and} \qquad
\cos\alpha = \frac{1}{\sqrt{1 +  4\pi^2 g^2}} \ .
\ee
The first identity means, in particular, that the BMN modes stay within one Brillouin zone ($0<p(n)\ll 1$). Since the
symmetric orbifold is dual to the background with $q_{\rm NS}=1$ and $q_{\rm R}=0$, we find from the second identity that our coupling constant is to be identified with
\be
2 \pi g \equiv g_s \, q_{\rm R} \ .
\ee
Thus the parameter $g$ plays the role of the 't Hooft coupling in super Yang-Mills theory.
In particular, the symmetric orbifold deformation therefore describes switching on R-R flux.

Finally, we note that with the above identifications, the zero-momentum condition (\ref{peri1}) corresponds to the orbifold invariance condition, and so the zero-momentum BMN states are accounted for in terms of orbifold invariant (or physical) magnon states.

Next we can use that the BMN answer is exact in the R-R flux, i.e.\ exact in $g$. Recall that, as argued in Section~\ref{subsec:dispersion},  the constraint that the anti-commutators of the form (\ref{SStilde}) have to vanish to all orders, fixes the momentum dependence of the coefficients $a(p)$, $b(p)$, $c(p)$ and $d(p) $  --- see eq.~(\ref{coeff-mom-dep}). Taken together, these arguments therefore suggest that our result is actually true to all orders in $g$. (This is analogous to the argument made in \cite{Beisert:2005tm}.)

Finally, the massless torus BMN modes $c^i_n$ are to be identified schematically with the fractional torus modes as \footnote{We are following here the suggestion of  \cite{Frolov:2023pjw}. The compatibility with the BMN limit was, however, not checked there.}
\be
c^{i\, \dagger}_n\leftrightarrow \alpha^i_{-\frac{n}{w}} \ .
\ee
In terms of $p=\frac{n}{w}$, the dispersion relation of the fractional symmetric orbifold modes is
\begin{align}
\hat{\omega}(p) & = \sqrt{p^2 + 4 g^2 \sin^2(\pi p)}  = p \, \sqrt{1 + 4 \, \frac{g^2 \sin^2(\pi p)}{p^2} }  \\
& = p \sqrt{1+ 4\, g^2 \pi^2}  + {\cal O}(p^3) \ .
\end{align}
The first factor accounts for the fact that the size of ${\rm AdS}_3$ and ${\rm S}^3$ has been modified as a function of $g$, but there is no correction term at order ${\cal O}(p^2)$. This is then compatible with the fact that the BMN dispersion relation for the massless torus modes is trivial.

\section{S-matrix}\label{sec:s-matrix}

In Section~\ref{sec:effective_description} we have shown that, in the large $w$ limit, we can effectively describe the action of the various supercharges on the multi-magnon states by eq.~(\ref{eq:magnon_commutators}), together with the prescription of how the length-changing operators ${\cal Z}_{\pm}$ can be moved past the magnon operators, see eq.~(\ref{Zal1}). As a consequence the states now depend on the ordering of the magnons, but one would expect that different orderings are related to one another by a $2$-particle `$S$-matrix'. It is the aim of this section to construct this $S$-matrix by requiring that it commutes with the action of the symmetry generators of eq.~(\ref{eq:symm_alg_modes}), again following the strategy of \cite{Beisert:2005tm}.\footnote{In the context of $\text{AdS}_3$, worldsheet $S$-matrices have been calculated using the same methods, for example, in \cite{Borsato:2014hja}, and our results were subsequently argued to agree with those found in these papers \cite{Frolov:2023pjw}.} As we shall see, these constraints  fix the $S$-matrix up to a few overall factors, and the resulting solution automatically satisfies the Yang-Baxter equation.

While for $\mcl{N}=4$ SYM, one can think of the $S$-matrix as completing the asymptotic states into eigenstates of the spin-chain Hamiltonian, this interpretation is not yet available to us since we have not yet identified the underlying (integrable) spin chain. However, the results of this paper suggest that such a description should be possible, and it would be interesting to work it out in detail.

\subsection{Ansatz}

To start with we make the most general ansatz for the $S$-matrix when acting on two magnon states, and we will take it to interchange the two momenta. We will also assume that the $S$-matrix does not change left- into right-movers or vice versa, i.e.\ we can separately consider the four `sectors'  LL, LR, RL, and RR, where L and R stands for a left- or a right-mover, respectively.\footnote{Furthermore, we can also have scatterings between barred and unbarred modes, and thus get additional sectors $\text{L}\bar{\text{L}}$, etc.~These sectors behave analogously to the ones we study explicitly here.}  As we shall see, the condition to commute with the supercharges fixes the $S$-matrix in each sector up to one overall coefficient.

\subsubsection{LL-Scattering}

To be more explicit, let us consider the LL sector, for which we make the most general ansatz
\begin{subequations}\label{eq:LL_S-matrix_form}
\begin{align}
\mscr{S}\, \bigl| \alpha^i(p),\alpha^j(q)\bigr\rangle  =& A_\mrm{LL}^{ij}\!(p,q)\, \bigl|\alpha^j(q),\alpha^i(p)\bigr\rangle
+ \bigl(1-\delta^{ij}\bigr)\, B_\mrm{LL}^{ij}\!(p,q)\,\bigl|\alpha^i(q),\alpha^j(p)\bigr\rangle\nonumber\\
&\hspace*{-2cm} + \bigl(1-\delta^{ij}\bigr)\,K_\mrm{LL}^{ij}\!(p,q)\,\bigl|\psi^+(q),\psi^-(p)\bigr\rangle
+ \bigl(1-\delta^{ij}\bigr)\,L_\mrm{LL}^{ij}\!(p,q)\,\bigl|\psi^-(q),\psi^+(p)\bigr\rangle\ ,\\
\mscr{S}\, \bigl|\alpha^i(p),\psi^b(q)\bigr\rangle =& C_\mrm{LL}^{ib}\!(p,q)\,\bigl|\psi^b(q),\alpha^i(p)\bigr\rangle + D_\mrm{LL}^{ib}\!(p,q)\, \bigl|\alpha^i(q),\psi^b(p)\bigr\rangle\ ,\\
\mscr{S}\, \bigl|\psi^a(p),\alpha^j(q)\bigr\rangle =& E_\mrm{LL}^{aj}\!(p,q)\,\bigl|\alpha^j(q),\psi^a(p)\bigr\rangle +F_\mrm{LL}^{aj}\!(p,q)\,\bigl|\psi^a(q),\alpha^j(p)\bigr\rangle\ ,\\
\mscr{S}\,\bigl|\psi^a(p),\psi^b(q)\bigr\rangle =& H_\mrm{LL}^{ab}\!(p,q)\, \bigl|\psi^b(q),\psi^a(p)\bigr\rangle + \bigl(1-\delta^{ab}\bigr)\,J_\mrm{LL}^{ab}\!(p,q)\,\bigl|\psi^a(q),\psi^b(p)\bigr\rangle \nonumber\\
&\hspace*{-2cm}  +\bigl(1-\delta^{ab}\bigr)\,M_\mrm{LL}^{ab}\!(p,q)\,\bigl|\alpha^1(q),\alpha^2(p)\bigr\rangle + \bigl(1-\delta^{ab}\bigr)\,N_\mrm{LL}^{ab}\!(p,q)\,\bigl|\alpha^2(q),\alpha^1(p)\bigr\rangle\ .
\end{align}
\end{subequations}
Here, $i,j\in\{1,2\}$ and $a,b\in\{+,-\}$, and we have only considered terms the states on the two sides have the same $\mathfrak{su}(2)$ charge and the same conformal dimension --- this is a consequence of the fact that the $S$-matrix must also commute with the bosonic symmetries.

Next we impose the constraint that the $S$-matrix must commute with the supercharge generators,
\begin{equation}
0=[O,\mscr{S}]\ ,\qquad O\in\{Q_i,S_i,\wt{Q}_i,\wt{S}_i\}\ .
\end{equation}
For example, using eq.~(\ref{eq:magnon_commutators}), the requirement that
\begin{equation}
0 = [\wt{Q}_1,\mscr{S}]\, \bigl|\alpha^2(p),\psi^-(q)\bigr\rangle
=[S_1,\mscr{S}]\, \bigl|\alpha^2(p),\psi^-(q)\bigr\rangle
\end{equation}
leads to four equations for each commutator. To be specific, let us focus on those equations relating $A_\text{LL}^{21}$, $C_\text{LL}^{2\,-}$, and $M_\text{LL}^{+-}$, which equal
\begin{align}
0 &= \eta_q\, A_\text{LL}^{21}(p,q) - \eta_q \,C_\text{LL}^{2\,-}(p,q) +\eta_p\, M_\text{LL}^{+-}(p,q)\ ,\nonumber\\
0 &= \frac{\eta_q}{x_q}e^{i\pi q} \,A_\text{LL}^{21}(p,q) -\frac{\eta_q}{x_q}e^{i\pi(q+2p)} \,C_\text{LL}^{2\,-}(p,q) + \frac{\eta_p}{x_p}e^{i\pi (p+2 q)} \,M_\text{LL}^{+-}(p,q)\ .
\end{align}
The other equations have a similar form. As it turns out, they can all be simultaneously solved, using the identity\footnote{The fact that a solution exists hinges again on the fact that there are phase factors coming from the ordering of modes, see eq.~(\ref{Zal1}).}
\begin{equation}
\eta_p^2=g\,x_p\,\sin(\pi p)\ ,
\end{equation}
and the solution is unique, up to one overall scale factor which we may take to be $A^{11}_\text{LL}(p,q)$. The explicit solution is listed in appendix \ref{app:sLL-matrix_elements}.

\subsubsection{LR-Scattering}

Let us next turn to the LR sector where new features emerge. In particular, now length-changing terms can appear on the right-hand-side. More specifically, the most general ansatz respecting the conservation of the charges is
\begin{subequations}\label{eq:LR_S-matrix_form}
\begin{align}
\mscr{S}\,\bigl|\alpha^i(p),\wt{\alpha}^j(\tilde{q})\bigr\rangle =& A_\mrm{LR}^{ij}\!(p,\tilde{q})\,\bigl|\wt{\alpha}^j(\tilde{q}),\alpha^i(p)\bigr\rangle+ \bigl(1-\delta^{ij}\bigr) B_\mrm{LR}^{ij}\!(p,\tilde{q})\,\bigl|\wt{\psi}^-(\tilde{q}),\psi^-(p)\,\mcl{Z}_+\bigr\rangle\nonumber\\
&\hspace*{-2.5cm} +\bigl(1-\delta^{ij}\bigr)\,C_\mrm{LR}^{ij}\!(p,\tilde{q})\,\bigl|\wt{\psi}^+(\tilde{q}),\psi^+(p)\,\mcl{Z}_-\bigr\rangle + \bigl(1-\delta^{ij}\bigr)\,M_\mrm{LR}^{ij}\!(p,\tilde{q})\,\bigl|\wt{\alpha}^i(\tilde{q}),\alpha^j(p)\bigr\rangle\ ,\\
\mscr{S}\,\bigl|\alpha^i(p),\wt{\psi}^\pm(\tilde{q})\bigr\rangle =& D_\mrm{LR}^{i\pm}\!(p,\tilde{q})\,\bigl|\wt{\psi}^\pm(\tilde{q}),\alpha^i(p)\bigr\rangle + E_\mrm{LR}^{i\pm}\!(p,\tilde{q})\,\bigl|\wt{\alpha}^i(\tilde{q}),\psi^\mp(p)\,\mcl{Z}_\pm\bigr\rangle\ ,\\
\mscr{S}\,\bigl|\psi^\pm(p),\wt{\alpha}^j(\tilde{q})\bigr\rangle =& K_\mrm{LR}^{\pm j}\!(p,\tilde{q})\,\bigl|\wt{\alpha}^j(\tilde{q}),\psi^\pm(p)\bigr\rangle + L_\mrm{LR}^{\pm j}\!(p,\tilde{q})\,\bigl|\wt{\psi}^\mp(\tilde{q}),\alpha^j(p)\,\mcl{Z}_\pm\bigr\rangle\ ,\\
\mscr{S}\,\bigl|\psi^a(p),\wt{\psi}^b(\tilde{q})\bigr\rangle =& F_\mrm{LR}^{ab}\!(p,\tilde{q})\,\bigl|\wt{\psi}^b(\tilde{q}),\psi^a(p)\bigr\rangle + \delta^{ab}\delta^{a\pm}\big( H_\mrm{LR}^{ab}\!(p,\tilde{q})\,\bigl| \wt{\alpha}^1(\tilde{q}),\alpha^2(p)\,\mcl{Z}_\pm\bigr\rangle \nonumber\\
&+ J_\mrm{LR}^{ab}\!(p,\tilde{q})\,\bigl|\wt{\alpha}^2(\tilde{q}),\alpha^1(p)\,\mcl{Z}_\pm\bigr\rangle + N_\mrm{LR}^{ab}\!(p,\tilde{q})\,\bigl|\wt{\psi}^\mp(\tilde{q}),\psi^\mp(p)\,\mcl{Z}^2_\pm\bigr\rangle\big)\ .
\end{align}
\end{subequations}
The coefficients can again be solved by similar methods since the structure of the equations is as before. There is a unique solution, up to an overall factor which we may take to be $A_\mrm{LR}^{11}(p,q)$, see Appendix~\ref{app:sLR-matrix_elements}. We should mention that a non-trivial solution only exists provided all the length-changing terms, including the one involving ${\cal Z}_\pm^2$, are present.

\subsubsection{Left-Right symmetry}

The analysis for the RL and RR sectors is essentially the same. Indeed, the conditions one obtains are almost identical, up to the important difference that right-movers pick up the opposite phases relative to left-movers when they are moved past the operator ${\cal Z}_\pm$.  This is reflected in the solution of the $S$-matrix in that the different coefficients are related to one another as
\begin{equation}
X_\mrm{RR}(p,q) = \bigl(X_\mrm{LL}(p,q)\bigr)^*\ ,\qquad X_\mrm{RL}(p,q) = \bigl(X_\mrm{LR}(p,q)\bigr)^*\ ,
\end{equation}
if we identify
\begin{equation}
A_\mrm{RR}^{11}(p,q) = \bigl(A_\mrm{LL}^{11}(p,q)\bigr)^*\ ,\qquad A_\mrm{RL}^{11}(p,q) = \bigl(A_\mrm{LR}^{11}(p,q)\bigr)^*\ .
\end{equation}
Here, $X_\mrm{RR}$ and $X_\mrm{RL}$ denote arbitrary coefficients in the $S$-matrix action of eqs.~(\ref{eq:LL_S-matrix_form}) and (\ref{eq:LR_S-matrix_form}), respectively.

In the same way, one can also work out the $S$-matrix on states made of one unbarred and one barred magnon. The ansatz and solution are completely analogous.

\subsection{The Yang-Baxter equation}

As we have seen, the $S$-matrix is essentially fixed by these symmetry constraints. We can then ask whether it satisfies the Yang-Baxter equation (YBE). The YBE guarantees that all scattering processes are described by iterated 2-particle elastic scattering, and it is usually taken as a sign that the problem is integrable. Explicitly, the YBE is an identity concerning certain tensor products of the $S$-matrix when acting on three particle states,
\begin{equation}
\bigl(\mscr{S}\otimes \mbb{I}\bigr)\bigl(\mbb{I}\otimes\mscr{S}\bigr)\bigl(\mscr{S}\otimes \mbb{I}\bigr) = \bigl(\mbb{I}\otimes\mscr{S}\bigr)\bigl(\mscr{S}\otimes \mbb{I}\bigr)\bigl(\mbb{I}\otimes\mscr{S}\bigr)\ .
\end{equation}
As we have determined the $S$-matrix explicitly, we can directly check this identity on the 512 possible three-magnon states. We have done this using Mathematica, and we have found that all of these identities are automatically correct! In checking these equations we made use of the identity
\begin{equation}
\eta_p^2=g\,x_p\,\sin(\pi p) \ ,
\end{equation}
which is satisfied by the Zhukovski variables. (One also needs to keep track of the additional phases that arise when the magnons are moved past the length-changing operators $\mcl{Z}_\pm$, see eqs.~(\ref{Zal}) and (\ref{Zal1}).)

\subsection{Unitarity}

We can also ask whether the $S$-matrix is unitary, i.e.\ whether it defines an involution in the sense of
\begin{equation}
\mscr{S}^2 = \mbb{I}\ .
\end{equation}
We have checked, again using the explicit expressions from Appendix~\ref{app:s-matrix_elements}, that this is the case provided that we relate the so far undetermined coefficients as
\begin{equation}
A_\mrm{LL}^{11}(q,p)\,A_\mrm{LL}^{11}(p,q) = A_\mrm{RR}^{11}(q,p)\,A_\mrm{RR}^{11}(p,q) = A_\mrm{RL}^{11}(q,p)\,A_{\mrm{LR}}^{11}(p,q) = 1 \ .
\end{equation}

\subsection{Bound states}

Following the analysis of \cite{Dorey:2006dq} for $\mcl{N}=4$ SYM, we can also analyse the bound state spectrum of the left-moving magnons, say.\footnote{Related bound states were also very recently analysed in \cite{OhlssonSax:2023qrk}.} The LL S-matrix in Appendix~\ref{app:sLL-matrix_elements} develops a pole when
\begin{equation}
e^{i\pi p_2}\,x_2 - e^{-i\pi p_1}\,x_1 = 0\ ,
\end{equation}
where $x_i$ are the Zhukovski variables. Such a pole corresponds to a two-magnon bound state. Analogously, a $Q$-bound state consists of $Q$ magnons with complex momenta $p_i$, for which the $S$-matrix of the system has a pole of order $Q-1$; the total momentum of the bound state is then $p=\sum_{i=1}^Q p_i$, which we take to be in the interval $[0,1]$. Since the $S$-matrix factorises, the pole condition for the $Q$ magnon bound state is that
\begin{equation}\label{eq1}
e^{i\pi p_{i+1}}\,x_{i+1} - e^{-i\pi p_i}\,x_i = 0\ ,\qquad i=1,\dots,Q-1\ .
\end{equation}
We will find is that for each positive $Q$, there is a bound state of $Q$ magnons with dispersion relation, see Appendix~\ref{app:bound} for a derivation
\begin{equation}\label{eq:bound_state_dispersion}
\epsilon_Q(p) = \sqrt{(Q-p)^2+4g^2\,\sin^2(\pi p)}\ .
\end{equation}
This result has a rather natural interpretation. If we think of our system as being described by a spin-chain, the momenta should only take values in the range $[0,1]$, which we can think of as being the first Brillouin zone. On the other hand, in the orbifold theory the fractional modes do not satisfy any such constraint, and all the relations we have derived should be true for arbitrary $p=\frac{n}{w}$, which runs over the range $(-\infty,1]$ in our conventions. The above calculation thus suggests that the orbifold modes for which $\left|\frac{n}{w}\right| >1$ can be thought of as corresponding to bound states of magnons coming from the first Brillouin zone.

\section{Conclusion}\label{sec:conclusion}

In this paper we have studied the dynamics of the symmetric orbifold theory under the perturbation that corresponds to switching on R-R flux in the ${\rm AdS}_3 \times {\rm S}^3$ space. We have determined how the individual magnon operators (associated to the fractional torus modes) transform under the global ${\cal N}=(4,4)$  supersymmetry algebra to first order in perturbation theory. In the $w$-cycle twisted sector with large $w$, this action allowed us to deduce the anomalous dimensions of the operators, using similar arguments as for the dynamical spin chain of \cite{Beisert:2005tm}. We also compared our results with the BMN limit, and argued that our results are actually exact to all orders in the perturbation.

We have found convincing evidence that the resulting system is integrable: in particular, there is an essentially unique  2-body $S$-matrix that commutes with the action of the supersymmetry generators (to all orders in perturbation theory), and it satisfies the Yang-Baxter equations, the hallmark of integrability. Finally, the poles of the $S$-matrix allow us to read off the dispersion relations of the bound states, and they have a rather simple (and natural) structure.
\smallskip

There are many interesting questions and open problems that should now become accessible. First of all, it would be very interesting to understand how our description fits into the various proposals for ${\rm AdS}_3 \times {\rm S}^3$ that have been made over the years, see e.g.\ \cite{Borsato:2013qpa,Hoare:2013lja,Lloyd:2014bsa,OhlssonSax:2014jtq,Borsato:2016xns,Cavaglia:2021eqr}.\footnote{After this paper was posted on the {\tt arXiv},  \cite{Frolov:2023pjw} appeared in which it is argued that  our $S$-matrix agrees with that given in \cite{Lloyd:2014bsa}. In order to see the correspondence explicitly, the dictionary explained in   \cite{Borsato:2013qpa} and \cite{Borsato:2014hja} is useful.} Our analysis makes a specific prediction for the anomalous conformal dimensions of essentially all states in any large $w$ sector, and it would be interesting to confirm this independently, e.g.\ by direct computations to higher order in perturbation theory. It would also be interesting to relate our results to other explicit computations that have been performed over the years, see e.g.\ \cite{Burrington:2012yq,Gaberdiel:2015uca,Hampton:2018ygz,Guo:2019ady,Benjamin:2021zkn,Lima:2020boh,Guo:2020gxm,Apolo:2022fya,Guo:2022ifr,Hughes:2023fot}. However, these papers have mostly analysed the anomalous conformal dimensions in the untwisted (or $w=2$ twisted \cite{Guo:2019ady,Guo:2020gxm}) sector, and hence it is not possible to  compare their results directly to our large $w$ analysis. On the other hand, it is interesting that also our individual magnons acquire anomalous conformal dimensions proportional to the square root of the conformal weight $\sqrt{h}$, c.f.\ \cite{Guo:2022ifr,Hughes:2023fot}.

Another interesting direction would be to try to start from our $S$-matrix to determine the (asymptotic) Bethe equations and thus the spectrum using integrability methods. In this context it would also be very interesting to analyse the finite $w$ corrections; given that we have actually done the computation at finite $w$, see Section~\ref{strategy}, it would be very enlightening to understand how they arise from our asymptotic formulae by some sort of wrapping corrections. It would also be instructive to connect our results with the quantum spectral curve proposal of \cite{Cavaglia:2022xld}.

As we have mentioned at various stages of the paper, given that the symmetric orbifold calculation was done in terms of the covering surface, it should be possible to interpret it directly in terms of the worldsheet theory. The spin-chain like contractions that we have found should have a simple description in terms of the Feynman diagram systematics of \cite{Gaberdiel:2020ycd}. Furthermore, this should explain in which sense the spin-chain/string-bit degrees of freedom that should arise upon going to `position space' have only local interactions. Among other things, this should help towards identifying the underlying microscopic spin-chain description.

Finally, our calculations have in effect determined R-R flux corrections from a worldsheet prespective, and it would be interesting to learn the general lessons from this specific example. This should also help in informing us about how R-R flux should be incorporated in the ${\rm AdS}_5\times {\rm S}^5$ context, see in particular the proposal of \cite{Gaberdiel:2021qbb,Gaberdiel:2021jrv}.

\section*{Acknowledgements}  We thank Sujay Ashok, Niklas Beisert, Frank Coronado, Justin David, Lorenz Eberhardt, Abhijit Gadde, Gin Guo, Ben Hoare, Anthony Houppe, Bob Knighton, Shota Komatsu, Samir Mathur, Edward Mazenc, Shiraz Minwalla, Kiarash Naderi, Leonardo Rastelli, Shlomo Razamat, Ashoke Sen, Alessandro Sfondrini, Vit Sriprachyakul, Bogdan Stefanski, Arkady Tseytlin, Pedro Vieira, and Spenta Wadia for useful conversations.
MRG and RG thank the KITP in Santa Barbara for hospitality during the program on `Bootstrapping Quantum Gravity' at an early stage of the project; this was supported by the grant NSF PHY-1748958 to the Kavli Institute for Theoretical Physics (KITP). We also gratefully acknowledge the hospitality of the Pollica Physics workshop on `New connections between Physics and Number Theory' at an important intermediate stage of the work. The work of BN is supported through a personal grant of MRG from the Swiss National Science Foundation. The work of the group at ETH is supported in part by the Simons Foundation grant 994306  (Simons Collaboration on Confinement and QCD Strings), as well the NCCR SwissMAP that is also funded by the Swiss National Science Foundation. The work of RG is supported in part by the J.C.~Bose Fellowship of the DST-SERB. RG acknowledges the support of the Department of Atomic Energy, Government of India, under project no.~RTI4001, as well as the framework of support for the basic sciences by the people of India.

\appendix

\section{Conventions}\label{app:conventions}

\begingroup
\allowdisplaybreaks

In this appendix we collect our conventions for the torus fields and the ${\cal N}=4$ superconformal generators.

\subsection{\texorpdfstring{$\mcl{N}=4$ algebra}{N=4 algebra}}\label{app:algebra}

We denote the left-moving four bosons and four fermions of the $\mbb{T}^4$ by $\alpha^i,\bar{\alpha}^i$, $i=1,2$ and $\psi^\pm,\bar{\psi}^\pm$, respectively. They satisfy the OPE relations
\begin{equation}
\bar{\alpha}^i(x)\alpha^j(y)\sim \frac{\epsilon^{ij}}{\big(x-y\big)^2}\ ,\qquad \bar{\psi}^\pm(x)\psi^\mp(y)\sim \frac{\pm 1}{x-y} \ .
\end{equation}
Right-movers are always denoted by a tilde. The (left-moving) $\mcl{N}=4$ generators are built out of these fields as
\begin{equation}\label{N4fields}
\begin{array}{cclccl}
G^+ &= & \bar{\alpha}^2\,\psi^++\alpha^2\,\bar{\psi}^+\ , \qquad & K^+ &= & \bar{\psi}^+\,\psi^+\ , \\
G'^+ &= & -\bar{\alpha}^1\,\psi^+-\alpha^1\,\bar{\psi}^+\ ,\qquad & K^- &= & -\bar{\psi}^-\,\psi^-\ , \\
G^- &= & \bar{\alpha}^1\,\psi^-+\alpha^1\,\bar{\psi}^-\ , \qquad & K^3 &= & \frac{1}{2}:\bar{\psi}^+\,\psi^-+\bar{\psi}^-\,\psi^+:\ , \\
G'^- &= & \bar{\alpha}^2\,\psi^-+\alpha^2\,\bar{\psi}^-\ , \qquad & &&
\end{array}
\end{equation}
and
\begin{equation}
L =  \,:\bar{\alpha}^1\,\alpha^2-\bar{\alpha}^2\,\alpha^1:+\,\frac{1}{2}:\psi^+\,\partial \bar{\psi}^- + \bar{\psi}^-\,\partial \psi^+ - \bar{\psi}^+\,\partial\psi^- - \psi^-\,\partial\bar{\psi}^+:\ .
\end{equation}
These fields generate the $c=6$ ${\cal N}=4$ superconformal algebra, whose modes satisfy
\begin{equation}
\begin{array}{rcl}
{}   [L_m,L_n]&=& (m-n)L_{m+n}+\tfrac{1}{2}(m^3-m)\delta_{m+n,0}\ ,\\
{}   [L_m,K^a_n]&= & -nK^a_{m+n}\ ,\\
{}   [L_m,G^a_r]&= & \big(\tfrac{m}{2}-r\big)G^a_{m+r}\ ,\\
{}   [L_m,G'^a_r]&= & \big(\tfrac{m}{2}-r\big)G'^a_{m+r}\ ,\\
{}   \{G^+_r,G^-_s\}&= & \{G'^+_r,G'^-_s\}=L_{r+s}+(r-s)K^3_{r+s}+\big(r^2-\tfrac{1}{4}\big)\delta_{r+s,0}\ ,\\
{}   \{G^\pm_r,G'^\pm_s\}&= & \mp(r-s)K^\pm_{r+s}\ ,
\end{array}
\end{equation}
as well as
\begin{equation}
\begin{array}{cclccl}
{}[K^3_m,K^3_n] &=& \tfrac{m}{2}\delta_{m+n,0}\ , \qquad   & [K^3_m,G^\pm_r]&=& \pm\tfrac{1}{2}G^\pm_{m+r}\ , \\
{} [K^3_m,K^\pm_n]&=& \pm K^\pm_{m+n}\ ,\qquad & [K^3_m,G'^\pm_r]&= & \pm\tfrac{1}{2}G'^\pm_{m+r}\ , \\
{}[K^+_m,K^-_n]&= & 2K^3_{m+n}+m\,\delta_{m+n,0}\ ,\qquad & & &
\end{array}
\end{equation}
and
\begin{equation}
\begin{array}{cclccl}
{}[K^+_m,G^-_r] &=& G'^+_{m+r}\ , \qquad & [K^+_m,G'^-_r] &=& -G^+_{m+r}\ ,\\
{}[K^-_m,G^+_r] &=& -G'^-_{m+r}\ , \qquad & [K^-_m,G'^+_r] &=& G^-_{m+r}\ .
\end{array}
\end{equation}
The other (anti-)commutators vanish. It is sometimes convenient to bosonise the free fermions as
\begin{equation}
\psi^+=e^{i\phi}\ ,\quad \bar{\psi}^- =-e^{-i\phi}\ ,\quad \bar{\psi}^+ = e^{i\phi'}\ ,\quad \psi^- = e^{-i\phi'}\ ,
\end{equation}
where $\phi,\phi'$ are free bosons with OPEs
\begin{equation}
\phi(x)\phi(y)\sim -\log(x-y)\ ,\quad \phi'(x)\phi'(y)\sim -\log(x-y)\ .
\end{equation}

\endgroup

\subsection{States}\label{app:states}

In the symmetric product orbifold of $\mbb{T}^4$, we denote the $w$-twisted sector ground-state by $\sigma_w$. For odd $w$ it has conformal dimension $h=\tilde{h}=\tfrac{w^2-1}{4w}$ and $\mfr{su}(2)$ charge $j=\tilde{\jmath} = 0$. For even $w$, there is actually a doublet of Ramond ground states, and we denote by $\sigma_w$ the state with $h=\tilde{h}=\tfrac{w}{4}$ and $j=\tilde{\jmath}=-\tfrac{1}{2}$.

In each twisted sector, there are four (left and right) BPS states with charge and dimension $\frac{w-1}{2},\frac{w}{2},\frac{w+1}{2}$. We denote the BPS state with $h=\tilde{h}=j=\tilde{\jmath}=\tfrac{w-1}{2}$ by $\ket{\text{BPS}_-}_w$. Explicitly, this state is
\begin{align}
\ket{\mrm{BPS}_-}_w &= \big(\bar{\psi}^+_{-\frac{w-2}{2w}}\psi^+_{-\frac{w-2}{2w}}\cdots\bar{\psi}^+_{-\frac{1}{2w}}\psi^+_{-\frac{1}{2w}}\big)\big(\tilde{\bar{\psi}}^+_{-\frac{w-2}{2w}}\tilde{\psi}^+_{-\frac{w-2}{2w}}\cdots\tilde{\bar{\psi}}^+_{-\frac{1}{2w}}\tilde{\psi}^+_{-\frac{1}{2w}}\big)\sigma_w\ ,\quad\text{for $w$ odd}\ ,\nonumber \\
\ket{\mrm{BPS}_-}_w &= \big(\bar{\psi}^+_{-\frac{w-2}{2w}}\psi^+_{-\frac{w-2}{2w}}\cdots\bar{\psi}^+_{0}\psi^+_{0}\big)\big(\tilde{\bar{\psi}}^+_{-\frac{w-2}{2w}}\tilde{\psi}^+_{-\frac{w-2}{2w}}\cdots\tilde{\bar{\psi}}^+_{0}\tilde{\psi}^+_{0}\big)\sigma_w\ ,\quad\text{for $w$ even}\ .
\end{align}
The other BPS states can be obtained by applications of $\bar{\psi}^+_{-1/2}$, $\psi^+_{-1/2}$ and the corresponding right-movers. The standard BPS state we work with has $h=\tilde{h}=j=\tilde{\jmath}=\frac{w+1}{2}$ and is given by
\begin{equation}
\ket{\mrm{BPS}}_w ={\psi}^+_{-\frac{1}{2}}\bar{\psi}^+_{-\frac{1}{2}}\tilde{{\psi}}{}^+_{-\frac{1}{2}}\tilde{\bar{\psi}}^+_{-\frac{1}{2}}\ket{\mrm{BPS}_-}_w\ .
\end{equation}
On the covering surface, this state corresponds to
\begin{equation}
:e^{i\tfrac{w+1}{2}(\phi+\phi'+\tilde{\phi}+\tilde{\phi}')}:\ ,
\end{equation}
in the bosonised form of the free fermions.

The perturbing field $\Phi$ is a descendant of the lower BPS state in the two-twisted sector, given by
\begin{equation}\label{perturbing}
\ket{\Phi} = \frac{i}{\sqrt{2}}\,\big(G^-_{-\frac{1}{2}}\tilde{G}'^-_{-\frac{1}{2}}-G'^-_{-\frac{1}{2}}\tilde{G}^-_{-\frac{1}{2}}\big)\ket{\mrm{BPS}_-}_2\ .
\end{equation}
This field has $h=\tilde{h}=1$ and $j=\tilde{\jmath}=0$.

\section{Contractions}

In this Appendix we provide some more details regarding the calculations of the various Wick contractions at finite $w$. We first explain the forms of the amplitudes (\ref{amp1}) and (\ref{amp2}), and then show how to evaluate the integrals in eq.~(\ref{cintegral}).

\subsection{The fermion-boson amplitude (\ref{amp1})}\label{app:amp1}

In this Appendix we calculate the amplitude
\be\label{amp1p}
{}_{w+1}\bigl\langle \wh{\rm BPS}| \, \hat{\bar{\psi}}^+(\zeta) \, \hat{\Phi}_{\rm L}(1) \, \hat{\alpha}{}^2(z)\,\ket{\wh{\mrm{BPS}}}_w =
\bigl(\zeta-1\bigr)^{-\frac{1}{2}}\, \zeta^{\frac{w+1}{2}}\, (1-z)^{-2} \ .
\ee
The left-moving part $\hat{\Phi}_\mrm{L}$ of the perturbation is given by
\begin{equation}
\hat{\Phi}_\mrm{L}(1) = \hat{\bar{\alpha}}^1(1)\,\hat{\phi}_2^\dagger(1)\ ,
\end{equation}
where $\hat{\phi}_2^\dagger$ is the bottom component of the spin field on the covering surface. Hence, there is only one other boson in the correlator, which contracts with $\hat{\alpha}^2(z)$ to give the factor
\begin{equation}
\frac{1}{(1-z)^2}\ .
\end{equation}
The remaining correlator is
\begin{equation}
{}_{w+1}\bigl\langle \wh{\rm BPS}| \, \hat{\bar{\psi}}^+(\zeta) \, \hat{\phi}_2^\dagger(1)\,\ket{\wh{\mrm{BPS}}}_w\ .
\end{equation}
The fermion $\hat{\bar{\psi}}^+$ has the following OPEs with the lifts of the BPS states,
\begin{align}\label{app:eq:ferm-BPS_OPE}
\hat{\bar{\psi}}^+(\zeta)\,\ket{\wh{\mrm{BPS}}}_w &= \mcl{O}\bigl(\zeta^{\frac{w+1}{2}}\bigr)\ ,\nonumber\\
\hat{\bar{\psi}}^+(\zeta)\,\ket{\wh{\mrm{BPS}}}^\dagger_w &= \mcl{O}\bigl(\zeta^{-\frac{w+1}{2}}\bigr)\ ,\nonumber\\
\hat{\bar{\psi}}^+(\zeta)\,\hat{\phi}_2^\dagger(0) &= \zeta^{-\frac{1}{2}}\,\hat{\phi}_2(0) + \mcl{O}\bigl(\zeta^{\frac{1}{2}}\bigr)\ ,
\end{align}
and hence the correlator as a function of $\zeta$ has a branch-cut singularity $(\zeta-1)^{-\frac{1}{2}}$ at $1$, a zero of order $\frac{w+1}{2}$ at zero, and grows as $\zeta^{\frac{w}{2}}$ for $\zeta\to\infty$. We look at the function
\begin{equation}
\mrm{D}(\zeta) = (\zeta-1)^{-\frac{1}{2}}\,\zeta^{-\frac{w+1}{2}}\, {}_{w+1}\bigl\langle \wh{\rm BPS}| \, \hat{\bar{\psi}}^+(\zeta) \, \hat{\phi}_2^\dagger(1)\,\ket{\wh{\mrm{BPS}}}_w\ ,
\end{equation}
which has the following properties. It is a single-valued meromorphic function in $\zeta$ with a first order pole at $\zeta=1$ and a regular point at $\zeta=0$. Furthermore, it decays as $\zeta^{-1}$ for $\zeta\to \infty$. By Liouville's theorem, it is then given by
\begin{equation}
\mrm{D}(\zeta) = \frac{\mrm{res}_{\zeta=1} D(\zeta)}{\zeta-1} = \frac{1}{\zeta-1}\,{}_{w+1}\bigl\langle \wh{\rm BPS}| \, \hat{\phi}_2(1)\,\ket{\wh{\mrm{BPS}}}_w\ ,
\end{equation}
where we have used the OPE in eq.~(\ref{app:eq:ferm-BPS_OPE}). We normalise $ {}_{w+1}\bigl\langle \wh{\rm BPS}| \, \hat{\phi}_2(1)\,\ket{\wh{\mrm{BPS}}}_w=1$  (for the subtleties of this normalisation see appendix \ref{app:orbifold_averaging}). Then we find
\begin{equation}
{}_{w+1}\bigl\langle \wh{\rm BPS}| \, \hat{\bar{\psi}}^+(\zeta) \, \hat{\phi}_2^\dagger(1)\,\ket{\wh{\mrm{BPS}}}_w = (\zeta-1)^{\frac{1}{2}}\,\zeta^{\frac{w+1}{2}}\, \mrm{D}(\zeta) = (\zeta-1)^{-\frac{1}{2}}\,\zeta^{\frac{w+1}{2}}\ .
\end{equation}

\subsection{The fermion-fermion amplitude (\ref{amp2})}\label{app:amp2}

In this appendix we calculate the amplitude
\be
{}_{w+1}\bigl\langle \wh{\rm BPS}| \, \hat{\bar\psi}{}^+(\zeta)\,  \hat{\phi}_2(1) \, \hat{\psi}{}^-(z)\,\ket{\wh{\mrm{BPS}}}_w
\ee
in a similar way as above. We define
\begin{equation}
\mrm{D}(\zeta,z) = \zeta^{-\frac{w+1}{2}}(\zeta-1)^{-\frac{1}{2}}\,(1-z)^{\frac{1}{2}}z^{\frac{w+1}{2}}\,{}_{w+1}\bigl\langle \wh{\rm BPS}| \, \hat{\bar\psi}{}^+(\zeta)\,  \hat{\phi}_2(1) \, \hat{\psi}{}^-(z)\,\ket{\wh{\mrm{BPS}}}_w \ ,
\end{equation}
which is single valued, decays at infinity in both coordinates, and only has a pole at $\zeta=z$. Thus,
\begin{equation}
\mrm{D}(\zeta,z) = \frac{1}{\zeta-z}\ ,
\end{equation}
giving
\begin{equation}
{}_{w+1}\bigl\langle \wh{\rm BPS}| \, \hat{\bar\psi}{}^+(\zeta)\,  \hat{\phi}_2(1) \, \hat{\psi}{}^-(z)\,\ket{\wh{\mrm{BPS}}}_w = \zeta^{\frac{w+1}{2}}(\zeta-1)^{\frac{1}{2}}\,z^{-\frac{w+1}{2}}(1-z)^{-\frac{1}{2}}\,(\zeta-z)^{-1}\ .
\end{equation}

\subsection{Performing the integrals for \texorpdfstring{$c_{m,n}$}{c(m,n)}}\label{cmnint}

In this appendix we show how to perform the integrals in eq.~(\ref{cintegral}), i.e.\ we calculate
\begin{equation}
\begin{tikzpicture}[baseline=-.5ex]
\node[](n1)at(0,0){$\bar{\psi}^+_m$};
\node[](n2)[right=-1ex of n1]{$\alpha^2_n$};
\draw[](n1.south)-- ([yshift=-1ex]n1.south)--([yshift=-1ex]n2.south)node[midway]{$\times$}--(n2.south);
\end{tikzpicture} = \tfrac{1}{\sqrt{(w+1)(w-n)}}\cint{C(\infty)}\!\!d\zeta\,\bigl(\partial\bm{\Gamma}(\zeta)\bigr)^{\frac{1}{2}}\bm{\Gamma}^{-\frac{m}{w+1}}(\zeta)\,\cint{C(0)}\!\!dz\,\bm{\Gamma}^{-1+\frac{n}{w}}(z)\,\,\frac{ \bigl(\zeta-1\bigr)^{-\frac{1}{2}}\,\zeta^{\frac{w+1}{2}}}{\bigl(1-z\bigr)^2}\ .
\end{equation}
Using the form of the covering map from eq.~(\ref{coveringww2}) we find
\begin{align}
\cint{C(\infty)}\!\!d\zeta\,\bigl(\partial\bm{\Gamma}(\zeta)\bigr)^{\frac{1}{2}}&\,\bm{\Gamma}^{-\frac{m}{w+1}}(\zeta)\,\bigl(\zeta-1\bigr)^{-\frac{1}{2}}\,\zeta^{\frac{w+1}{2}} \nonumber\\
&=\sqrt{w(w+1)}\,w^{-\frac{m}{w+1}}\cint{C(\infty)}\!\!d\zeta\,\,\zeta^{w-m}\,\Bigl(\frac{w+1}{w}\,\frac{1}{\zeta}-1\Bigr)^{-\frac{m}{w+1}}\ . \label{B.4}
\end{align}
The integrand is single-valued in the region $|\zeta|>\frac{w+1}{w}$.

This integral can now be evaluated either by using contour deformation techniques, or by a direct power series expansion.

\subsubsection*{Contour deformation} The integrand has a branch cut from $0$ to $\frac{w+1}{w}$. We can pull the contour around $\infty$ (oriented clockwise by convention) over the sphere around the branch-cut, as shown below.

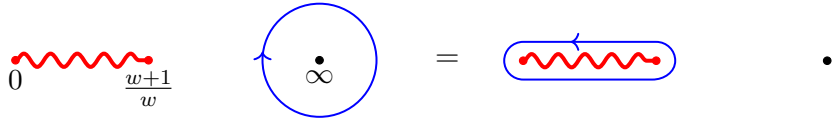
\begin{center}
\begin{tikzpicture}

\draw[draw=red,fill=red](-1,0)node[anchor=north]{$0$}circle(0.05);
\draw[draw=red,fill=red](0.75,0)node[anchor=north]{$\frac{w+1}{w}$}circle(0.05);
\draw[style={decorate, decoration=snake},draw = red, line width = 1.5](-1,0)--(0.75,0);

\draw[fill=black](3,0)node[anchor=north]{$\infty$}circle(0.05);

\draw[draw = blue, decoration={markings, mark=at position 0.5 with {\arrow{<}}},
postaction={decorate},line width = 0.75](3,0)circle(0.75);

\end{tikzpicture}\hspace{0.5cm}
\begin{tikzpicture}

\node[]at(-2,0){\large $=$};

\draw[draw=red,fill=red](-1,0)circle(0.05);
\draw[draw=red,fill=red](0.75,0)circle(0.05);
\draw[style={decorate, decoration=snake},draw = red, line width = 1.5](-1,0)--(0.75,0);

\draw[fill=black](3,0)circle(0.05);

\draw[draw = blue, decoration={markings, mark=at position 0.8 with {\arrow{<}}},
postaction={decorate},line width = 0.75](0.75,0.25)to[out=0,in=90](1,0)to[out=270,in=0](0.7,-0.25)--(-1,-0.25)to[out=180,in=270](-1.25,0)to[out=90,in=180](-1,0.25)--(0.75,0.25);

\draw[draw=white](0,-0.75)--(1,-0.75);

\end{tikzpicture}
\end{center}
The small semicircles around the ends of the branch cut do not contribute to the integral, and thus we are left with an integral just above and just below the branch cut in opposite directions. They do not cancel, because of the relative phase, and we get
\begin{align}
\cint{C(\infty)}\!\!d\zeta\,\,\zeta^{w-m}&\,\Bigl(\frac{w+1}{w}\,\frac{1}{\zeta}-1\Bigr)^{-\frac{m}{w+1}} \nonumber\\
&= \cint{C(\infty)}\!\!d\zeta\,\,\zeta^{w(1-\frac{m}{w+1})}\,\Bigl(\frac{w+1}{w}-\zeta\Bigr)^{-\frac{m}{w+1}} \nonumber\\
&= \frac{1-e^{-2\pi i \frac{m}{w+1}}}{2\pi i}\,\int_0^{\frac{w+1}{w}}du\,\,u^{w(1-\frac{m}{w+1})}\Bigl(\frac{w+1}{w}-u\Bigr)^{-\frac{m}{w+1}}\nonumber\\
&= \bigl(\tfrac{w+1}{w}\bigr)^{w+1-m} e^{-i\pi \frac{m}{w+1}}\frac{\sin\bigl(\pi \frac{m}{w+1}\bigr)}{\pi}\,\int_0^1 dv\,\,v^{w(1-\frac{m}{w+1})}\bigl(1-v\bigr)^{-\frac{m}{w+1}}\ .
\end{align}
The relative phase factor $(1-e^{-2\pi i \frac{m}{w+1}})$ appears in the following way. We choose the branch of $\zeta^{w(1-\frac{m}{w+1})}$ such that the phase is $1$ just below the branch cut, where we have the positively oriented integral. For the negatively oriented integral along the top of the branch cut, we then have the phase
\begin{equation}
e^{2\pi i \frac{w m}{w+1}} = e^{-2\pi i \frac{m}{w+1}}\ ,
\end{equation}
and we have to subtract the two contributions because of the orientation. We have substituted $v=\frac{w}{w+1}u$ to go from the third to the fourth line. The integral over $v$ can now be recognised as the Beta function and gives
\begin{equation}
\int_0^1 dv\,\,v^{w(1-\frac{m}{w+1})}\bigl(1-v\bigr)^{-\frac{m}{w+1}} = \frac{\Gamma\bigl(w(1-\frac{m}{w+1})+1\bigr)\,\Gamma\bigl(1-\frac{m}{w+1}\bigr)}{\Gamma\bigl(w+2-m\bigr)}\ .
\end{equation}
Thus we obtain altogether
\begin{align}\label{app:eq:contour_result}
\cint{C(\infty)}\!\!d\zeta\,\,\zeta^{w-m}&\,\Bigl(\frac{w+1}{w}\,\frac{1}{\zeta}-1\Bigr)^{-\frac{m}{w+1}} \nonumber\\
&= \bigl(\tfrac{w+1}{w}\bigr)^{w-m} e^{-i\pi \frac{m}{w+1}}\frac{\sin\bigl(\pi \frac{m}{w+1}\bigr)}{\pi}\,\frac{\Gamma\bigl(w(1-\frac{m}{w+1})\bigr)\,\Gamma\bigl(1-\frac{m}{w+1}\bigr)}{\Gamma\bigl((w+1)(1-\frac{m}{w+1})\bigr)}\ .
\end{align}

\subsubsection*{Series expansion} Alternatively, we can arrive at the same result by expanding the factor $\bigl(1-\frac{w+1}{w}\frac{1}{\zeta}\bigr)^{-\frac{m}{w+1}}$ in a power series as
\begin{equation}
\Bigl(1-\frac{w+1}{w}\,\frac{1}{\zeta}\Bigr)^{-\frac{m}{w+1}} = \sum_{k=0}^\infty \bigl(-\tfrac{w+1}{w}\bigr)^{k}\,\binom{-\frac{m}{w+1}}{k}\,\zeta^{-k}\ ,
\end{equation}
which converges absolutely for $|\zeta|>\frac{w+1}{w}$. As
\begin{equation}
\cint{C(\infty)}\!\!d\zeta\,\, \zeta^{-r} = \delta_{r,1}\ ,
\end{equation}
the contour integral in (\ref{B.4}) simply picks out the $(w+1-m)$-th term from this series, and is thus given by
\begin{equation}
\cint{C(\infty)}\!\!d\zeta\,\,\zeta^{w-m}\,\Bigl(1-\frac{w+1}{w}\,\frac{1}{\zeta}\Bigr)^{-\frac{m}{w+1}} = \bigl(-\tfrac{w+1}{w}\bigr)^{w+1-m}\,\binom{-\frac{m}{w+1}}{w+1-m}\ .
\end{equation}
The generalised binomial coefficient arising in this expression can be written in terms of Gamma functions as
\begin{align}
\binom{-\frac{m}{w+1}}{w+1-m}&=\frac{\Gamma\bigl(1-\frac{m}{w+1}\bigr)}{\Gamma\bigl(w+2-m\bigr)\,\Gamma\bigl(-w+\frac{m\,w}{w+1}\bigr)}\ .
\end{align}
Using repeatedly the identities
\begin{equation}
\Gamma(x+1) = x\,\Gamma(x)\ ,\qquad \Gamma(x)\,\Gamma(1-x) = \frac{\pi}{\sin(\pi x)}\ ,
\end{equation}
as well as
\begin{equation}
(-1)^{m+1}\,\sin\bigl(\pi\frac{w m}{w+1}\bigr) = \sin\bigl(\pi\frac{m}{w+1}\bigr)
\end{equation}
this then also leads to eq.~(\ref{app:eq:contour_result}).\bigskip

\noindent In a similar manner, although slightly more complicated due to the appearance of a hypergeometric function, we can calculate the $z$ contour integral
\begin{equation}
\cint{C(0)}\!\!dz\,\bm{\Gamma}^{-1+\frac{n}{w}}(z)\,\,\frac{1}{\bigl(1-z\bigr)^2} = (w+1)^{-1+\frac{n}{w}}\,\bigl(-\frac{w}{w+1}\bigr)^{w-n}(w+1)(w-n)\,\binom{-1+\frac{n}{w}}{w-n}\ ,
\end{equation}
which can again be shown to equal eq.~(\ref{eq:ferm-bos_contraction}).

\subsection{Orbifold averaging and universal prefactors}\label{app:orbifold_averaging}

In this appendix we determine the overall normalisation factor $N_w$ of the correlator, see eq.~(\ref{3.27}). In particular, we shall keep track of the symmetry factors that come from the systematics of the symmetric orbifold. Essentially all of our calculations refer to three-point functions of the form
\begin{equation}
{}_{w+1}\bigl\langle {\rm BPS}| \, {\phi}_2(1)\,\ket{{\mrm{BPS}}}_w\ ,
\end{equation}
which equal, see e.g.~\cite[eq.~(5.26)]{Lunin:2001pw},
\begin{equation}
{}_{w+1}\bigl\langle {\rm BPS}| \, {\phi}_2(1)\,\ket{{\mrm{BPS}}}_w =  \frac{i}{\sqrt{2}}\, \frac{w}{2(w+1)}\ ,
\end{equation}
where the prefactor $ \frac{i}{\sqrt{2}}$ comes from the normalisation of the perturbing field, see eq.~(\ref{perturbing}).
When calculating contractions of left-moving magnons, we contract the right-moving supercharge in the base space, see Section~\ref{genpert}, but the left-moving supercharge descendant is lifted to the covering surface. As it is given by terms of the form
\begin{equation}
G^-_{-\frac{1}{2}}\,\phi_2 \supset \psi^-_0\,\bar{\alpha}^1_{-\frac{1}{2}}\,\phi_2\ ,
\end{equation}
this lifting contributes a factor \cite{Gaberdiel:2022oeu}
\begin{equation}
\sqrt{\frac{2}{a_1}}\ ,
\end{equation}
where $a_1$ is the leading coefficient in the expansion of the covering map around the ramified pre-image of the insertion of $\phi_2$. For our situation, the covering map is
\begin{equation}
\bm{\Gamma}(z) = (w+1)z^w-wz^{w+1}\ ,
\end{equation}
and hence
\begin{equation}
a_1 = -\frac{1}{2}w(w+1) \ .
\end{equation}
Overall, we thus get a factor
\begin{equation}\label{app:eq:extra_factor}
\frac{i}{\sqrt{2}}\,\frac{w}{2(w+1)}\,\frac{2}{i\,\sqrt{w(w+1)}}\ .
\end{equation}
Finally, we need to take the symmetry factor of the symmetric orbifold into account. For this we need
to average over all the conjugacy classes, and account for the fact that the cycle colours can be permuted. For a correlator with $s$ insertions of twist operators with cycle length $w_j$, one gets a normalisation factor \cite{Lunin:2000yv,Pakman:2009mi}
\begin{equation}
\sum_{\msf{g}}N^{1-\msf{g}-\frac{s}{2}}\,\prod_{j=1}^{s}\sqrt{w_j}\,\left(1+\mcl{O}\bigl(N^{-1}\bigr)\right)\,\sum_{\text{configurations}}\llangle\cdots\rrangle\ .
\end{equation}
For our three-point function, there is only one configuration (all of the others are related by conjugation) with genus $\msf{g}=0$. In the planar limit $N\to\infty$, we thus obtain the additional prefactor due to the orbifold averaging
\begin{equation}
N^{-\frac{1}{2}}\,\sqrt{2w(w+1)}\ ,
\end{equation}
where the $N$ dependent factor just reflects the fact that we consider a sphere correlator with $g_s= N^{-\frac{1}{2}}$. Combining this with eq.~(\ref{app:eq:extra_factor}), we therefore obtain altogether in the large $N$ limit
\begin{equation}\label{B.35}
N_w =    \frac{w}{w+1}\ .
\end{equation}
Note that $N_w$ goes to $1$ in the infinite $w$ limit.

\section{Consistency of the algebra at finite \texorpdfstring{$w$}{w}}\label{app:vanishing_central_extension}

In this appendix we show that the central term of the superconformal algebra vanishes
\begin{equation}
\{G^+_{-\frac{1}{2}},\wt{G}'^+_{-\frac{1}{2}}\} \,\,  \alpha^2(\tfrac{n_1}{w})\,\alpha^2(\tfrac{n_2}{w})\,\ket{\mrm{BPS}}_w = 0 \ .
\end{equation}
Using eq.~(\ref{eq:2-magnon_charge_action0}) as well as (\ref{deformedaction}), this is equivalent to
\begin{equation}\label{app:eq:coeff_to_vanish}
c^{m_1,m_2}_{n_1,n_2}\,\sqrt{1-\tfrac{m_1}{w+1}}+c^{m_2,m_1}_{n_1,n_2}\,\sqrt{1-\tfrac{m_2}{w+1}} = 0\ .
\end{equation}
Using
\begin{equation}
c^{m_1,m_2}_{n_1,n_2} = g\Big(
\begin{tikzpicture}[baseline=-.5ex]
\node[](n1)at(0,0){$\bar{\psi}^+_{m_1}$};
\node[](n2)[right=-1ex of n1]{$\bar{\alpha}^1_{m_2}$};
\node[](n3)[right=-1ex of n2]{$\alpha^2_{n_1}$};
\node[](n4)[right=-1ex of n3]{$\alpha^2_{n_2}$};
\draw[](n1.south)-- ([yshift=-1ex]n1.south)--([yshift=-1ex]n3.south)node[midway]{$\times$}--(n3.south);
\draw[](n2.north)-- ([yshift=1ex]n2.north)--([yshift=1ex]n4.north)--(n4.north);
\end{tikzpicture}
+
\begin{tikzpicture}[baseline=-.5ex]
\node[](n1)at(0,0){$\bar{\psi}^+_{m_1}$};
\node[](n2)[right=-1ex of n1]{$\bar{\alpha}^1_{m_2}$};
\node[](n3)[right=-1ex of n2]{$\alpha^2_{n_1}$};
\node[](n4)[right=-1ex of n3]{$\alpha^2_{n_2}$};
\draw[](n1.south)-- ([yshift=-1ex]n1.south)--([yshift=-1ex]n4.south)node[midway]{$\times$}--(n4.south);
\draw[](n2.north)-- ([yshift=1ex]n2.north)--([yshift=1ex]n3.north)--(n3.north);
\end{tikzpicture}
\Big)\times N_w\cdot \pi\cdot \delta_{\frac{n_1+n_2}{w},\frac{m_1+m_2}{w+1}}+\mcl{O}\bigl(g^2\bigr) \ ,
\end{equation}
and the expression for the contractions given in eqs.~(\ref{eq:ferm-bos_contraction}) and (\ref{eq:bos-bos_contraction}), one finds that the left-hand-side of eq.~(\ref{app:eq:coeff_to_vanish}) is proportional to
\begin{align}
\left((-1)^{-m_1+n_1}\frac{\sin\big(\pi \frac{n_1}{w}\big)\sin\big(\pi \frac{m_1}{w+1}\big)}{\sin\big(\pi \frac{(w+1)n_1}{w}\big)\sin\big(\pi \frac{w m_1}{w+1}\big)}\frac{1}{wm_2-(w+1)n_2}+(n_1\leftrightarrow n_2)\right)+(m_1\leftrightarrow m_2)\ .
\end{align}
For integer $n_i,m_j$, the prefactor of the `propagator' $\frac{1}{wm_2-(w+1)n_2}$ is always $1$. Therefore, the expression is simply
\begin{align}
\frac{1}{wm_2-(w+1)n_2}+&\frac{1}{wm_1-(w+1)n_1}+\frac{1}{wm_2-(w+1)n_1}+\frac{1}{wm_1-(w+1)n_2}  \nonumber\\
&\hspace{-3cm}= \Biggl[w(m_1+m_2)-(w+1)(n_1+n_2)\Biggr]\,\bigg(\frac{1}{(wm_1-(w+1)n_1)(wm_2-(w+1)n_2)}\nonumber \\
&\hspace{4cm}+\frac{1}{(wm_1-(w+1)n_2)(wm_2-(w+1)n_1)}\bigg)\ .
\end{align}
The square bracket is now zero by conservation of the conformal dimension,
\begin{equation}
\frac{m_1}{w+1}+\frac{m_2}{w+1} = \frac{n_1}{w}+\frac{n_2}{w}\ ,
\end{equation}
and thus the anti-commutator is identically zero.

The argument for the case involving $M$ boson magnon excitations works similarly. In general there will be a sum of all possible contractions between the in- and out-states. All but one of these contractions are of `propagator type', i.e.\ they are given by (\ref{eq:bos-bos_contraction}) and hence contain a factor $\frac{1}{wm_i-(w+1)n_{\sigma(i)}}$, where $\sigma\in S_M$ is the perturbation specifying which oscillators are contracted with one another. (The remaining factor, on the other hand, is of type (\ref{eq:ferm-bos_contraction}).) For a fixed permutation $\sigma$ there are $M$ such terms, depending on which of the $M$ legs is contracted via (\ref{eq:ferm-bos_contraction}). It is then easy to see that the sum over these $M$ terms is proportional to the momentum constraint.

\section{The BMN dispersion relation for \texorpdfstring{${\rm AdS}_3$}{AdS3}}\label{BMNdis}

In this appendix we explain the derivation of the dispersion relation (\ref{eq:BMN_dispersion}) for the left-movers. We begin with the light cone action for ${\rm AdS}_3\times {\rm S}^3$, see \cite[eq.~(C.2)]{Berenstein:2002jq}
\be
S = \frac{1}{2\pi \alpha'} \, \int dt \, \int_0^{2\pi \alpha' p^+} d\sigma \, \frac{1}{2} \Bigl[ |\dot{Z}_i|^2  - |Z_i' + i\,\cos\alpha\, Z_i|^2 - \sin^2\alpha\,|Z_i|^2 + \hbox{fermions} \Bigr] \ .
\ee
Here $Z_1 = y_1 + i y_2$ is the complex coordinate associated to the physical ${\rm AdS}_3$ directions, while $Z_2 = y_3 + i y_4$ describes the two physical bosonic ${\rm S}^3$ directions --- the remaining two directions, one from ${\rm AdS}_3$ and one from ${\rm S}^3$, make up the spacetime light-cone and hence do not describe physical excitations in light-cone gauge. The angle $\alpha$ parametrises the fluxes as
\begin{equation}
\cos\alpha = \frac{q_{\rm NS}}{\sqrt{q_{\rm NS}^2 + g_s^2 q_{\rm R}^2}}\ .
\end{equation}
The associated equations of motion are then
\begin{align}
0 & = \Bigl[ - \partial^2_\tau + \partial^2_\sigma + 2 i\,\cos\alpha\, \partial_\sigma -1 \Bigr] Z_i \\
0 & = \Bigl[ - \partial^2_\tau + \partial^2_\sigma - 2 i\,\cos\alpha\, \partial_\sigma -1 \Bigr] \bar{Z}_i \ .
\end{align}
If we make the ansatz
\begin{align}\label{ansatz}
Z_i  & = \sum_{n\in\mathbb{Z}}  \Bigl( \frac{1}{\sqrt{2\omega_n}} \, a^{i \, \dagger}_n \, e^{-i (\omega_n t + n \sigma)} + \frac{1}{\sqrt{2\bar\omega_n}} \, b^i_n \, e^{i (\bar{\omega}_n t + n \sigma)}  \Bigr) \\
\bar{Z}_i & = \sum_{n\in\mathbb{Z}}  \Bigl(\frac{1}{\sqrt{2\omega_n}} \,  a^{i}_n \, e^{i (\omega_n t + n \sigma)} + \frac{1}{\sqrt{2\bar\omega_n}} \, b^{i\, \dagger}_n \, e^{-i (\bar{\omega}_n t + n \sigma)}  \Bigr) \ ,
\end{align}
the equations of motion imply that
\be
\omega_n = \sqrt{\sin^2\alpha + \Bigl( \cos\alpha+ \frac{n}{\alpha' p^+}  \Bigr)^2}  \ , \qquad
\bar{\omega}_n = \sqrt{\sin^2\alpha + \Bigl( \cos\alpha- \frac{n}{\alpha' p^+}  \Bigr)^2} \ .
\ee
The canonical commutation relations then imply that
\be\label{acomm0}
\bigl[\, a^{i}_n, a^{j\dagger}_m\, \bigr] = \delta^{ij}\, \delta_{nm} \ , \qquad \bigl[\, b^{i}_n, b^{{j}\dagger}_m\, \bigr] = \delta^{ij}\, \delta_{nm} \ .
\ee
Note that it follows from the $\sigma$-dependence of the solution in (\ref{ansatz}) that the zero momentum condition is, see  \cite[eq.~(3.3)]{Berenstein:2002jq},
\be\label{peri0}
\sum_{i=1,2} \sum_{n\neq 0 } n \bigl( a^{i\, \dagger}_n\, a^i_n  + {b}^{i\, \dagger}_n\, {b}^{i}_n \bigr) = 0 \ .
\ee

\section{S-matrix elements}\label{app:s-matrix_elements}

In this appendix we list explicit expressions for the various $S$-matrix elements.

\begingroup
\allowdisplaybreaks

\subsection{LL S-matrix}\label{app:sLL-matrix_elements}
Let us start by giving explicit expressions for the coefficients of the LL S-matrix in eq.~(\ref{eq:LL_S-matrix_form}). The coefficients are determined up to an overall factor which we choose to be $A_\mrm{LL}^{11}(p,q)$. Then, the expressions for the coefficients can be obtained in terms of the following subset,
\begin{align}
A_\mrm{LL}^{12}(p,q) &= A_\mrm{LL}^{11}(p,q)\,\frac{(e^{i\pi q}x_q-e^{i\pi p}x_p)(e^{-i\pi q}x_q-e^{-i\pi p}x_p)}{(e^{i\pi q}x_q-e^{-i\pi p}x_p)(e^{-i\pi q}x_q-e^{i\pi p}x_p)}\ ,\nonumber\\
B_\mrm{LL}^{12}(p,q) &= -A_\mrm{LL}^{11}(p,q)\,\frac{x_p(e^{i\pi p}-e^{-i\pi p})x_q(e^{i\pi q}-e^{-i\pi q})}{(e^{i\pi q}x_q-e^{-i\pi p}x_p)(e^{-i\pi q}x_q-e^{i\pi p}x_p)}\ ,\nonumber\\
C_\mrm{LL}^{1+}(p,q) &= A_\mrm{LL}^{11}(p,q)\,\frac{(e^{i\pi q}x_q-e^{i\pi p}x_p)}{(e^{i\pi q}x_q-e^{-i\pi p}x_p)}\ ,\nonumber\\
D_\mrm{LL}^{1+}(p,q) &= A_\mrm{LL}^{11}(p,q)\,\frac{\eta_p}{\eta_q}\frac{x_q(e^{i\pi q}-e^{-i\pi q})}{(e^{i\pi q}x_q-e^{-i\pi p}x_p)}\ ,\nonumber\\
E_\mrm{LL}^{+1}(p,q) &= A_\mrm{LL}^{11}(p,q)\,\frac{(e^{-i\pi q}x_q-e^{-i\pi p}x_p)}{(e^{i\pi q}x_q-e^{-i\pi p}x_p)}\ ,\nonumber\\
H_\mrm{LL}^{++}(p,q) &= -A_\mrm{LL}^{11}(p,q)\,\frac{(e^{-i\pi q}x_q-e^{i\pi p}x_p)}{(e^{i\pi q}x_q-e^{-i\pi p}x_p)}\ ,\nonumber\\
H_\mrm{LL}^{+-}(p,q) &= -A_\mrm{LL}^{11}(p,q)\,\frac{(e^{-i\pi q}x_q-e^{-i\pi p}x_p)^2}{(e^{i\pi q}x_q-e^{-i\pi p}x_p)(e^{-i\pi q}x_q-e^{i\pi p}x_p)}\ ,\nonumber\\
K_\mrm{LL}^{12}(p,q) &= A_\mrm{LL}^{11}(p,q)\,\frac{\eta_p}{\eta_q}\frac{x_q(e^{i\pi q}-e^{-i\pi q})(e^{i\pi q}x_q-e^{i\pi p}x_p)}{(e^{i\pi q}x_q-e^{-i\pi p}x_p)(e^{-i\pi q}x_q-e^{i\pi p}x_p)}\ .
\end{align}
The other coefficients can be related to the ones above as\footnote{Note that $\eta_p$, $x_p$ are real. Furthermore, we assume that $A_\mrm{LL}^{11}(p,q)$ is real so that the complex conjugation does not change the overall pre-factor.}
\begin{align}
A_\mrm{LL}^{21}(p,q) &= A_\mrm{LL}^{12}(p,q)\ , & A_\mrm{LL}^{22}(p,q) &= A_\mrm{LL}^{11}(p,q)\ , \nonumber\\
B_\mrm{LL}^{21}(p,q) &= B_\mrm{LL}^{12}(p,q)\ ,&&\nonumber\\
C_\mrm{LL}^{2+}(p,q) &= C_\mrm{LL}^{1+}(p,q)\ , & C_\mrm{LL}^{i-}(p,q) &= \bigl(C_\mrm{LL}^{1+}(p,q)\bigr)^*\ , \nonumber\\
D_\mrm{LL}^{2+}(p,q) &= D_\mrm{LL}^{1+}(p,q)\ , & D_\mrm{LL}^{i-}(p,q) &= \bigl(D_\mrm{LL}^{1+}(p,q)\bigr)^*\ , \nonumber\\
E_\mrm{LL}^{+2}(p,q) &= E_\mrm{LL}^{+1}(p,q)\ , & E_\mrm{LL}^{-i}(p,q) &= \bigl(E_\mrm{LL}^{+1}(p,q)\bigr)^*\ , \nonumber\\
F_\mrm{LL}^{+i}(p,q) &= D_\mrm{LL}^{1+}(p,q)\ , & F_\mrm{LL}^{-i}(p,q) &= \bigl(D_\mrm{LL}^{1+}(p,q)\bigr)^*\ , \nonumber\\
H_\mrm{LL}^{--}(p,q) &= \bigl(H_\mrm{LL}^{++}(p,q)\bigr)^*\ , & H_\mrm{LL}^{-+}(p,q) &= \bigl(H_\mrm{LL}^{+-}(p,q)\bigr)^*\ , \nonumber\\
J_\mrm{LL}^{ij}(p,q) &= B_\mrm{LL}^{12}(p,q)\ ,&&\nonumber\\
K_\mrm{LL}^{21}(p,q) &= -K_\mrm{LL}^{12}(p,q)\ ,&&\nonumber\\
L_\mrm{LL}^{ij}(p,q) &= -\bigl(K_\mrm{LL}^{ij}(p,q)\bigr)^*\ ,&&\nonumber\\
M_\mrm{LL}^{+-}(p,q) &= -\bigl(K_\mrm{LL}^{12}(p,q)\bigr)^*\ , & M_\mrm{LL}^{-+}(p,q) &= K_\mrm{LL}^{12}(p,q)\ , \nonumber\\
N_\mrm{LL}^{+-}(p,q) &= \bigl(K_\mrm{LL}^{12}(p,q)\bigr)^*\ , & N_\mrm{LL}^{-+}(p,q) &= -K_\mrm{LL}^{12}(p,q)\ .
\end{align}

\subsection{LR S-matrix}\label{app:sLR-matrix_elements}
Here, we write down the coefficients of the LR S-matrix in eq.~(\ref{eq:LR_S-matrix_form}) up to the overall factor $A_\mrm{LR}^{11}(p,q)$. The S-matrix is given by

\begin{align}
A_\mrm{LR}^{12}(p,q) &= A_\mrm{LR}^{11}(p,q)\,\frac{(1+e^{i\pi(p+q)}x_p x_q)(1+e^{-i\pi (p+q)}x_p x_q)}{(1+e^{i\pi(p-q)}x_p x_q)(1+e^{-i\pi(p-q)}x_p x_q)}\ ,\nonumber\\
B_\mrm{LR}^{12}(p,q) &= A_\mrm{LR}^{11}(p,q)\,\frac{\eta_p}{\eta_q}\,e^{-2\pi i p}\,\frac{x_q(e^{i\pi q}-e^{-i\pi q})(1+e^{i\pi(p+q)}x_p x_q)}{(1+e^{i\pi(p-q)}x_p x_q)(1+e^{-i\pi(p-q)}x_p x_q)}\ ,\nonumber\\
D_\mrm{LR}^{1+}(p,q) &= A_\mrm{LR}^{11}(p,q)\,e^{2\pi i p}\,\frac{(1+e^{-i\pi (p+q)}x_p x_q)}{(1+e^{i\pi(p-q)}x_p x_q)}\ ,\nonumber\\
E_\mrm{LR}^{1+}(p,q) &= A_\mrm{LR}^{11}(p,q)\,\frac{\eta_p}{\eta_q}\frac{x_q(e^{i\pi q}-e^{-i\pi q})}{(1+e^{i\pi(p-q)}x_p x_q)}\ ,\nonumber\\
F_\mrm{LR}^{++}(p,q) &= -A_\mrm{LR}^{11}(p,q)\,e^{2\pi i(p+q)}\,\frac{(1+e^{-i\pi (p+q)}x_p x_q)^2}{(1+e^{i\pi(p-q)}x_p x_q)(1+e^{-i\pi(p-q)}x_p x_q)}\ ,\nonumber\\
F_\mrm{LR}^{+-}(p,q) &= -A_\mrm{LR}^{11}(p,q)\,e^{-2\pi i(p-q)}\,\frac{(1+e^{i\pi (p-q)}x_p x_q)}{(1+e^{-i\pi(p-q)}x_p x_q)}\ ,\nonumber\\
H_\mrm{LR}^{++}(p,q) &= -A_\mrm{LR}^{11}(p,q)\,\frac{\eta_p}{\eta_q}\,e^{2\pi i q}\,\frac{x_q(e^{i\pi q}-e^{-i\pi q})(1+e^{-i\pi(p+q)}x_p x_q)}{(1+e^{i\pi(p-q)}x_p x_q)(1+e^{-i\pi(p-q)}x_p x_q)}\ ,\nonumber\\
K_\mrm{LR}^{+1}(p,q) &= A_\mrm{LR}^{11}(p,q)\,e^{2\pi i q}\,\frac{(1+e^{-i\pi (p+q)}x_p x_q)}{(1+e^{-i\pi(p-q)}x_p x_q)}\ ,\nonumber\\
L_\mrm{LR}^{+1}(p,q) &= A_\mrm{LR}^{11}(p,q)\,\frac{\eta_p}{\eta_q}\,e^{-2\pi i(p-q)}\,\frac{x_q(e^{i\pi q}-e^{-i\pi q})}{(1+e^{-i\pi(p-q)}x_p x_q)}\ ,\nonumber\\
M_\mrm{LR}^{12}(p,q) &= -A_\mrm{LR}^{11}\,\frac{x_p(e^{i\pi p}-e^{-i\pi p})x_q(e^{i\pi q}-e^{-i\pi q})}{(1+e^{i\pi(p-q)}x_p x_q)(1+e^{-i\pi(p-q)}x_p x_q)}\ ,\nonumber\\
N_\mrm{LR}^{++}(p,q) &= A_\mrm{LR}^{11}\,e^{-2\pi i(p-q)}\,\frac{x_p(e^{i\pi p}-e^{-i\pi p})x_q(e^{i\pi q}-e^{-i\pi q})}{(1+e^{i\pi(p-q)}x_p x_q)(1+e^{-i\pi(p-q)}x_p x_q)}\ .
\end{align}
The rest of the coefficients are related to these by
\begin{align}
A_\mrm{LR}^{21}(p,q) &= A_\mrm{LR}^{12}(p,q)\ , & A_\mrm{LR}^{22}(p,q) &= A_\mrm{LR}^{11}(p,q)\ , \nonumber\\
B_\mrm{LR}^{21}(p,q) &= -B_\mrm{LR}^{12}(p,q)\ ,&&\nonumber\\
C_\mrm{LR}^{ij}(p,q) &= \bigl(B_\mrm{LR}^{ij}(p,q)\bigr)^*\ , && \nonumber\\
D_\mrm{LR}^{2+}(p,q) &= D_\mrm{LR}^{1+}(p,q)\ , & D_\mrm{LR}^{i-}(p,q) &= \bigl(D_\mrm{LR}^{1+}(p,q)\bigr)^*\ , \nonumber\\
E_\mrm{LR}^{2+}(p,q) &= E_\mrm{LR}^{1+}(p,q)\ , & E_\mrm{LR}^{i-}(p,q) &= \bigl(E_\mrm{LR}^{1+}(p,q)\bigr)^*\ , \nonumber\\
F_\mrm{LR}^{--}(p,q) &= \bigl(F_\mrm{LR}^{++}(p,q)\bigr)^*\ , & F_\mrm{LR}^{-+}(p,q) &= \bigl(F_\mrm{LR}^{+-}(p,q)\bigr)^*\ , \nonumber\\
H_\mrm{LR}^{--}(p,q) &= \bigl(H_\mrm{LR}^{++}(p,q)\bigr)^*\ , && \nonumber\\
J_\mrm{LR}^{ab}(p,q) &= -H_\mrm{LR}^{ab}(p,q)\ ,&&\nonumber\\
K_\mrm{LR}^{+2}(p,q) &= K_\mrm{LR}^{+1}(p,q)\ , & K_\mrm{LR}^{-i}(p,q) &= \bigl(K_\mrm{LR}^{+1}(p,q)\bigr)^*\ , \nonumber\\
L_\mrm{LR}^{+2}(p,q) &= L_\mrm{LR}^{+1}(p,q)\ , & L_\mrm{LR}^{-i}(p,q) &= \bigl(L_\mrm{LR}^{+1}(p,q)\bigr)^*\ , \nonumber\\
M_\mrm{LR}^{21}(p,q) &= M_\mrm{LR}^{12}(p,q)\ , && \nonumber\\
N_\mrm{LR}^{--}(p,q) &= \bigl(N_\mrm{LR}^{++}(p,q)\bigr)^*\ .&&
\end{align}

\endgroup

\section{Bound state dispersion relation}\label{app:bound}

In this appendix we outline how to derive the dispersion relation eq.~(\ref{eq:bound_state_dispersion}). We define the variables
\begin{equation}
x_p^\pm := e^{\pm i \pi p}\,x_p\ ,
\end{equation}
such that
\begin{equation}
\frac{x_p^+}{x_p^-} = e^{2\pi i p}\ .
\end{equation}
Because of the relation
\begin{equation}
x_p - \frac{1}{x_p} = \frac{1-p}{g\,\sin(\pi p)}\ ,
\end{equation}
these variables satisfy
\begin{equation}\label{eq:pm_zhukovski_rel}
\Bigl(x_p^++\frac{1}{x_p^+}\Bigr)-\Bigl(x_p^-+\frac{1}{x_p^-}\Bigr) = \frac{2i}{g}\,(1-p)\ .
\end{equation}
Furthermore, because the dispersion relation of a single magnon is given by
\begin{equation}
\epsilon_1(p) = \sqrt{(1-p)^2 + 4g^2\,\sin^2(\pi p)} = \frac{\eta_p^2}{x_p}\,\Bigl(x_p + \frac{1}{x_p}\Bigr) = g\,\sin(\pi p)\,\Bigl(x_p + \frac{1}{x_p}\Bigr)\ ,
\end{equation}
they additionally satisfy
\begin{equation}\label{eq:pm_magnon_disp_rel}
\Bigl(x_p^+-\frac{1}{x_p^+}\Bigr)-\Bigl(x_p^--\frac{1}{x_p^-}\Bigr) = \frac{2i}{g}\,\epsilon_1(p)\ .
\end{equation}
The bound state condition eq.~(\ref{eq1}) becomes
\begin{equation}\label{eq:bound_state_condition}
x_{i+1}^+ - x_i^- = 0\ ,\qquad i=1,\dots,Q-1\ .
\end{equation}
Using eq.~(\ref{eq:bound_state_condition}) and adding the relations of eq.~(\ref{eq:pm_zhukovski_rel}) for each $p_i$, the sum telescopes and we obtain
\begin{equation} \label{6.23}
\Bigl(x_1^++\frac{1}{x_1^+}\Bigr)-\Bigl(x_Q^-+\frac{1}{x_Q^-}\Bigr)= \frac{2i}{g}\,(Q-p)\ ,
\end{equation}
where $p=\sum_{i} p_i$. Furthermore, we have
\begin{equation}
x_1^+ = e^{2\pi i p_1}\,x_1^- = e^{2\pi i p_1}\,x_2^+ = e^{2\pi i (p_1+p_2)}\,x_2^-=\dots =e^{2\pi i p}\,x_Q^-\ ,
\end{equation}
so that the relation (\ref{6.23}) becomes
\begin{equation}
e^{i\pi p}\,x_Q^- - \frac{e^{-i\pi p}}{x_Q^-} = \frac{Q-p}{g\,\sin(\pi p)}\ .
\end{equation}
By squaring both sides and adding $+4$, the left-hand-side becomes again a total square, and we deduce that
\begin{equation}
e^{i\pi p}\,x_Q^- + \frac{e^{-i\pi p}}{x_Q^-} = \frac{\sqrt{(Q-p)^2 + 4g^2\,\sin^2(\pi p)}}{g\,\sin(\pi p)}\ .
\end{equation}
On the other hand, adding up eqs.~(\ref{eq:pm_magnon_disp_rel}) for $p=p_i$ with $i=1,\ldots,Q$, we obtain upon using (\ref{eq:bound_state_condition}),
\be
\Bigl( x_1^+ - \frac{1}{x_1^+} \Bigr) -  \Bigl( x_Q^- - \frac{1}{x_Q^-} \Bigr) = \frac{2i}{g} \, \sum_{i=1}^{Q} \epsilon_1(p_i) \ ,
\ee
in a similar way to obtain
\begin{equation}
e^{i\pi p}\,x_Q^- + \frac{e^{-i\pi p}}{x_Q^-} = \frac{\sum_{i=1}^Q \epsilon_1(p_i)}{g\,\sin(\pi p)}\ .
\end{equation}
Comparing the last two equations we therefore conclude that the $Q$-bound state dispersion relation equals
\begin{equation}
\epsilon_Q(p) = \sum_{i=1}^Q \epsilon_1(p_i) = \sqrt{(Q-p)^2 + 4g^2\,\sin^2(\pi p)} \ .
\end{equation}

\end{document}